\documentclass[unsortedaddress,twocolumn,eqsecnum,aps]{revtex4}  % nice copy for distributing
\usepackage[dvips]{graphicx}
\usepackage{dcolumn}
\usepackage{bm}
\usepackage{color}
\usepackage{morefloats}
\begin{document}

% \input{invite.tex}

% -----------------------------------------------
% \input{title.tex}
% ===============================================
% FILE title.tex
% LAST MODIFIED: 18 February 2018
% ================================================
% ================================================
%              Title
% ================================================

\title{Davydov-Type Excitonic Effects on the Absorption Spectra of Parallel-Stacked
       and Herringbone Aggregates of Pentacene: Time-Dependent Density-Functional Theory
       and Time-Dependent Density-Functional Tight Binding}
\author{Ala Aldin M. H. M. Darghouth}
\email[Present address: Department of Chemistry, College of Science, University of Mosul, Iraq; e-mail: ]{ala.darghout@univ-grenoble-alpes.fr}
% \email[e-mail: ]{ala.darghout@univ-grenoble-alpes.fr}
% \altaffiliation[Present address: ]{Department of Chemistry, College of Science, University of Mosul, Mosul 41002, Iraq}
% \thanks{Present address: Department of Chemistry, College of Science, University of Mosul, Mosul 41002, Iraq}
\author{Gabriela Calinao Correa} % \cite{Ela},}
% \email[Present address: Doctoral student in the Department of Materials Science, Cornell University, Ithaca, New York; e-mail: ]{gcc73@cornell.edu}
\email[Present address: Doctoral student in the Department of Materials Science, Cornell University, USA; e-mail: ]{gcc73@cornell.edu}
% \thanks{Present address: Doctoral student in the Department of Materials Science, Cornell University, Ithaca, New York}
% \altaffiliation[Present address: ]{Doctoral student in the Department of Materials Science, Cornell University, Ithaca, New York}
\author{Sacha Juillard,}
% \email[]{sacha.juillard@univ-savoie.fr}
\email[e-mail: ]{juillard.sacha@gmail.com}
\author{Mark E. Casida,} 
\email[e-mail: ]{mark.casida@univ-grenoble-alpes.fr}
\affiliation{
Laboratoire de Chimie Th\'eorique, D\'epartement de Chimie Mol\'eculaire (DCM),
Institut de Chimie Mol\'eculaire de Grenoble (ICMG), Universit\'e Grenoble-Alpes,
301 rue de la Chimie, CS 40700, 38058 Grenoble Cedex 9, France}

\author{Alexander Humeniuk,}
\email[e-mail: ]{alexander.humeniuk@gmail.com}
\author{Roland Mitri\'c}
\email[e-mail: ]{roland.mitric@uni-wuerzburg.de}
\affiliation{
   Institut f\"ur Physikalische und Theoretische Chemie,
   Julius-Maximilians-Universit\"at W\"urzburg,
   Emil-Fischer-Stra{\ss}e 42,
   D-97074 W\"urzburg, Germany}

% \date{\today $\mbox{ }$ [Version 5.00]}

\begin{abstract}
%mec % ----------------------------------------------------------------------
%mec \begin{center}
%mec \includegraphics[width=0.95\textwidth]{./graphics/GraphicalAbstract.eps}
%mec \end{center}
%mec % ----------------------------------------------------------------------
Exciton formation leads to J-bands in solid pentacene. 
Describing these exciton bands represents a challenge
for both time-dependent (TD) density-functional theory (DFT) and for its
semi-empirical analogue, namely for TD density-functional tight binding (DFTB)
for three reasons: (i) solid pentacene and pentacene aggregates are bound only
by van der Waals forces which are notoriously difficult to describe with DFT 
and DFTB, (ii) the proper description of the long-range coupling between
molecules, needed to describe Davydov splitting, is not easy to include in
TD-DFT with traditional functionals and in TD-DFTB, and (iii) mixing may occur
between local and charge transfer excitons, which may, in turn, require special 
functionals.   We assess how far TD-DFTB has progressed towards 
a correct description of this type of exciton by including both a dispersion 
correction for the ground state and a range-separated hybrid functional for 
the excited state  and comparing the results against corresponding
TD-CAM-B3LYP/CAM-B3LYP+D3 results.  Analytic results for parallel-stacked ethylene are derived
which go beyond Kasha's exciton model [Kasha, Rawls, and {El-Bayoumi}, {\em Pure Appl. Chem.}
{\bf 11}, 371 (1965)] in that we are able to make a clear distinction between 
charge transfer and energy transfer excitons.  This is further confirmed when it
is shown that range-separated hybrids have a markedly greater effect on charge-transfer
excitons than on energy-transfer excitons in the case of parallel-stacked 
pentacenes.  TD-DFT calculations with the CAM-B3LYP functional and TD-lc-DFT
calculations lead to negligeable
excitonic corrections for the herringbone crystal structure, possibly because
of an overcorrection of charge-transfer effects.  In this case,
TD-DFT calculations with the B3LYP functional or TD-DFTB calculations parameterized
to B3LYP give the best results for excitonic corrections for the herringbone
crystal structure as judged from comparison with experimental spectra and with
Bethe-Salpeter equation calculations from the literature.
% Oddly enough Kasha's original formulation only seems to work when a range-separated
% functional is used in the case of off-set parallel pentacenes.  This is traced back
% to a lack of consideration of avoided crossings in Kasha's original formulation.  
% Our improved model based upon nearest neighbor interactions does not suffer from this
% difficulty.
\end{abstract}

\maketitle

% -----------------
% THE END 
% -----------------
% -----------------------------------------------
% \newpage
% \tableofcontents
% -----------------------------------------------
\section{Introduction}
\label{sec:intro}
% \input{intro.tex}
% \begin{verbatim}
% ================================
% File: intro.tex
% Last update: 11 January 2018
% ================================
% \end{verbatim}

Organic electronics \cite{MS90b,PS99,JWK07,A08,A09} has emerged in recent years as an important niche
market, notably for organic light emitting diodes used in lighting, television,
computer, and telephone display screens, and for organic solar cells.
Part of the appeal of organic electronics is the ease
of design of new materials via tools from the organic chemist's large
and diverse toolbox and the ease of fabrication of ``plastic'' and ``printable''
electronics.  At the heart of the functioning of organic electronic devices are 
energy-transfer (ET) and charge-transfer (CT) processes \cite{MK00} whose
understanding could benefit from better modeling.  However the size and complexity
of organic materials and the need to treat electonic excited states puts severe
limitations on the modeling methods that can be used.  These limitations become even
more severe if the goal is to model exciton dynamics or charge transport.  Even standard
methods for ``large'' systems such as density-functional theory (DFT) and time-dependent
(TD) DFT may need to be approximated by their semi-empirical analogues, notably by density-functional
tight-binding (DFTB) and TD-DFTB, in order to treat large-enough systems to be of practical interest
in organic electronics.  Moreover ``ordinary'' DFT(B) and TD-DFT(B) is not good enough
for treating organic molecular solids and excitations in these systems because of the
need for at least a minimally-correct treatment of van der Waals (vdW) forces and CT excitations.
In this article, we evaluate the ability of TD-DFTB and of state-of-the-art TD long-range
corrected (lc) DFTB to simulate the results of TD-B3LYP and TD-CAM-B3LYP calculations for describing 
exciton structure in the spectra of pentacene aggregates.  In the process, we revisit 
Kasha's exciton model \cite{KRE65}, which is often used in analyzing experimental results,
and point out some of its strengths and weaknesses.

Organic materials are typically bound together by some combination of 
hydrogen bonding and vdW forces.  In the case of pentacene, the
forces binding the molecules together are purely vdW in nature.
It is thus imperative to be able to include dispersion forces.  Traditional
density functionals, such as the local density approximation (LDA), generalized
gradient approximations (GGAs), meta-GGAs, and hybrid functionals fail to include
the ``action at a distance'' aspect of dispersion forces because of their
inability to give an accurate description of forces between molecules with
nonoverlapping densities.  Perhaps ironically, $C_6$ van der Waals coefficients --- and
hence dispersion forces --- may be calculated accurately by TD-DFT.  At this time, the most
popular way to include dispersion forces in DFT calculations is to add on
a semi-empirical correction \cite{GAEK10} which is designed to interpolate
between the DFT description of the charge density and the TD-DFT description of
$C_6$ coefficients.

Organic electronics relies upon charge transport.  However positive and 
negative carriers may be transported together in a charge-neutral packet
called an exciton.  From the condensed-matter point of view, excitons
are born as local excitations.  In fact, it is useful to make a distinction
between ``exciton structure'' and ``exciton dynamics'' (p.~5 Ref.~\cite{K63}).
Although related to each other, exciton structure is more directly related
to absorption spectra --- the subject of the present article --- while
exciton dynamics falls more conveniently under the heading of charge and
energy transport \cite{MK00}.  Even within the seemingly narrow subject of
exciton structure, excitons seem to mean different things to different people.  
In particular, solid-state physicists may seem to require periodic (crystal)
boundary conditions \cite{K63} in their definition of excitons, while chemists
\cite{K59,KRE65} and biochemists \cite{AVG00} do not.

An important distinction between solids and molecules has to do with what we
informally call the ``size of a photon.''  In the usual way of thinking, monochromatic
light has a well-defined momentum and hence, by Heisenberg's well-known uncertainty
principle, must be infinitely delocalized in space.  The theory of molecular spectroscopy
is almost (but not really) contradictory regarding the size of the photon in the sense that 
it usually assumes monochromatic light modeled by an electronic field at fixed frequency, 
but one whose interaction with the molecule is sufficiently sudden to be able to use 
Fermi's golden rule.  That is, the moleule is small compared to the size of of the photon 
but the photon interacts with the molecule for only a short period of time while it passes by.

Here we focus on molecular
solids where intermolecular interactions are important.  
Frenkel introduced the term ``excitation packets'' in his early study of the conversion of light into heat in solids \cite{F31,F32}.  
Unlike molecules which may often be considered to be small
enough compared to the size of a photon that the photon may be approximated by an 
oscillating electric field, a solid is large compared to a photon.  
In particular, solids typically only interact locally with light, say, only near the surface
where illuminated by a laser beam.  Yet 
the crystal molecular orbitals should in principle extend over the entire crystal and
so must also be large compared to the size of a photon.  
This is why the proper way to calculate the macroscopic dielectric function $\epsilon_M(\omega)$, 
including local field effects, is as
\begin{equation}
  \epsilon_M(\omega) = \lim_{{\vec q} \rightarrow 0} 
   \frac{1}{\epsilon_{00}^{-1}({\vec q}, \omega)} \, ,
\end{equation}
where $\epsilon_{{\vec G}_1,{\vec G}_2}({\vec q},0)$ is the microscopic dielectric
function.
Yet experimental observations, and indeed common sense, suggests that it is not always
the macroscopic quantity which is important because it is sometimes useful to think of 
photons as being absorbed locally microscopically by single molecules or groups of 
molecules, with energy propagating out from the microscopic region of absorption.
% Yet experimental observations, and indeed common sense, suggests that photons
% may be absorbed locally.  
How may this observation be reconciled with the
well-established concept of crystal molecular orbitals in periodic systems?
Frenkel's excitation packets resolved this 
apparent paradox by allowing the nearly degenerate crystal molecular orbitals to
form wave packets whose size is on the order of one or several molecules and so
for which photon absorption may be treated much like that of a molecule.  
This, in modern language, is
the Frenkel exciton (FR).  Another type of exciton --- the Wannier-Mott 
exciton (WM) \cite{W37} --- may
be constructed for metals and semiconductors.  
Although not critical for the present work, it should be noted that FRs
and WMs in periodic systems may be regarded as delocalized crystal states
with a high conditional probability that, having specified the position of
one charge, the other charge will then be found somewhere in the local
neighborhood.
If the electron-hole distance that emerges from the conditional probability
is small compared to the size of a molecule, then we have the case of a FR
exciton.  If this distance is comparable to the entire size of the wave function
envelope, then we have the case of a WM exciton.
The FRs and WMs form limiting cases, with real excitons being somewhere 
inbetween \cite{K63}.  Thus, for
a solid-state physicist, an exciton is a localized excitation which is small compared
to a solid.  A variation upon Frenkel excitons are Davydov excitons 
\cite{D62,D08} which will be discussed in the next paragraph.

Physical chemists and chemical physicists seem to have come across 
the exciton idea in a different way than did solid-state physicists, namely 
by noticing the appearance of new spectral features when certain dyes 
aggregate in concentrated solution.  If new very narrow peaks appear 
at lower energies, they are referred to as J-bands \cite{M95} 
(J for Jelly \cite{J36,J37} who, along with Scheibe \cite{S37,WKS11} 
were some of the first to investigate this phenomenon); if the new peaks 
appear at higher energies, they are referred to as H-bands (H for 
hypsochromic).  Kasha and coworkers were able to give a convincing 
description of the origin of these bands in terms of the
same ideas used by Davydov for solids \cite{K59,KRE65}.
In particular, local excitons on different molecules interact in 
such a way as to lead to Davydov splitting (DS) of otherwise degenerate
excitations.  It is thus important to describe, not just {\em intra}molecular
interactions correctly, but also to describe {\em inter}molecular
interactions correctly, if the goal is to model J- or H-band DS.  Several
ways to improve the description of intermolecular interactions in DFT
are available, including GGAs, global hybrids, and range-separated hybrids.

Yet another complication can arise as excitations need not be only
within a single molecule (local excitation, LE), but rather may include 
excitations transfering charge from one molecule to a nearby molecule
(charge transfer, CT).  As FRs result from interacting LEs, one might
think that CT could be ignored when modeling J- or H-bands.  However this
is not the case when CT and LEs mix, as is thought to occur in 
crystalline pentacene \cite{CSRG15}.  

A valid question is how well these may be described with modern quantum chemistry
tools.  DFT has largely supplanted the older Hartree-Fock (HF) theory, except in 
cases where HF calculations are followed up by sophisticated post-HF correlated 
calculations.  Although hybrid methods which integrate HF exchange into DFT have 
become popular, major Achilles heels of DFT have been dispersion forces and charge 
transfer phenomena.  Likewise time-dependent (TD) DFT has become the dominant 
single-determinant-based approach for describing the excited states of
medium- and large-sized molecules.  But TD-DFT inherits many of the
same problems as DFT, with a few more of its own \cite{CH12,CH15}.
Time-dependent DFT with conventional functionals is notorious for underestimating 
CT excitations.  
This problem was clearly explained by Dreuw, Weisman, and Head-Gordon
in their paper of 2003 \cite{DWH03} but was already apparent in an earlier paper
by Tozer {\em et al.} in 1999 \cite{TAH+99}.  Later several diagnostic criteria
were suggested to know when CT was likely to lead to a problem with
TD-DFT with the best known one being the $\Lambda$ criterion 
\cite{PBHT08,PT09,PSGT09,KMMS09,WWT09,DT10,LZG12}.
(Ref.~\onlinecite{M17} provides a recent review of the 
CT problem in TD-DFT.)
It should also be born in mind that CT excitations are not necessarily
handled correctly by TD-HF which may over-estimate CT excitations
as HF lacks important many-body screening effects present in more sophisticated
methods such as the $GW$ and Bethe-Salpeter equation (BSE) approaches.

The situation in solid-state physics evolved somewhat differently beginning with
the observation that the exact density functional must have some sort of ultranonlocality
if atoms in the middle of a dielectric are to feel the effect of the field induced
by charges on the surface of a dielectric.  This led, for example, to the incorporation
of current into TD-DFT \cite{VK96}.  The lack of ultranonlocality is often invoked
to explain why TD-DFT calculations do not show exciton peaks in solid argon 
\cite{SMOR07}.  Sharma {\em et al.} proposed a bootstrap appproximation to
improve TD-DFT spectra for solids \cite{SDSG11}.
Ullrich and coworkers have discussed the problem of improving
functionals for better description of excitons in crystal spectra \cite{TLU09,YU13,UY16}.
The current recommendation to avoid the underestimation of CT excitations in TD-DFT
is to use range-separated hybrids (RSHs) as these can ``meet the challenge of CT
excitations'' \cite{K17}.  In fact, in their article of 2010, Wong 
and Hsieh argued strongly for the use of RSHs for the improvement of the description
of excitons in the spectra of oligoacenes \cite{WH10}.  While encouraging, that study
is also a bit misleading in the present context because it refers to excitons within
a single covalently bonded molecule while our concern here is with excitonic effects
on the spectra of aggregates held together by vdW forces.  Nevertheless
we concur on the importance of RSHs for describing excitons.

Another advance has been in the development of optimally-tuned (OT) RSHs 
\cite{SKB09,BLS10,KKK13,K17}.
OT-RSHs improve the description of CT excitations by adjusting the range-separation
parameter so that frontier molecular orbital energies agree with ionization potentials
and electron affinities calculation in a $\Delta$SCF manner --- that is, as the
difference between the $N\pm1$- and $N$-electron self-consistent field (SCF) 
ground state energies.
They are highly recommended within their range of application.  However they
are expected to fail for large systems with delocalized states when the
$\Delta$SCF ionization potential and electron affinities go to zero.
Moreover na\"{\i}ve use of OT-RSHs can lead to discontinuities in potential
energy surfaces \cite{KKK13} which can be fatal for photochemical modeling and
especially for photochemical dynamics.

Two other approaches to describing excitonic effects with TD-DFT should be mentionned.
This is subsystem TD-DFT \cite{JN14,RP16} which grew out of a little article by
Casida and Weso{\l}owski \cite{CW04} showing how TD-DFT could be done on a
subsystem of a larger system.  The advantage of this method is that it incorporates
the ideas of the exciton model from the very begining as the system is
viewed as made up of interacting chromophores.  A different approach, albeit
incorporating the exciton model from the very beginning, is used in Ref.~\onlinecite{KVSI17}.  Note, however, that neither of these approaches are used in the 
present article.  Instead, we emphasize obtaining
excitonic effects from a supermolecule approach to vdW aggregates.
We will apply the results from our analysis to analyze the results from various
TD-DFT(B) calculations.

Density-functional tight-binding (DFTB) and TD-DFTB are semi-empirical
versions, respectively, of DFT and of TD-DFT which (as should be expected from
good approximations to DFT and to TD-DFT) inherit many of the
problems from their first principles counterparts.  
DFTB was first developed in the mid-1990s as an approximation to DFT \cite{PFK+95}.
It is now part of many, if not most, quantum chemistry packages.  TD-DFTB
was introduced in 2001 \cite{NSD+01} and has been gradually improving.
In general DFTB requires large amounts of effort in order to obtain a good
set of parameters.  This effort depends upon which density functional
is to be emulated and so must be repeated for each new functional.  The present
work uses the very recent {\sc DFTBaby} program \cite{HM15,HM17} which is 
specifically defined for TD-DFTB fewest switches surface-hopping photodynamics.
It includes both a vdW correctoin and a DFTB analogue of a RSH.  Understandably,
given the problems of OT-RSHs for calculating PESs, we have not used a DFTB versin 
of an OT-RSH with {\sc DFTBaby}.

In the interest of future (and on-going) work on large and complex systems,
our primary interest is in TD-DFTB.
The present study seeks to find out how well state-of-the-art TD-DFTB calculations
can mimic state-of-the-art TD-DFT calculations, including dispersion corrections
and RSHs, for describing excitonic effects in pentacene aggregates.  

As an important
goal is also understanding, we focus primarily on the overly simple case of
parallel stacked pentacene molecules.  However we then do go on and extend
our tests to the known herringbone structure of solid pentacene which is
an old, but still fairly popular, system in organic electronics \cite{A08}
and which is known to show J-bands.

This paper is organized as follows.  
Hartree atomic units ($\hbar=m_e=e=1$) are used throughout this paper.
A series of appendices have been included with a brief 
review of DFT (Appendix~\ref{sec:DFT}), 
TD-DFT (Appendix~\ref{sec:TD-DFT}), DFTB (Appendix~\ref{sec:DFTB}), and 
TD-DFTB (Appendix~\ref{sec:TD-DFTB}) in order to keep this article
at least somewhat self-contained.  These appendices are intended only 
to present the basic ideas of the methods used in this article in a 
relatively schematic way, but include appropriate 
references to the original literature or to important review articles
for those seeking more information.
The next section reviews a minimum
of basic theory needed for this paper.  Sec.~\ref{sec:details} presents
the details of how our computations were carried out and Sec.~\ref{sec:results}
presents and discusses our results.  Sec.~\ref{sec:conclude} concludes.

\section{Exciton Analysis}
\label{sec:theory}
% \input{theory.tex}
% \begin{verbatim}
% ==========================================
% File: theory.tex
% Last modified: 11 February 2018
% ==========================================
% \end{verbatim}

% ------------------------------------------------------
\begin{figure}
\includegraphics[width=0.5\textwidth]{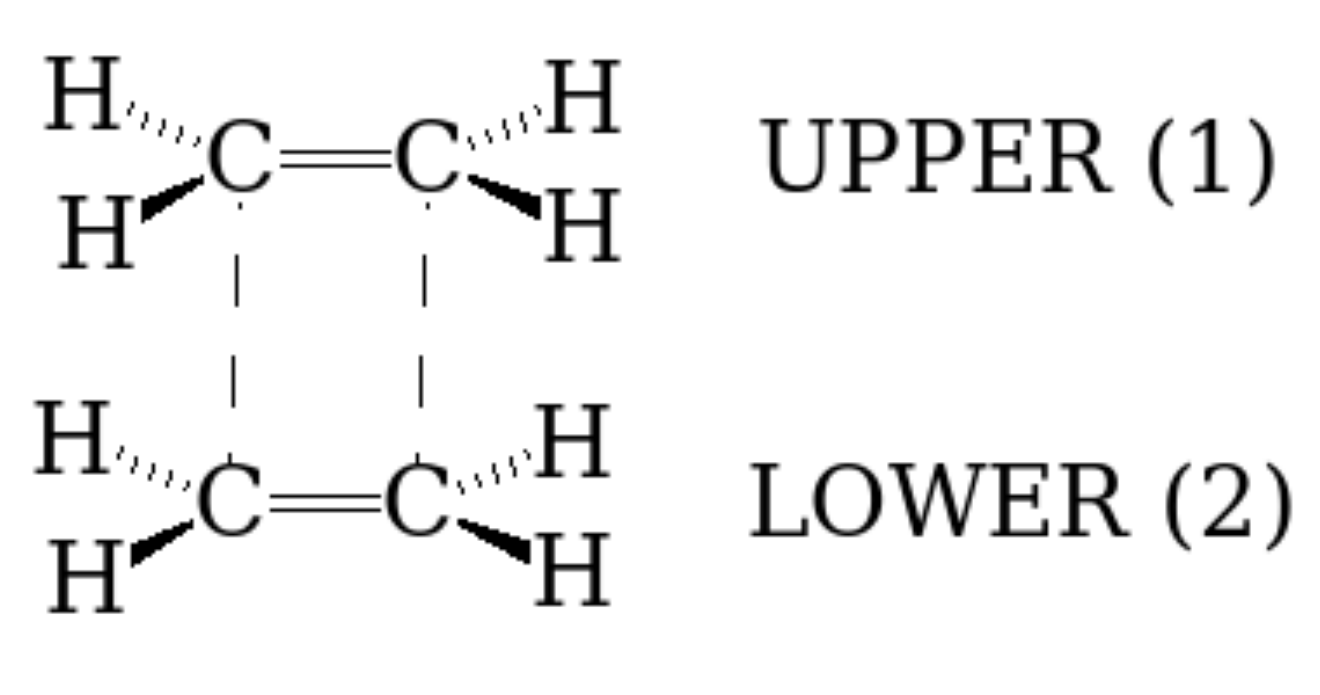}
\caption{
\label{fig:stacked_ethylene}
Two vertically-stacked ethylene molecules.
}
\end{figure}
% -------------------------------------------------------
We have noticed that there seems to be a great deal of confusion in the literature
regarding charge transfer in excitonic systems (e.g., see Ref.~\onlinecite{ZBCH09}).  
Indeed delocalization of electron density over several molecules does not necessarily imply excitonic charge transfer; what may be taken at first as an indication of charge transfer, may turn out to be a manifestation of energy transfer.
For this reason, we wish to be especially careful to define these terms within 
the context of excitonic theory and, in particular, we seek to explain via an 
algebraic example how excitations described using MOs, delocalized over 
several molecules in a supermolecule, may be analyzed and understood in terms 
of the (mainly) pairwise interaction of excitations localized on different 
molecules to create ET and CT excitons. 
In the process of this work, we shall see that single molecule spectra are 
shifted and single peaks may be split into multiplets when molecules form 
aggregates.  We will call these {\em observed multiplets} Davydov multiplets
and the {\em observed splittings} Davydov splittings (DSs).  We emphasize that
the explanation for these splittings can be different than the traditional
explanation proposed by Davydov \cite{D62} and by Kasha \cite{KRE65} as 
CT effects were not present in their models.  Note, however, that including
both ET and CT effects is in keeping with state-of-the-art
practice in the Bethe-Salpeter equation Green's function \cite{CSRG15,BDJ18} 
treatment of 
the spectra of molecular solids \cite{TNL03,CGR12,CSRG15} though our terminology
differs from theirs (their classification of an exciton as being of FR type
is analogous to our classification as ET type, while we both agree in 
referring to CT type excitons).
 
In particular, we wish to show algebraically
and using chemical intuition to what extent Kasha's exciton model \cite{KRE65}
does or does not emerge from a linear combination of 
singly-excited determinants over MOs.
For concreteness, we will treat the $\pi$ system of vertically-stacked molecules 
of ethylene (Fig.~\ref{fig:stacked_ethylene}).  
This is close enough to the case 
treated numerically in Sec.~\ref{sec:results} that we will be able to use the equations 
developed for vertically-stacked ethylene to help understand the exciton physics of 
vertically-stacked pentacenes.

% -----------------------------------------
\subsection{Kasha's Exciton Model}
% -----------------------------------------

A biography highlighting key scientific achievements of Michael Kasha (1920-2013) appeared
one year after his death \cite{DHW+14}.  It rightly notes that Kasha % , who was also the
% last doctoral student of Gilbert Newton Lewis, 
was ``a key founder of modern photophysics, photochemistry, and molecular spectroscopy in 
condensed phases'' and gives a number of examples justifying this claim \cite{DHW+14}.  
Kasha's exciton model is a venerable but still much-used model of how weak interactions between
molecules lead to new features in the spectra of aggregates.  It may also be applied to 
weakly interacting excitations between different parts of the same molecule.

Kasha's work on excitons spans a period from 1959 to 1965 \cite{K59,K63a,HK64c,MK64,KRE65}
with an isolated contribution in 1976 \cite{K76}.  It apparently began when Albert Szent-Gy\"orgyi
pointed out that some molecules phosphoresce in solid matrices but fluoresce in solution
\cite{DHW+14}.  In 1963, Kasha and Robert Oppenheimer published a translation of Alexander
Sergeevich Davydov's 1951 book {\em Theory of Molecular Excitons} from the original Russian 
into English.  Davydov had focused on the spectroscopy of molecular solids, an area for which
there were very few well-resolved spectra at the time.  Given the pre-computer epoch when
Davydov developed his ideas, he necessarily made many simplifying approximations.  
Davydov's examples focused mainly on cases where there are two molecules per unit cell, but 
his last chapter, entitled ``Excitation Calculation of States of Molecules'' included a 
discussion of what happens to the spectra of the biphenyl molecule when the quasi-independent 
excitations of the two phenyl molecules interact.  This is the theory that Kasha's group would 
develop as the exciton model \cite{KRE65} which we briefly review here.

We follow the classic 1965 paper of Kasha, Rawls, and El-Bayoumi \cite{KRE65} very closely,
albeit with some differences in notation, and consider a van der Waals (vdW) dimer of
two identical molecules which will be labeled as 1 and 2.  {\em Key approximations will be emphasized 
in itallics}.  The ground $\Psi_{1/2}^0$ and excited state $\Psi_{1/2}^I$ of the isolated molecule 
1 or 2 satisfies the electronic Schr\"odinger equation,
\begin{equation}
  {\hat H}_{1/2} \Psi_{1/2} = E_{1/2} \Psi_{1/2} \, .
  \label{eq:kasha.1}
\end{equation}
The monomer excitation energy is,
\begin{equation}
  \omega_{1/2}^I = E_{1/2}^I - E_{1/2}^0 \, .
  \label{eq:kasha.2}
\end{equation}
The dimer Hamiltonian is,
\begin{equation}
  {\hat H} = {\hat H}_1 + {\hat H}_2 + {\hat V}_{1,2} \, ,
  \label{eq:kasha.3}
\end{equation}
where ${\hat V}_{1,2}$ is the interaction potential.  As the interaction potential is assumed
small, {\em we treat the problem perturbatively using the zero-order dimer wave function},
\begin{equation}
  \Psi^0 = \Psi_1^0 \Psi_2^0 \, .
  \label{eq:kasha.4}
\end{equation}
The ground-state energy is then,
\begin{equation}
  E^0 = E_1^0 + E_2^0 + E^0_{\text{vdW}} \, ,
  \label{eq:kasha.5}
\end{equation}
where
\begin{equation}
  E^0_{\text{vdW}} = \langle \Psi_1^0 \Psi_2^0 \vert {\hat V}_{1,2} \vert \Psi_1^0 \Psi_2^0 \rangle
  \label{eq:kasha.6}
\end{equation}
is interpretted as the vdW energy binding the two monomers into a dimer.

The dimer excited-state wave function is assumed to be of the form,
\begin{equation}
  \Psi^I = C_1 \Psi_1^I \Psi_2^0 + C_2 \Psi_1^0 \Psi_2^I  \, .
  \label{eq:kasha.7}
\end{equation}
That is, we consider that only one monomer is excited at a time and that monomer is excited to its
$I$th excited state.  {\em Note that charge transfer excitations have been neglected in this
exciton model.}  It is easy to set up the small configuration interaction problem,
\begin{eqnarray}
  & &  \left[ \begin{array}{cc} A & B \\ B & A \end{array} \right] 
   \left( \begin{array}{c} C_1 \\ C_2 \end{array} \right)  =  E^I 
   \left( \begin{array}{c} C_1 \\ C_2 \end{array} \right) \nonumber \\
  & &  A  =  \langle \Psi_1^I \Psi_2^0 \vert {\hat H} \vert \Psi_1^I \Psi_2^0 \rangle
       =   \langle \Psi_1^0 \Psi_2^I \vert {\hat H} \vert \Psi_1^0 \Psi_2^I \rangle \nonumber \\
  & & B =  \langle \Psi_1^I \Psi_2^0 \vert {\hat H} \vert \Psi_1^0 \Psi_2^I \rangle
       =   \langle \Psi_1^0 \Psi_2^I \vert {\hat H} \vert \Psi_1^I \Psi_2^0 \rangle \, ,
   \label{eq:kasha.8}
\end{eqnarray}
whose solutions are,
\begin{eqnarray}
  \Psi^I_{\pm} & = & \frac{1}{\sqrt{2}} \left( \Psi_1^I \Psi_2^0 \pm \Psi_1^0 \Psi_2^I \right) 
  \nonumber \\
  E^I_{\pm} & = & A \pm B \, .
  \label{eq:kasha.9}
\end{eqnarray}
The terms $A$ and $B$ are,
\begin{eqnarray}
  & & A  =  E_1^I + E_2^0 + E^I_{\text{vdW}} \nonumber \\
  & & E^I_{\text{vdW}}  =  \langle \Psi_1^I \Psi_2^0 \vert {\hat V}_{1,2} \vert \Psi_1^I \Psi_2^0 \rangle
       \nonumber \\
  & & B  =   E_{\text{exciton splitting}} \nonumber \\
  & & E_{\text{exciton splitting}}  =  
    \langle \Psi_1^I \Psi_2^0 \vert {\hat V}_{1,2} \vert \Psi_1^0 \Psi_2^I \rangle 
  \label{eq:kasha.10}
\end{eqnarray}
Hence there are two excitation energies,
\begin{equation}
  \omega^I_{\pm} = \omega^I_1 + \left( E^I_{\text{vdW}} - E^0_{\text{vdW}} \right)
  \pm E_{\text{exciton splitting}} \, .
  \label{eq:kasha.11}
\end{equation}
The oscillator strength,
\begin{equation}
  f_{\pm}^I = \frac{\omega^I_{\pm}}{3} \vert {\vec \mu}^{I0}_1 \pm {\vec \mu}^{I0}_2 \vert^2 \, ,
  \label{eq:kasha.12}
\end{equation}
where the transition dipole moment,
\begin{eqnarray}
  {\vec \mu}^{I0}_{\pm} & = & \langle \Psi^I \vert {\vec r} \vert \Psi_0 \rangle \nonumber \\
           & = & \frac{1}{\sqrt{2}} \left( \langle \Psi_1^I \Psi_2^0 \vert {\vec r} 
           \vert \Psi_1^0 \Psi_2^0 \rangle \pm \langle \Psi_1^0 \Psi_2^I \vert {\vec r} 
           \vert \Psi_1^0 \Psi_2^0 \rangle \right) \nonumber \\
           & = & \frac{1}{\sqrt{2}} \left( {\vec \mu}^{I0}_1 \pm {\vec \mu}^{I0}_2 \right)
  \label{eq:kasha.13}
\end{eqnarray}
Therefore each peak in the monomer absorption spectrum will be split into two exciton peaks
with different intensities.  {\em Similar results emerge in the following subsections, but
in a completely different way as we analyze aggregate molecular orbitals in terms of monomer
molecular orbitals instead of deriving aggregate absorption spectra from monomer absorption
spectra.  In particular, the Davydov splitting will emerge, not as the difference between
coupled excitation energies but rather as the difference between an energy-transfer (ET) 
excitation energy and a charge-transfer (CT) excitation energy.} 

The final approximation made in Kasha's exciton model is the 
{\em point dipole/point dipole approximation},
\begin{equation}
  E_{\text{exciton splitting}} = \frac{{\vec \mu}^{I0}_1 \cdot {\vec \mu}^{I0}_2}{r^3}
  - 3 
   \frac{\left( {\vec \mu}^{I0}_1 \cdot {\vec r} \right) \left( {\vec \mu}^{I0}_2 \cdot {\vec r} \right)}{r^5} \, .
  \label{eq:kasha.14}
\end{equation}
{\em This assumes molecules separated by a large distance relative to the size of the molecules.  
It is difficult to see how this can be applied in a quantitative fashion to the most common 
application of the excition model, namely to large dye molecules self-assembled 
into aggregates in solution, where the distance assumptions are hardly valid.}  In the next
subsection, we derive a similar but different theory of exciton splitting which will be 
further justified by explicit calculations in Sec.~\ref{sec:results}.   In particular, ET
and CT excitation energies behave as expected for different variations on TD-DFT and on TD-DFTB.

\subsection{Monomer}
% -----------------------------------------

% ------------------------------------------------------
\begin{figure}
\includegraphics[width=0.5\textwidth]{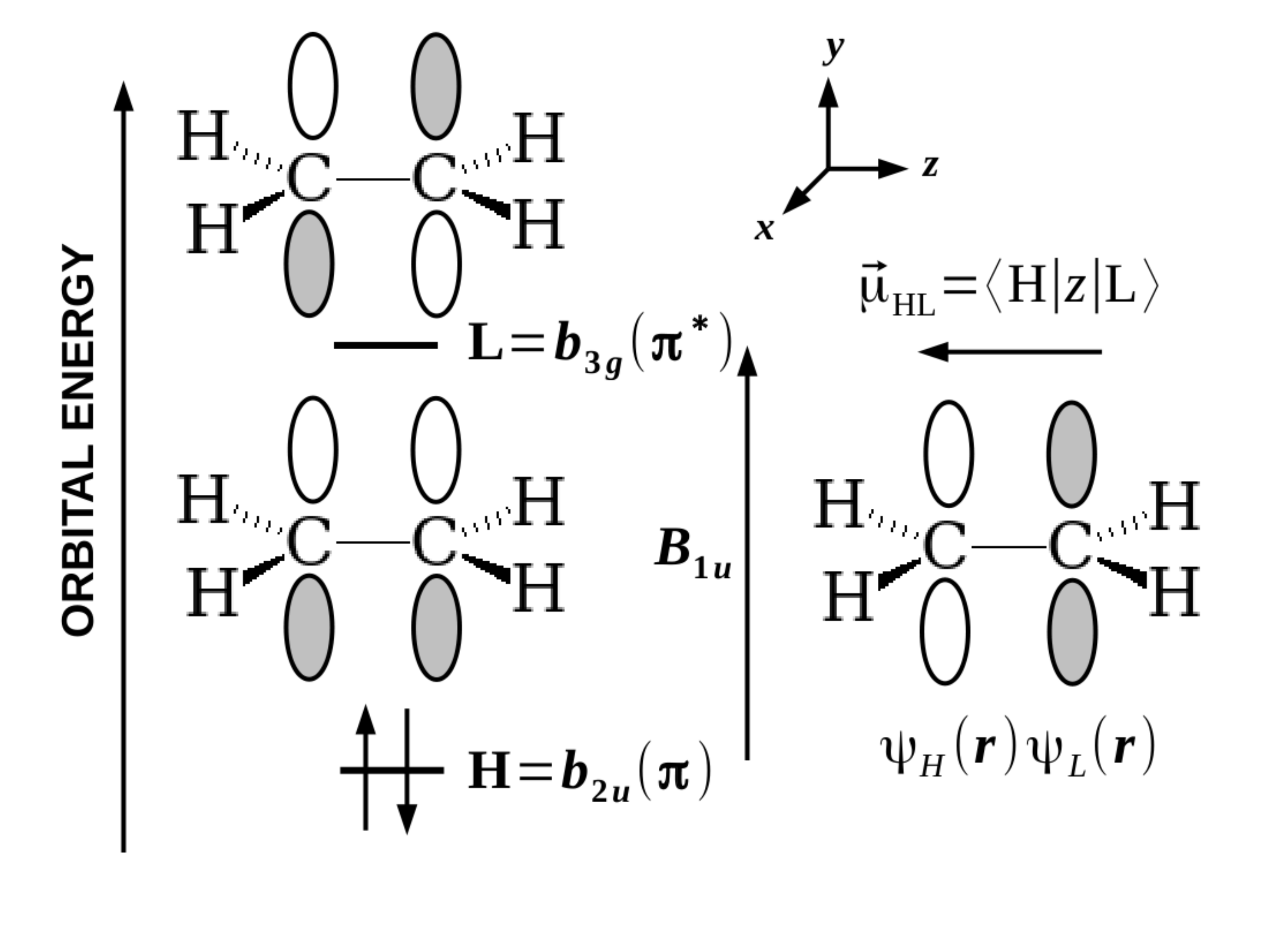}
\caption{
\label{fig:ethylene_MO}
Ethylene highest occupied molecular orbital (H) and lowest unoccupied molecular
orbital (L).
}
\end{figure}
% -------------------------------------------------------
The MOs of the $\pi$ system of ethylene are shown in Fig.~\ref{fig:ethylene_MO}.
MO symmetries have been assigned following the recommended International
Union of Pure and Applied Chemistry (IUPAC) nomenclature 
\cite{M55,M56} and the symmetry of the expected lowest energy excitations have 
been assigned.  Of particular importance for us is the sketch of
the transition density $\psi_H(\bf r) \psi_L(\bf r)$ on the right-hand side
of the figure with the associated transition dipole moment $\vec{\mu}_{HL}$.
Here H stands for the highest occupied molecular orbital, while L stands for
the lowest unoccupied molecular orbital.

% ------------------------------------------------------
\begin{figure}
\includegraphics[width=0.5\textwidth]{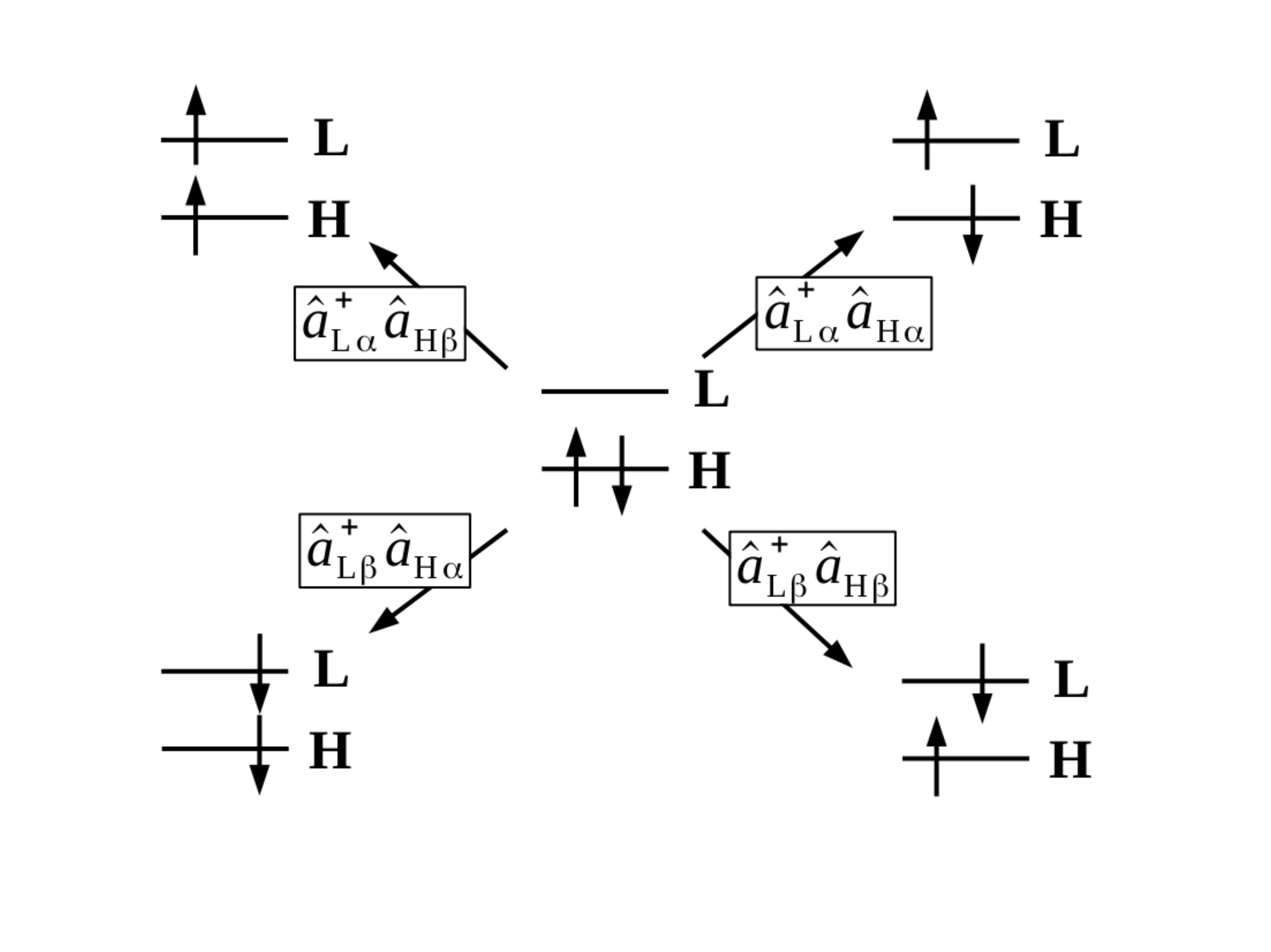}
\caption{
\label{fig:totem}
Two-orbital two-electron model (TOTEM).
}
\end{figure}
% -------------------------------------------------------
This is evidently a two-orbital two-electron model (TOTEM, 
Fig.~\ref{fig:totem}) and the excitations may be analyzed in this context.
There are four possible one-electron excitations for the TOTEM,
but spin symmetry must be taken properly into account.
We shall focus on the singlet transition which goes from the ground-state 
determinant $\Phi$ to the state,
\begin{equation}
  ^1(H,L) = \frac{1}{\sqrt{2}} \left( {\hat a}_{L\alpha}^\dagger {\hat a}_{H\alpha}
       + {\hat a}_{L\beta}^\dagger {\hat a}_{H\beta} \right) \Phi \, ,
  \label{eq:theory.1}
\end{equation}
where $\alpha$ and $\beta$ refer to spin states (i.e., spin up and spin down, 
respectively) and ${\hat a}_{L\alpha/\beta}^\dagger {\hat a}_{H\alpha\beta}$ is
an operator in second quantization formalism meaning to remove one electron
from $H\alpha/\beta$ and place it into $L\alpha/\beta$. In the specific case of the TOTEM, we may just write
\begin{eqnarray}
  ^1(H,L) & = & \frac{1}{\sqrt{2}} \left( \vert H , {\bar L} \vert +
                \vert L , {\bar H} \vert \right) \nonumber \\
          & = & \left[ \frac{1}{\sqrt{2}} \left( H(1) L(2) + L(1) H(2) \right) \right] 
                 \nonumber \\
          & \times &
                \left[ \frac{1}{\sqrt{2}} \left( \alpha(1) \beta(2) 
                 - \beta(1) \alpha(2) \right) \right] \, .
  \label{eq:theory.2}
\end{eqnarray}
There are also three triplet states which are degenerate in the absence
of spin-orbit coupling,
\begin{eqnarray}
  ^3(H,L)_{M_S=+1} & = &  \vert H , L \vert \nonumber \\
        & = & 
           \left[ \frac{1}{\sqrt{2}} \left( H(1) L(2) - L(1) H(2) \right) \right] 
                 \nonumber \\
          & \times &
                \left[ \alpha(1) \alpha(2) \right] \nonumber \\
  ^3(H,L)_{M_S=0} & = & \frac{1}{\sqrt{2}} \left( \vert H , \bar{L} \vert - \vert L , \bar{H} \vert \right) \nonumber \\
         & = &  
           \left[ \frac{1}{\sqrt{2}} \left( H(1) L(2) - L(1) H(2) \right) \right] 
                 \nonumber \\
          & \times &
                \left[ \frac{1}{\sqrt{2}} \left( \alpha(1) \beta(2) 
                 + \beta(1) \alpha(2) \right) \right] \nonumber \\
  ^3(H,L)_{M_S=-1} & = & \vert \bar{H} , \bar{L} \vert \nonumber \\
          & = &
           \left[ \frac{1}{\sqrt{2}} \left( H(1) L(2) - L(1) H(2) \right) \right] 
                 \nonumber \\
          & \times &
                \left[ \beta(1) \beta(2) \right] \, ,
  \label{eq:theory.3}
\end{eqnarray}
which will not concern us here.

% -----------------------------------------
\subsection{Dimer}
% -----------------------------------------

% ------------------------------------------------------
\begin{figure}
\includegraphics[width=0.5\textwidth]{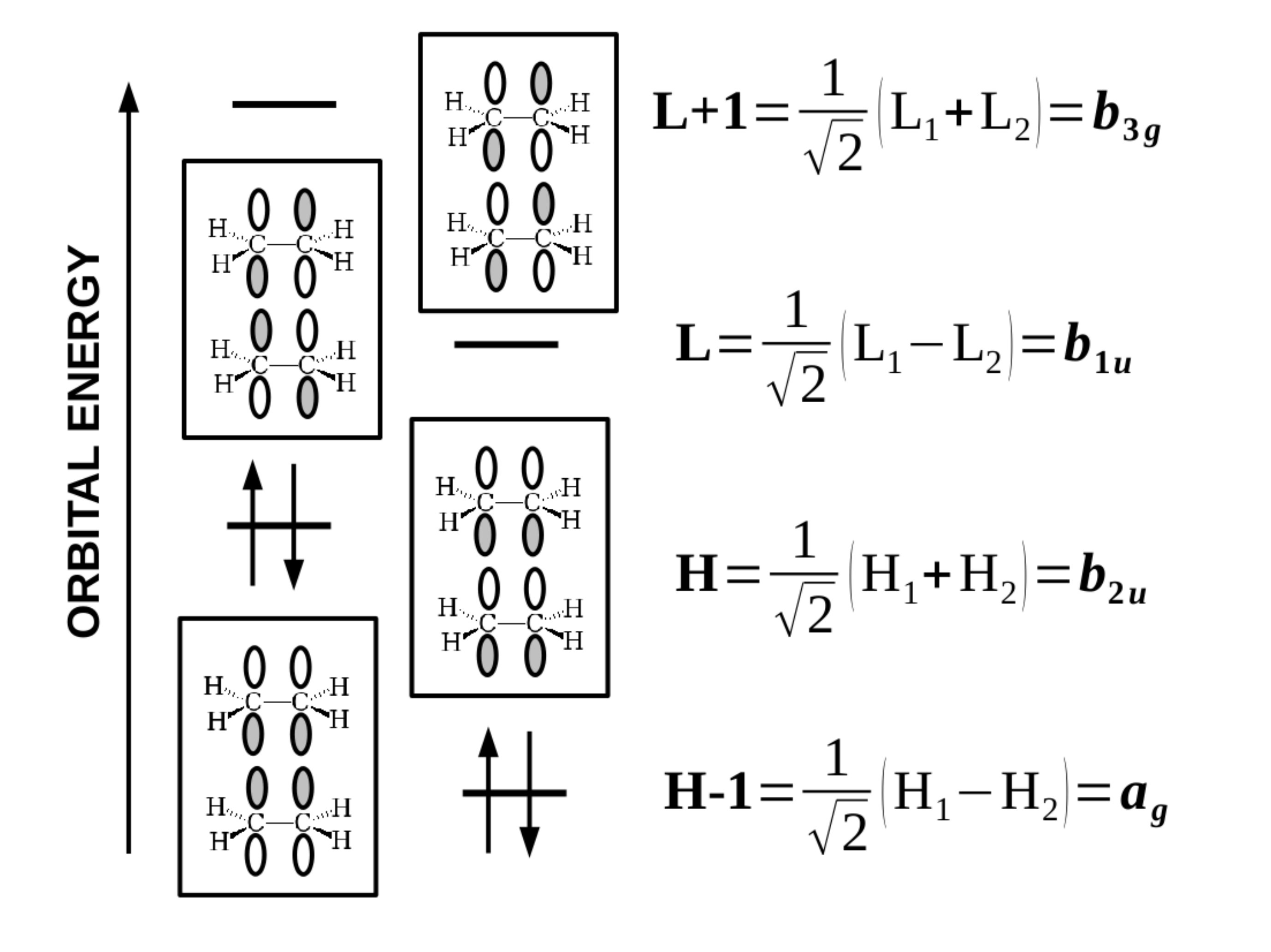}
\caption{
\label{fig:stacked_MOs}
MO diagram for two stacked ethylene molecules.  White indicates the positive phase
parts of the $p$ functions while grey indicates the negative phase parts.
Overlap between the MOs on different molecules have been neglected in
normalizing the supermolecule MOs.
}
\end{figure}
% -------------------------------------------------------
We are now ready to treat two interacting stacked ethylene molecules.
This system has been studied previously in the context of excitonic
effects \cite{KP63,SG94,HSG94,SHG95} and at a greater level of sophistication
than that needed here.  Instead, we try to keep our analysis as simple as 
possible by assuming weak interactions between the molecules so that we may 
go to trimers and oligomers.  Thus the analysis in the present section is most
correct only at large intermolecular distances.

The corresponding dimer MO diagram (Fig.~\ref{fig:stacked_MOs}) 
under the assumption of weak interactions between the molecules.
Here, after ordering MOs by energy, H$-n$ is the $n$th occupied MO
below H and L$+n$ is the $n$th unoccupied MO above L.
As expected the number of nodal planes also increases with MO
energy.  Although we might think of this as a four-orbital
four-electron model, we would like to think in terms of the 
exciton model, which we shall refer to as (TOTEM)$^2$ 
for evident reasons.  Both energy transfer (ET) and charge
transfer (CT) excitons will emerge from our analysis.

% ------------------------------------------------------
\begin{figure}
\includegraphics[width=0.5\textwidth]{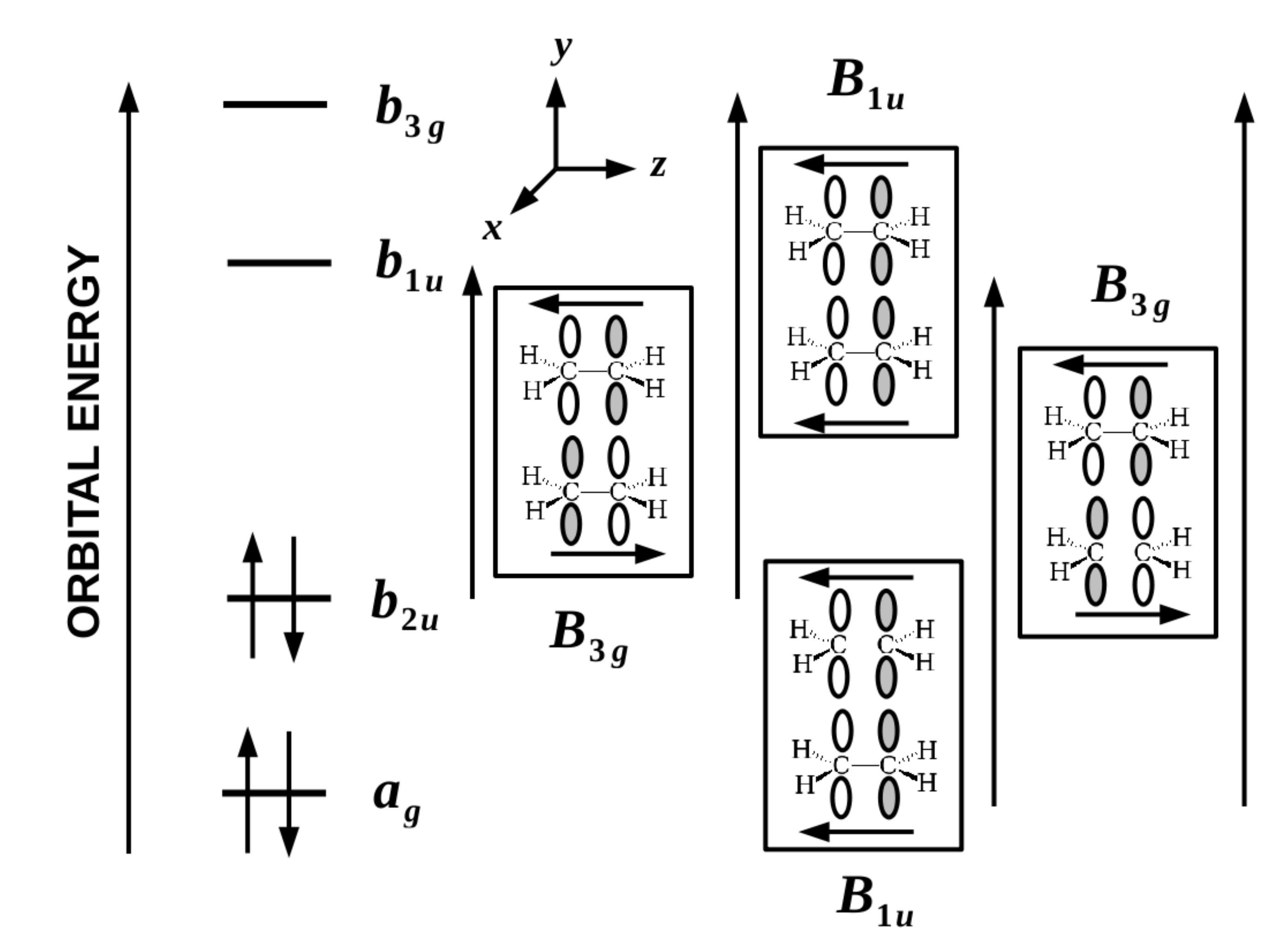}
\caption{
\label{fig:stacked_transitions}
The four transitions in (TOTEM)$^2$ and their associated
transition dipole moments.
}
\end{figure}
% -------------------------------------------------------
Figure~\ref{fig:stacked_transitions} shows the four possible singlet
transitions in (TOTEM)$^2$ from the point of view of the
MOs of the supermolecule composed of the two weakly-interacting
ethylene molecules.  Exciton analysis means that we want to re-express
the description of the excitations so that they are no longer expressed
in terms of the MOs of the supermolecule but rather are expressed in terms
of ET and CT excitons involving the MOs (H$_1$ and L$_1$) of molecule 1 and the
MOs (H$_2$ and L$_2$) of molecule 2.  Figure~\ref{fig:stacked_transitions}
shows that the transitions divide neatly into two symmetry types, namely
$B_{1u}$ and $B_{3g}$.  This simplifies our analysis as only orbitals
of the same symmetry may mix.  

% ------------------------------------------------------
\begin{figure}
\includegraphics[width=0.5\textwidth]{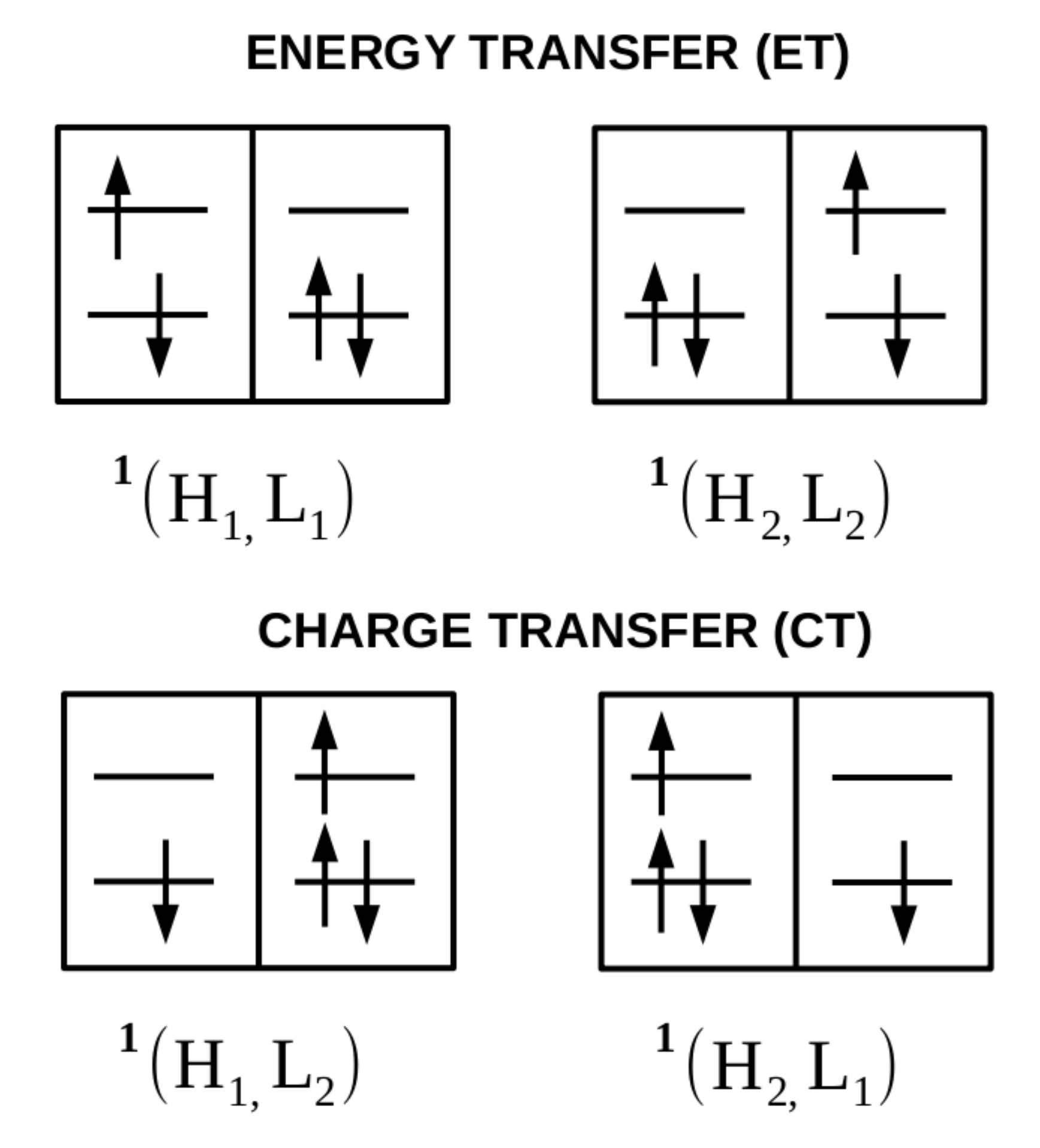}
\caption{
\label{fig:exciton_model}
Exciton model classification of transitions in the 
(TOTEM)$^2$ model.  In each double box, the left hand side shows the
orbital occupancy of the MOs in molecule 1 while the right hand
side shows the orbital occupancy of the MOs in molecule 2.
}
\end{figure}
% -------------------------------------------------------
Physically re-expressing supermolecule excitations in terms of ET and
CT excitons on individual molecules can only happen when
there are enough degrees of liberty --- and, in particular, quasidegenerate
states --- that delocalized orbitals can be re-expressed in terms of more
localized orbitals.  This does not happen for the $B_{3g}$ transitions. 
In fact, the first $B_{3g}$ transition is expected to be heavily dominated 
by the $^1(H,L)$ configuration and the remaining $B_{3g}$ transition
should be dominated by the $^1(H-1,L+1)$ configuration.  However the
$B_{3g}$ transitions are spectroscopically dark.  So we will just go 
directly on to the $B_{1u}$ orbitals.

\begin{figure}
\includegraphics[width=0.5\textwidth]{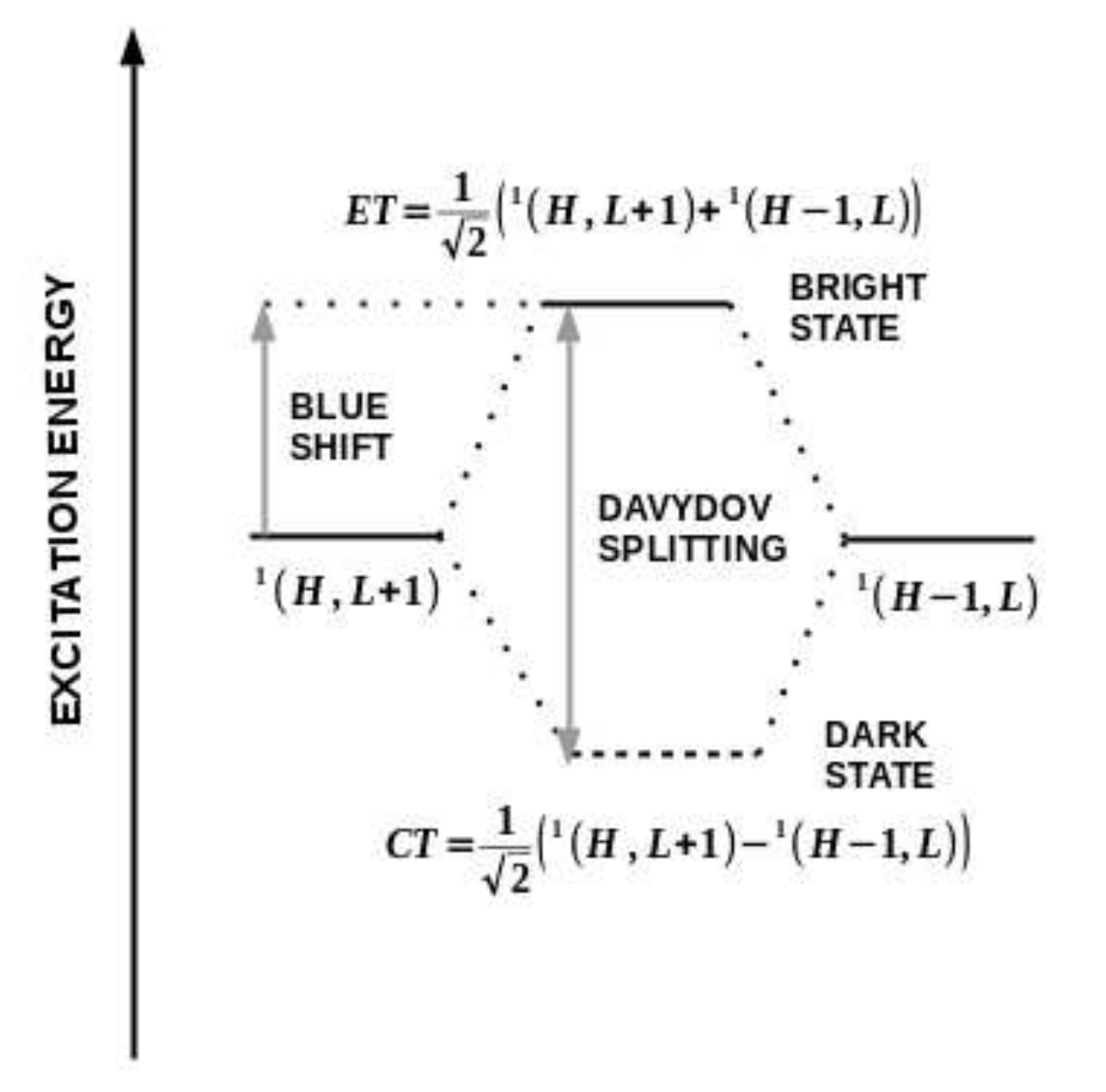}
\caption{
\label{fig:Kasha}
Schematic of the present theory for two parallel stacked molecules.
}
\end{figure}

The most general $B_{1u}$ transition is of the form,
\begin{equation}
  \Psi = c_1 \, ^1(H,L+1) + c_2 \, ^1(H-1,L) \, .
  \label{eq:theory.4.1}
\end{equation}
Expressing the aggregate MOs in terms of the local MOs of molecules
1 and 2 as given in Fig.~\ref{fig:stacked_MOs} leads to,
\begin{equation}
  \Psi=\frac{c_1+c_2}{\sqrt{2}} \mbox{ET}_{12} 
      +\frac{c_1-c_2}{\sqrt{2}} \mbox{CT}_{12} \, ,
  \label{eq:theory.4.2}
\end{equation}
where,
\begin{equation}
  \mbox{ET}_{12} = \frac{1}{\sqrt{2}} \left[ ^1(H_1,L_1) + ^1(H_2,L_2) \right]
  \label{eq:theory.4.3}
\end{equation}
is the pairwise ET exciton and,
\begin{equation}
  \mbox{CT}_{12} = \frac{1}{\sqrt{2}} \left[ ^1(H_1,L_2) + ^1(H_2,L_1) \right]
  \label{eq:theory.4.4}
\end{equation}
is the corresponding CT exciton.  (See Fig.~\ref{fig:exciton_model}. 
Note that we make no attempt to distinguish between F\"orster and Dexter ET excitons.)  
Physically we expect the $^1(H-1,L)$ and $^1(H,L+1)$ transitions to be
quasidegenerate (i.e., $c_1 \approx c_2$) as often happens in organic 
molecules with a conjugated $\pi$ system. 
Kasha's theory is recovered for exact degeneracy 
(i.e., when $c_1=\pm c_2=1/\sqrt{2}$).
More specifically, we would get a single ET peak corresponding to the 
bright state in Kasha's theory, but we do not see the dark ET peak
in our analysis because it has $B_{3g}$ symmetry.  Note however that
the $B_{1u}$ CT peak is likely to be less bright in practice than
the ET peak so that we will see something like the situation shown in 
Fig.~\ref{fig:Kasha}.  We refer to the difference of the ET and CT 
states as the Davydov splitting (DS = ET - CT).

An important aspect of applications of Kasha's theory is that it
can tell us something about the mutual orientation of molecules.
Thus Kasha's theory may be further developed to show a red shift
upon the formation of head-to-tail dimers when the CT state is bright and the 
ET state is dark \cite{KRE65}.  Other configurations yield two peaks with an 
experimentally observable DS whose relative intensities may be analyzed to 
give information about the relative orientation of the molecules in the 
dimer \cite{KRE65}.  It will be of interest once we have identified
the true nature of the exciton peaks to apply Kasha's na\"{\i}ve
theory {\em as if only ET excitonic effects were present} because this
remains a common way to analyze some experiments.  We will return to this
point at the end of Sec.~\ref{sec:results} by applying Kasha's na\"{\i}ve
theory to our calculated TD-DFT and TD-DFTB spectra.

% ------------------------------------------------------
% -------------------------------------------------------

% -----------------------------------------
\subsection{Trimer}
% -----------------------------------------

% ------------------------------------------------------
\begin{figure}
\includegraphics[width=0.5\textwidth]{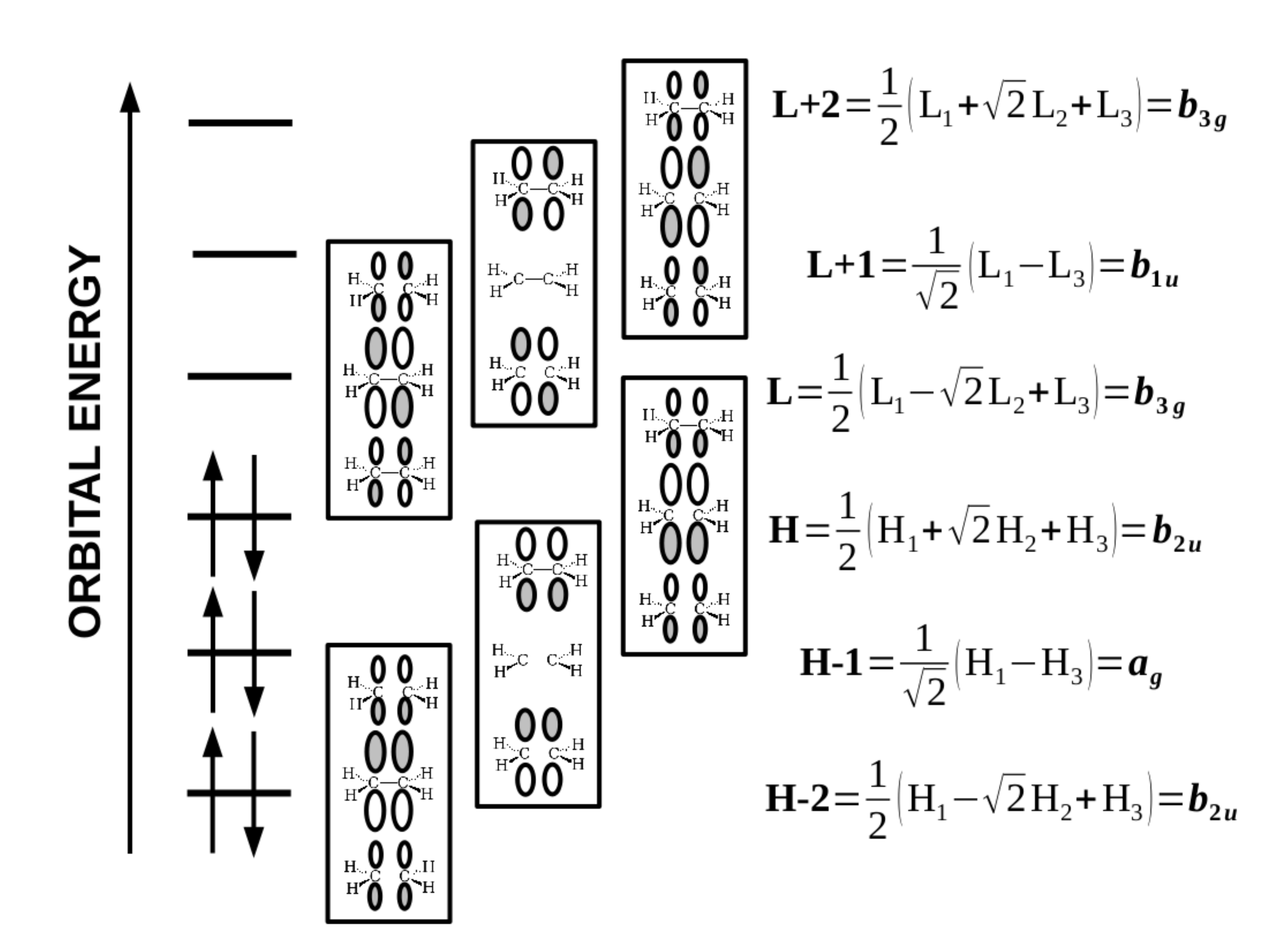}
\caption{
\label{fig:stacked_trimer_MOs}
MO diagram for three stacked ethylene molecules.  
White indicates the positive phase
parts of the $p$ functions while grey indicates the negative phase parts.
Overlap between the MOs on different molecules have been neglected in
normalizing the supermolecule MOs.
}
\end{figure}
% -------------------------------------------------------
Three stacked trimers introduce another key level of complexity in the
exciton model.  We now have an interior molecule interacting with two 
outer molecules.  This asymmetry means that transitions forbidden, and hence
dark, in the dimer may now be allowed, and hence bright, in the trimer.
Figure~\ref{fig:stacked_trimer_MOs} shows the (TOTEM)$^3$ MOs deduced by
analogy with the simple H\"uckel solution for the propenyl radical. 

% ------------------------------------------------------
\begin{figure}
\includegraphics[width=0.5\textwidth]{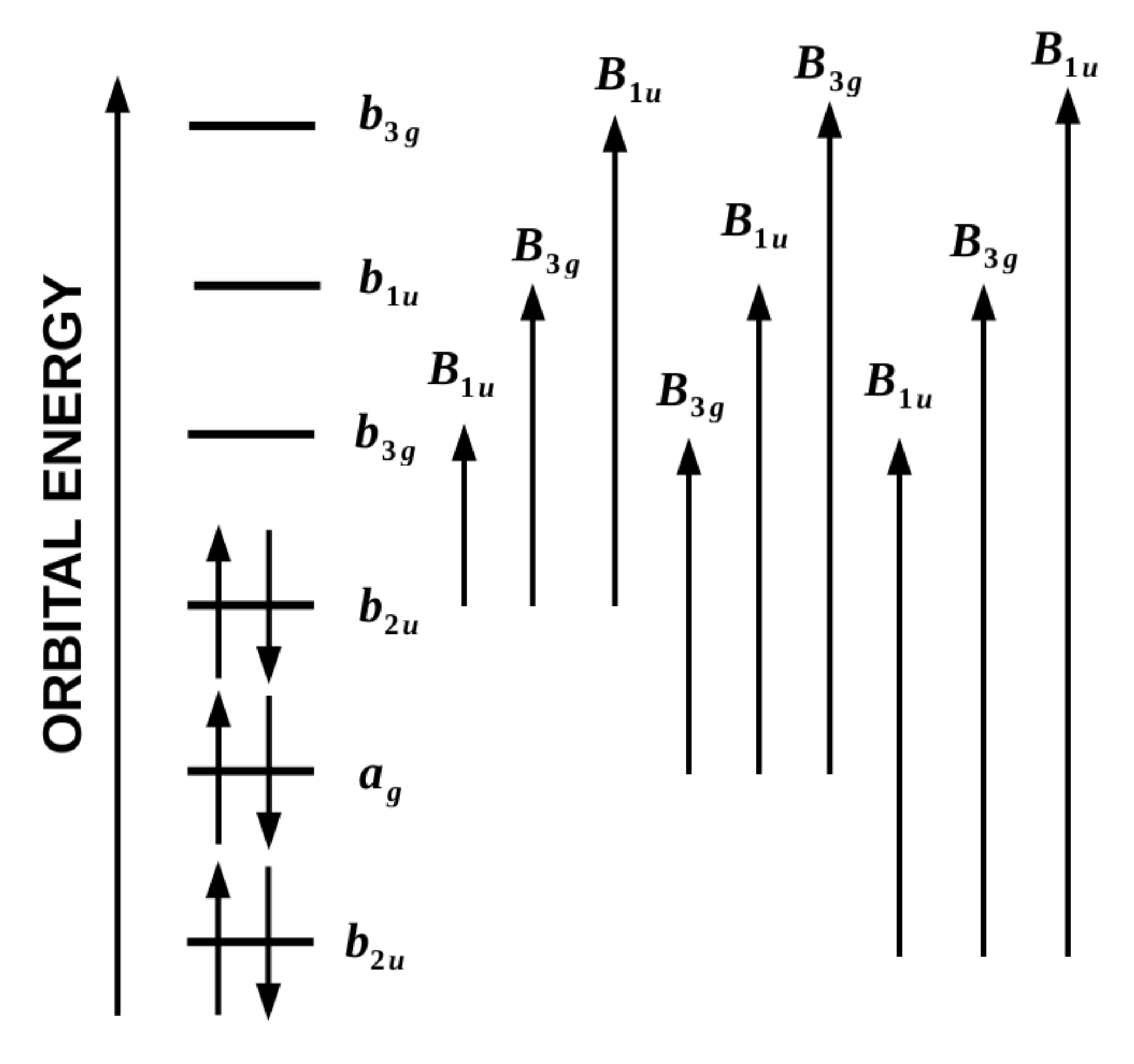}
\caption{
\label{fig:stacked_trimer_transitions}
The nine single excitations for three stacked ethylene molecules
along with their symmetry assignments.
}
\end{figure}
% -------------------------------------------------------
% ------------------------------------------------------
\begin{figure}
\includegraphics[width=0.5\textwidth]{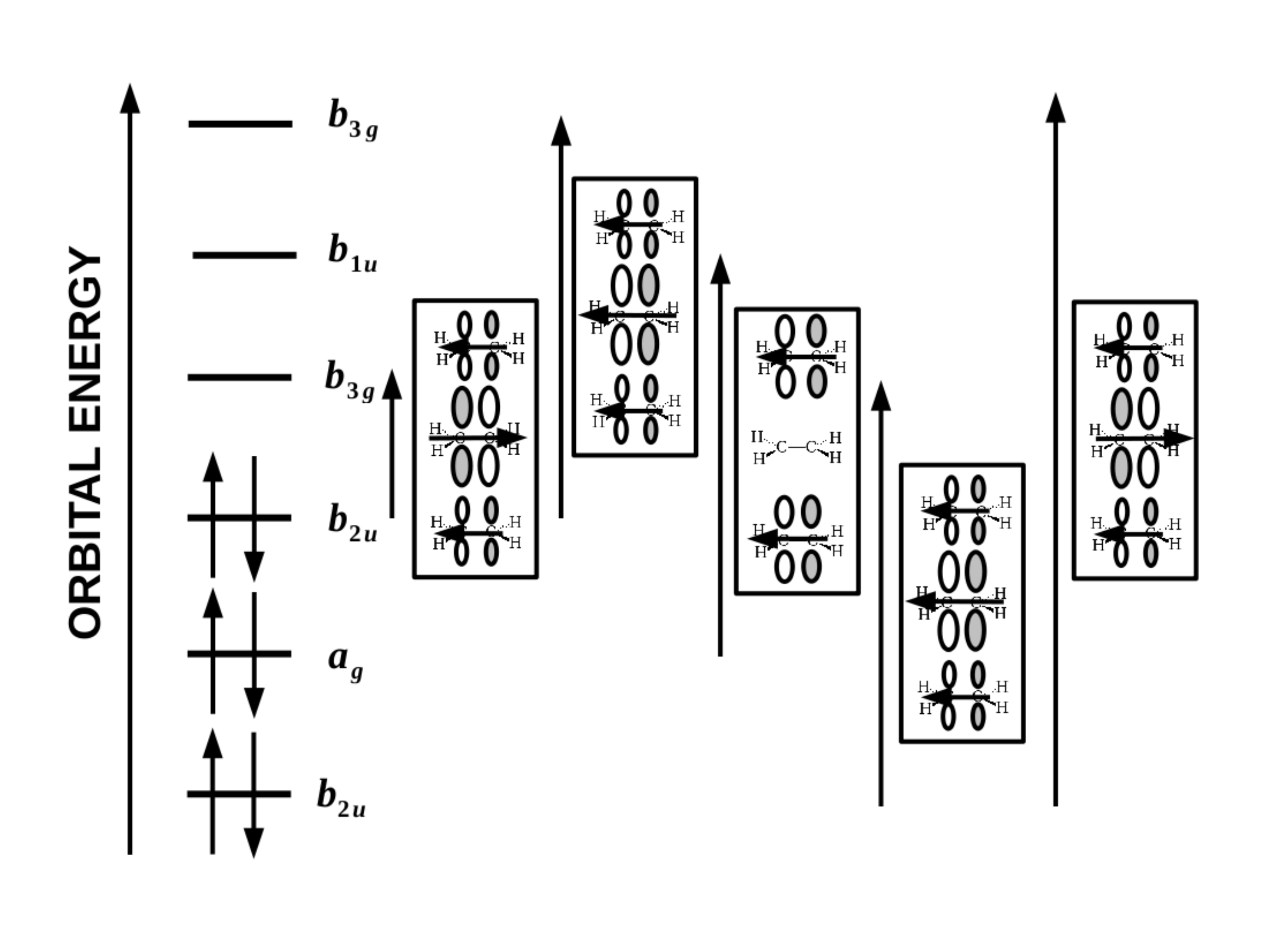}
\caption{
\label{fig:stacked_trimer_allowed}
The transition densities for the five symmetry allowed $B_{1u}$
singlet excitations.
}
\end{figure}
% -------------------------------------------------------
Figure~\ref{fig:stacked_trimer_transitions} show the nine single excitations.
Only the five $B_{1u}$ transitions are symmetry allowed for absorption 
spectroscopy.   Figure~\ref{fig:stacked_trimer_allowed} shows the 
transition densities for the five symmetry-allowed singlet transitions. 
We thus restrict our analysis to the linear combination
of only these states,
\begin{eqnarray}
  \Psi & = & c_1 \, ^1(H,L) + c_2 \, ^1(H,L+2) 
    +  c_3 \, ^1(H-1,L+1) \nonumber \\
  & + & c_4 \, ^1(H-2,L) + c_5 \, ^1(H-2,L+2) \, .
  \label{eq:theory.5}
\end{eqnarray}
Furthermore, we will use chemical intuition to predict the general 
form of these five allowed transitions.  In particular, only the 
$^1(H,L+2)$, $^1(H-1,H+1)$, and $^1(H-2,L)$ states are expected to 
be degenerate enough to mix to form pairwise ET and CT excitons.
This gives, after some algebra,
\begin{eqnarray}
  \Psi & = & c_1 \, ^1(H,L) \nonumber \\
        & + & \frac{3c_2+2c_3+3c_4}{2\sqrt{6}} \left( \frac{\mbox{ET}_{12} + \mbox{ET}_{23}}{\sqrt{3}} \right)
        \nonumber \\
        & + & \sqrt{\frac{2}{3}}c_3 \left\{ \frac{\sqrt{3}}{2} \left[ \mbox{ET}_{13}-\frac{1}{3}\left(\mbox{ET}_{12} + \mbox{ET}_{23} \right) \right] \right\}  \nonumber \\
        & + & \frac{c_2-c_4}{\sqrt{2}} \left( \frac{\mbox{CT}_{12} 
         +  \mbox{CT}_{23}}{\sqrt{2}} \right) \nonumber \\
        & + & \frac{c_2-2 c_3 +c_4}{2\sqrt{2}} \mbox{CT}_{13}
       \nonumber \\
        & + & c_5 \, ^1(H-2,L+2) \, ,
  \label{eq:theory.5b}
\end{eqnarray}
where the notation is an obvious generalization of that given in 
Eqs.~(\ref{eq:theory.4.3}) and (\ref{eq:theory.4.4}) and where the states
have been orthonormalized.  To a first approximation,
the $^1(H,L)$ is too low in energy to mix with the other terms and $^1(H-2,L+2)$
is too high in energy to mix with the other terms.  The ET and CT excitons
lie in-between these in energy.  Notice, however, that the CT terms vanish
if $c_3=c_4=c_5$ which is expected to be often approximately the case.
Note also that the pairwise ET terms are {\em not} orthogonal to each other as,
for example, $\langle \mbox{ET}_{12} \vert \mbox{ET}_{23} \rangle =
\langle \, ^1(H_2,L_2) \vert \, ^1(H_2,L_2) \rangle / 2 = 1/2$.  However the
ET terms have been grouped to reflect the symmetry of the stack and the 
distance over which energy must be transfered.

As we shall see numerically in Sec.~\ref{sec:results} for the trimer of stacked 
pentacene molecules, each term is a reasonably good first approximation to 
a calculated TD-DFT excitation.  Notice how the present model 
differs from Kasha's orginal model: 
(i) supermolecule MO excitations, such as $^1(H,L)$, which are better 
described in terms of supermolecule MOs than in terms of local MOs and (ii)
CT excitons are only expected to cancel approximately in practical calculations.
In reality ET and CT terms should mix.

The stacked trimer represents the simplest model where one molecule is interacting
with two surrounding molecules.  As such, it captures the basic physics of exciton
interactions between neighboring molecules ($1 \leftrightarrow 2$ and $2 \leftrightarrow 3$).
Equation~(\ref{eq:theory.5b}) shows that neglect of $1 \leftrightarrow 3$ interactions (i.e.,
CT$_{13}$ and ET$_{13}$ and their required orthogonalization to the other terms) leads to
only a single Davydov splitting into one ET and one CT peak.  However a more careful
analysis should include $1 \leftrightarrow 3$ interactions and Davydov multiplets may
also be expected to be observed.

% -----------------------------------------
\subsection{Higher Oligomers}
% -----------------------------------------

The extension of these ideas to (TOTEM)$^N$ for $N > 2$ is in principle
straightforward but becomes increasingly complicated.  However, it does
not seem unreasonable to expect the structure of the spectrum to stabilize
after a few layers, because the dominant interactions are expected to be
primarily only between adjacent molecules.  Thus we may anticipate that the 
numerical results in Sec.~\ref{sec:results} should already show most of 
the qualitatively important features when $N=3$ that are seen for still
larger values of $N$.

% ------------------------------------------------------
\begin{figure}
\includegraphics[width=0.5\textwidth]{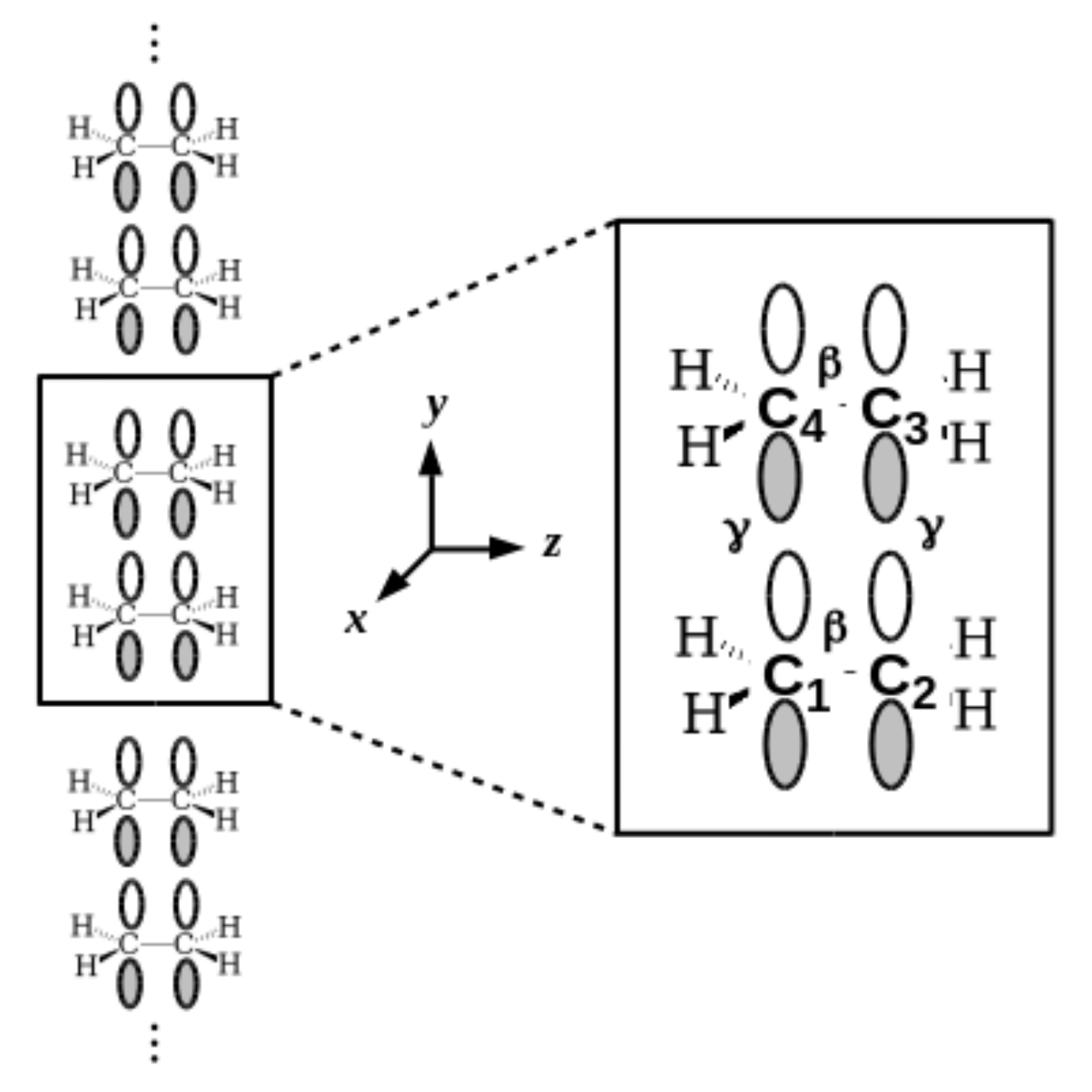}
\caption{
\label{fig:periodic_model}
Periodic model labeling and hopping parameters used for our stacked ethylene
tight-binding calculation.
}
\end{figure}
% -------------------------------------------------------
We may explore this further by a back of the envelope tight-binding calculation
for the periodic system of stacked ethylenes shown in Fig.~\ref{fig:periodic_model}.
This is basically just a periodic simple H\"uckel calculation and so should be
largely familiar to Quantum Chemists, even if the precise language and periodic
symmetry adapted linear combinations may take a little getting used to.

To carry out our tight-binding calculation, we must include two ethylene molecules 
in the unit cell.  In the exciton
model, the MOs of each ethylene molecule $\chi_\mu(\vec{r})$ are looked on much like 
local AOs (LAOs).  Combining them gives us a set of (TOTEM)$^2$ MOs which become 
local MOs (LMOs) 
\begin{equation}
  \psi_i(\vec{r}) = \sum_\mu \chi_\mu({\vec r}) c_{\mu,i} \, .
   \label{eq:theory.6}
\end{equation}
Periodic symmetry-adapted linear combinations have the form of crystal MOs (CMOs) 
\begin{equation}
  \psi_i(\vec{r};\vec{k}) = \sum_{\vec{R}} \psi_i(\vec{r}-\vec{R}) e^{i\vec{k}\cdot \vec{R}} \, ,
  \label{eq:theory.7}
\end{equation}
which may also be written as,
\begin{equation}
  \psi_i({\vec r};{\vec k}) = \sum_\mu \chi_\mu({\vec r};\vec{k}) c_{\mu,i}({\vec k}) \, ,
  \label{eq:theory.8}
\end{equation}
in terms of crystal AOs (CAOs),
\begin{equation}
  \chi_\mu({\vec r};{\vec k}) = \frac{1}{\sqrt{N}} \chi_\mu({\vec r}-{\vec R}) e^{i {\vec k} \cdot {\vec R}}
  \, .
  \label{eq:theory.9}
\end{equation}
The factor $N$ in this formalism represents the number of atoms in a fictitious ``finite crystal.''
It has been introduced for convenience, but it not really necessary.
The wave vector ${\vec k}$ serves both as a symmetry label and may also be viewed as a sort of electron
momentum which can be used in selection rules.  The $\vec{k}$-block of the CMO matrix equation is
\begin{equation}
  {\bf h}({\vec k}) \vec{c}_i({\vec k}) = \epsilon_i({\vec k}) {\bf s}({\vec k}) 
   \vec{c}_i({\vec k}) \, ,
  \label{eq:theory.10}
\end{equation}
where the matrix elements of the overlap matrix are given by,
\begin{eqnarray}
  s_{\mu,\nu}({\vec k}) & = & \sum_{\vec R} s^{({\vec R})}_{\mu,\nu} e^{i {\vec k} \cdot {\vec R}} 
  \nonumber \\
  s^{({\vec R})}_{\mu,\nu} & = & \int_{V=N\Omega} \chi^*_\mu({\vec r}+{\vec R}) \chi_\nu({\vec r}) \, d{\vec r}
  \, ,
  \label{eq:theory.11}
\end{eqnarray}
and the matrix elements of the hamiltonian matrix are given by,
\begin{eqnarray}
  h_{\mu,\nu}({\vec k}) & = & \sum_{\vec R} h^{({\vec R})}_{\mu,\nu} e^{i {\vec k} \cdot {\vec R}} 
  \nonumber \\
  h^{({\vec R})}_{\mu,\nu} & = & \int_{V=N\Omega} \chi^*_\mu({\vec r}+{\vec R}) \hat{h} \chi_\nu({\vec r}) \, d{\vec r}
  \, .
  \label{eq:theory.11b}
\end{eqnarray}
Here $\Omega$ is the volume of the unit cell and $N\Omega$ is the volume of the fictitious ``finite 
crystal.''

% ------------------------------------------------------
\begin{figure}
\includegraphics[width=0.5\textwidth]{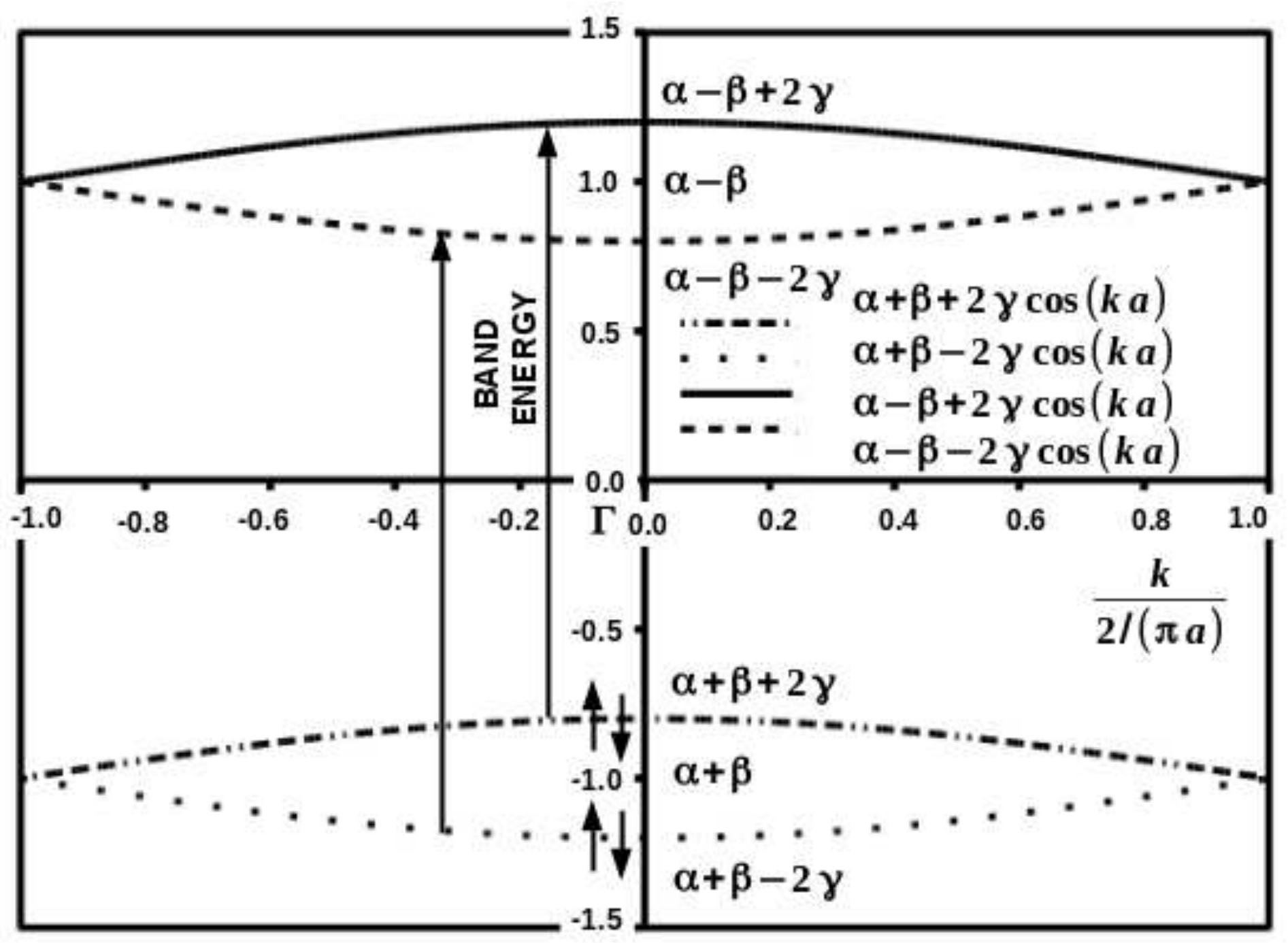}
\caption{
\label{fig:TBbands}
Bands for the stacked ethylene tight-binding model with $\alpha = 0$, $\beta = -1$,
and $\gamma = 0.1$.  The paired arrows ($\uparrow \downarrow$) are meant to
indicate filled bands below the fermi level.
}
\end{figure}
\begin{figure}
\includegraphics[width=0.5\textwidth]{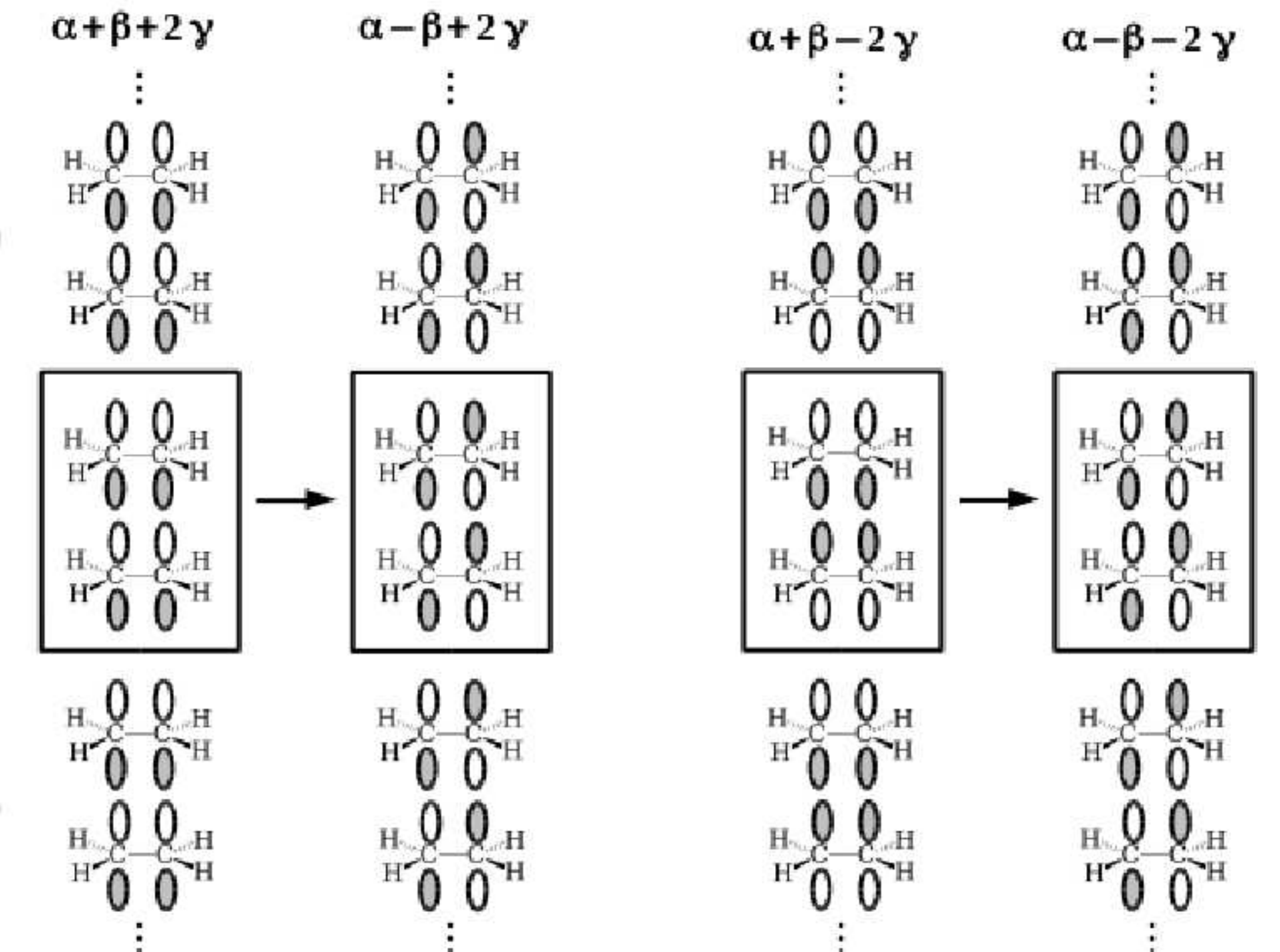}
\caption{
\label{fig:CMOs}
The CMOs for the optically-allowed transitions of (TOTEM)$^N$ at the $\Gamma$-point.
}
\end{figure}

% -------------------------------------------------------
Our model is subject to several simplifications.  For one, the wave vector is a number $k$ 
since our system is periodic in a single dimension ($y$). We will follow the common practice
of assuming that the overlap matrix ${\bf s}(k)$ is the identity.  The hamiltonian matrix ${\bf h}(k)$
is then constructed from the on-sight (i.e., coulomb) integral $\alpha$ and the hopping (i.e., 
resonance) integrals $\beta$ between the $p$ orbitals within each ethylenes and $\gamma$ between
adjacent $p$ orbitals in different ethylene molecules.  Note that $\alpha,\beta < 0$  but that 
$\gamma > 0$ for this particular configuration.  The position vector ${\vec R}$ is $Y$ so that
\begin{equation}
  {\bf h}(k) = \sum_Y {\bf h}^{(Y)} e^{i k Y} \, .
  \label{eq:theory.12}
\end{equation}
where,
\begin{eqnarray}
  {\bf h}^{(Y)} & = & \left[ \begin{array}{cc} {\bf A} & {\bf B} \\ 
                                           {\bf C} & {\bf A} \end{array} \right] \nonumber \\
  {\bf A} & = & \left[ \begin{array}{cc} \alpha \delta_{Y,0} & \beta \delta_{Y,0} \\
                                         \beta \delta_{Y,0} & \alpha \delta_{Y,0} \end{array} \right]
  \nonumber \\
  {\bf B} & = & \left[ \begin{array}{cc} 0 & \gamma\left( \delta_{Y,0} + \delta_{Y,+a}\right) \\
                       \gamma\left( \delta_{Y,0} + \delta_{Y,+a} \right) & 0 \end{array} \right]
  \nonumber \\
  {\bf C} & = & \left[ \begin{array}{cc} 0 & \gamma\left( \delta_{Y,0} + \delta_{Y,-a} \right) \\
                        \gamma\left( \delta_{Y,0} + \delta_{Y,-a} \right) & 0 \end{array} \right] \, ,
  \label{eq:theory.13}
\end{eqnarray}
and $a$ is the $y$-distance between ethylene molecules as opposed to the unit cell parameter which
is equal to $2a$.  Applying Eq.~(\ref{eq:theory.12}) to Eq.~(\ref{eq:theory.13}) then gives that
\begin{eqnarray}
  {\bf h}(k) & = & \left[ \begin{array}{cc} {\bf A} & {\bf B}(k) \\ 
                                           {\bf C}(k) & {\bf A} \end{array} \right] \nonumber \\
  {\bf A} & = & \left[ \begin{array}{cc} \alpha  & \beta  \\
                                         \beta  & \alpha  \end{array} \right]
  \nonumber \\
  {\bf B}(k) & = & \left[ \begin{array}{cc} 0 & \gamma\left( 1+ e^{i2ka} \right) \\
                       \gamma \left( 1+ e^{i2ka} \right)  & 0 \end{array} \right]
  \nonumber \\
  {\bf C}(k) & = & \left[ \begin{array}{cc} 0 & \gamma\left( 1+ e^{-i2ka} \right) \\
                        \gamma\left( 1+ e^{-i2ka} \right)  & 0 \end{array} \right] \, .
  \label{eq:theory.14}
\end{eqnarray}
This has four solutions, namely:
\begin{eqnarray}
  \epsilon_1(k) = \alpha+\beta+2\gamma \cos(ka) & \leftrightarrow & 
  \vec{c}_1(k)=\left( \begin{array}{c} 1 \\ 1 \\ +z^*/\vert z \vert 
   \\ +z^*\vert z \vert
  \end{array} \right) \nonumber \\
  \epsilon_2(k) = \alpha+\beta-2\gamma \cos(ka) & \leftrightarrow & 
  \vec{c}_2(k)=\left( \begin{array}{c} 1 \\ 1 \\ -z^*/\vert z \vert 
   \\ -z^*//\vert z \vert
  \end{array} \right) \nonumber \\
  \epsilon_3(k) = \alpha-\beta-2\gamma \cos(ka) & \leftrightarrow & 
  \vec{c}_3(k)=\left( \begin{array}{c} 1 \\ -1 \\ +z^*/\vert z \vert 
  \\ -z^*/\vert z \vert
  \end{array} \right) \nonumber \\
  \epsilon_4(k) = \alpha-\beta+2\gamma \cos(ka) & \leftrightarrow & 
  \vec{c}_4(k)=\left( \begin{array}{c} 1 \\ -1 \\ -z^*/\vert z \vert 
  \\ +z^*/\vert z \vert
  \end{array} \right) \, , \nonumber \\
  \label{eq:theory.15}
\end{eqnarray}
where,
\begin{equation}
  z = 1+ e^{i2ka} \, .
  \label{eq:theory.16}
\end{equation}
The associated band diagram is shown in Fig.~\ref{fig:TBbands}.  This is
a direct band system.  Assuming the $\Gamma$ point and no momentum
transfer to the lattice, then the two allowed transitions are those
shown by vertical arrows.  The corresponding CMOs (Fig.~\ref{fig:CMOs})
bear a close resemblence to the MOs of the (TOTEM)$^2$ dimer (Fig.~\ref{fig:stacked_transitions}),
showing that the (TOTEM)$^2$ analysis also applies in the periodic (TOTEM)$^N$ case.
% ------------------------------------------------------
% -------------------------------------------------------

Finally, it is illuminating to apply the same tight-binding model to the (TOTEM)$^2$ dimer.
The hamiltonian matrix to be diagonalized is then,
\begin{eqnarray}
  {\bf h}(k) & = & \left[ \begin{array}{cc} {\bf A} & {\bf B} \\ 
                                           {\bf B} & {\bf A} \end{array} \right] \nonumber \\
  {\bf A} & = & \left[ \begin{array}{cc} \alpha  & \beta  \\
                                         \beta  & \alpha  \end{array} \right]
  \nonumber \\
  {\bf B} & = & \left[ \begin{array}{cc} 0 & \gamma  \\
                       \gamma   & 0 \end{array} \right] \, ,
  \label{eq:theory.17}
\end{eqnarray}
which has the four solutions,
\begin{eqnarray}
  \epsilon_1 = \alpha+\beta+\gamma  & \leftrightarrow & 
  \vec{c}_1=\left( \begin{array}{c} 1 \\ 1 \\ 1 \\ 1 \end{array} \right) \nonumber \\
  \epsilon_2 = \alpha+\beta-\gamma & \leftrightarrow & 
  \vec{c}_2=\left( \begin{array}{c} 1 \\ 1 \\ -1 \\ -1 \end{array} \right) \nonumber \\
  \epsilon_3 = \alpha-\beta-\gamma  & \leftrightarrow & 
  \vec{c}_3=\left( \begin{array}{c} 1 \\ -1 \\ 1 \\ -1 \end{array} \right) \nonumber \\
  \epsilon_4(k) = \alpha-\beta+\gamma & \leftrightarrow & 
  \vec{c}_4=\left( \begin{array}{c} 1 \\ -1 \\ -1 \\ 1 \end{array} \right) \, .
  \label{eq:theory.18}
\end{eqnarray}
Comparing with the band solution for the periodic system (TOTEM)$^N$, we see that the energy
levels are displaced by $\gamma$ in (TOTEM)$^2$ rather than by $2\gamma$ in (TOTEM)$^N$, because
each ethylene in (TOTEM)$^2$ is only in contact with a single other ethylene, while each ethylene
in (TOTEM)$^N$ is in contact with two other ethylene molecules.  However {\em the key energy differences 
are the same for (TOTEM)$^2$ and (TOTEM)$^N$}, lending reassurance that the fundamental analysis
of the dimer model also applies for larger parallel stacks of ethylene molecules.  On the other hand,
inclusion of nonnearest neighbor interactions in the model is expected to yield small contributions from
higher-order Davydov multiplets, even if our simple model captures the main qualitative aspects 
of ET and CT % Frenkel
excitons in the stack of molecules.  

This completes our analytic study of stacked ethylene dimers.  In the Sec.~\ref{sec:results}, we will apply
this analysis to stacked pentamers and use it to gain a deeper insight into how different variations of
TD-DFT and TD-DFTB work.

%%%%%
% EOF
%%%%%
% -----------------------------------------------
\section{Computational Details}
\label{sec:details}
% \input{details.tex}
% \begin{verbatim}
% =================================
% File: details.tex
% Last modified: 11 February 2018
% =================================
% \end{verbatim}

Two programs were used to carry out the calculations reported in this paper, namely
{\sc Gaussian09}~\cite{g09} for DFT (Appendix~\ref{sec:DFT}) and TD-DFT 
(Appendix~\ref{sec:TD-DFT}) calculations and 
{\sc DFTBaby}~\cite{HM15,HM17} for DFTB (Appendix~\ref{sec:DFTB}) and TD-DFTB 
(Appendix~\ref{sec:TD-DFTB}) calculations.  
Note that, although {\sc Gaussian09} does have the ability to
carry out DFTB calculations, only {\sc DFTBaby} allows us to carry out state-of-the-art
TD-lc-DFTB calculations.  We will first describe the options used with each program in 
more detail.  We will then go on to describe how the programs were used in structural
and spectral studies.

{\sc Gaussian09} \cite{g09} calculations may be described in terms of a ``theoretical model'' 
(p.~5, Ref.~\onlinecite{HRSP86}) which is fully specified, in our case, by indicating
for an excited-state (i.e., TD) calculation, the the choice of functional and the 
orbital basis set. %  (pp.~65-88, Ref.~\onlinecite{HRSP86}).
This is conveniently expressed in expanded notation as
\begin{center}
    (TD-)DFA1/Basis1//DFA2/Basis2
\end{center}
(p.~96, Ref.~\onlinecite{HRSP86}), where DFA2 is the density-functional approximation
used for the geometry optimzation and Basis2 is the corresponding basis set used for the
geometry optimization, and DFA1 is the density-functional approximation used used for the
TD-DFT calculation and Basis1 is the corresponding orbital basis used in the TD-DFT
calculation.  The density-functionals used (LDA, B3LYP, HF, CAM-B3LYP with or without
Grimme's D3 correction) are described in Appendix~\ref{sec:DFT}. 
Two orbital basis sets were used here, namely the minimal STO-3G basis set \cite{HSP69,HDSP70}
and the much more flexible 6-31G(d,p) split-valence (hydrogen \cite{DHP71},
carbon \cite{HDP72}) plus polarization basis set \cite{HP72}.  An example of
the expanded notation is that TD-CAM-B3LYP/6-31G(d,p)//D3-CAM-B3LYP/6-31G(d,p) means 
that the geometry of the molecule was optimized using the 6-31G(d,p) basis set using the 
CAM-B3LYP functional with the D3 dispersion correction.  Then a TD-DFT calculation was 
carried out at that geometry using the 6-31G(d,p) orbital basis set and the CAM-B3LYP functional 
(Appendix~\ref{sec:TD-DFT}).
Often we will use a shorter nomenclature when the details of the theoretical model
are clear from context.

{\sc DFTBaby} \cite{HM15,HM17} was used to carry out lc-DFTB 
(Appendix~\ref{sec:DFTB}) 
and TD-lc-DFTB (Appendix~\ref{sec:TD-DFTB}) calculations.
The values for the confinement radius $r_0$ and the Hubbard parameter $U_H$ that were 
used to parameterize the electronic part of DFTB are shown in Table I of 
Ref.~\onlinecite{HM15}.  The parameter for the lc correction was set
to $R_{\text{lc}}$ = 3.03 bohr so 
$\mu=1/R_{\text{lc}}=0.33 \, \, \mbox{bohr$^{-1}$}$, which is a reasonable
compromise between $\mu=0.33 \, \,  \mbox{bohr$^{-1}$}$ used in the CAM-B3LYP functional 
and $\mu=0.4 \, \, \mbox{bohr$^{-1}$}$
used in the LRC family of functionals.

% ------------------------------------------------------
\begin{figure}
\includegraphics[width=0.3\textwidth]{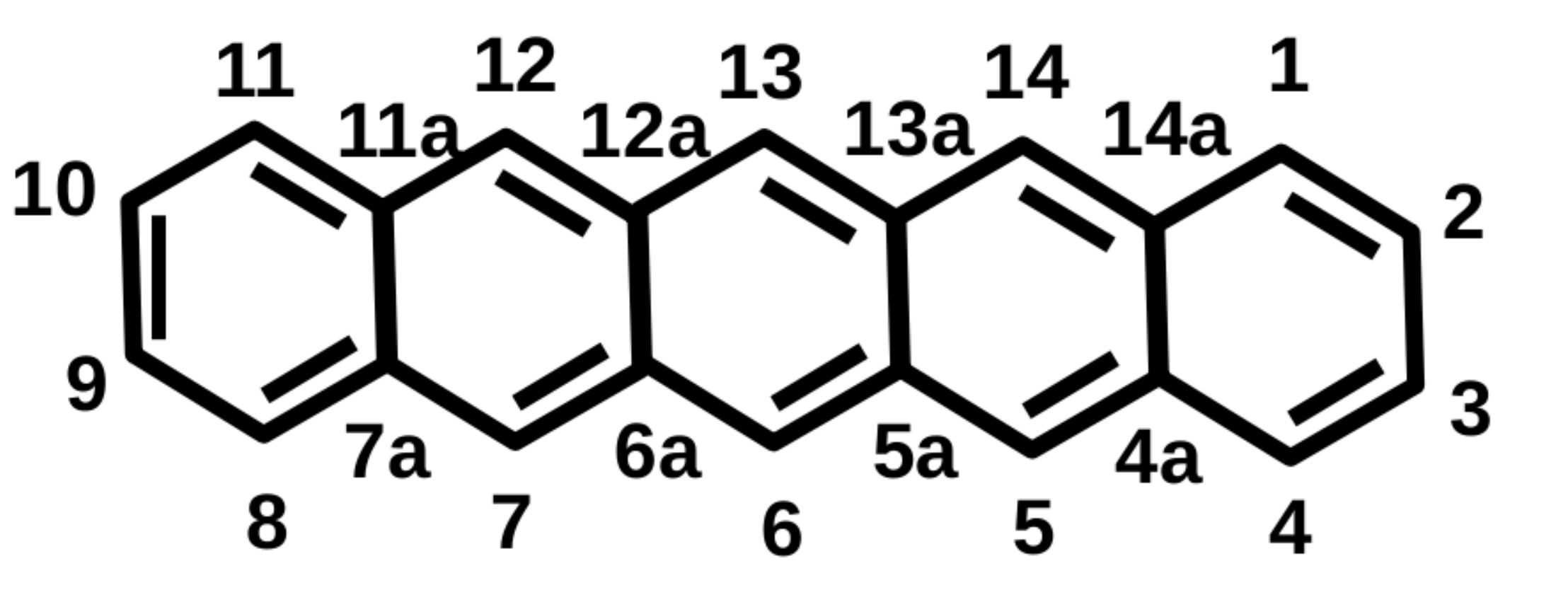}
\caption{
\label{fig:numbering}
Pentacene carbon numbering.
}
\end{figure}
\begin{figure}
\includegraphics[width=0.5\textwidth]{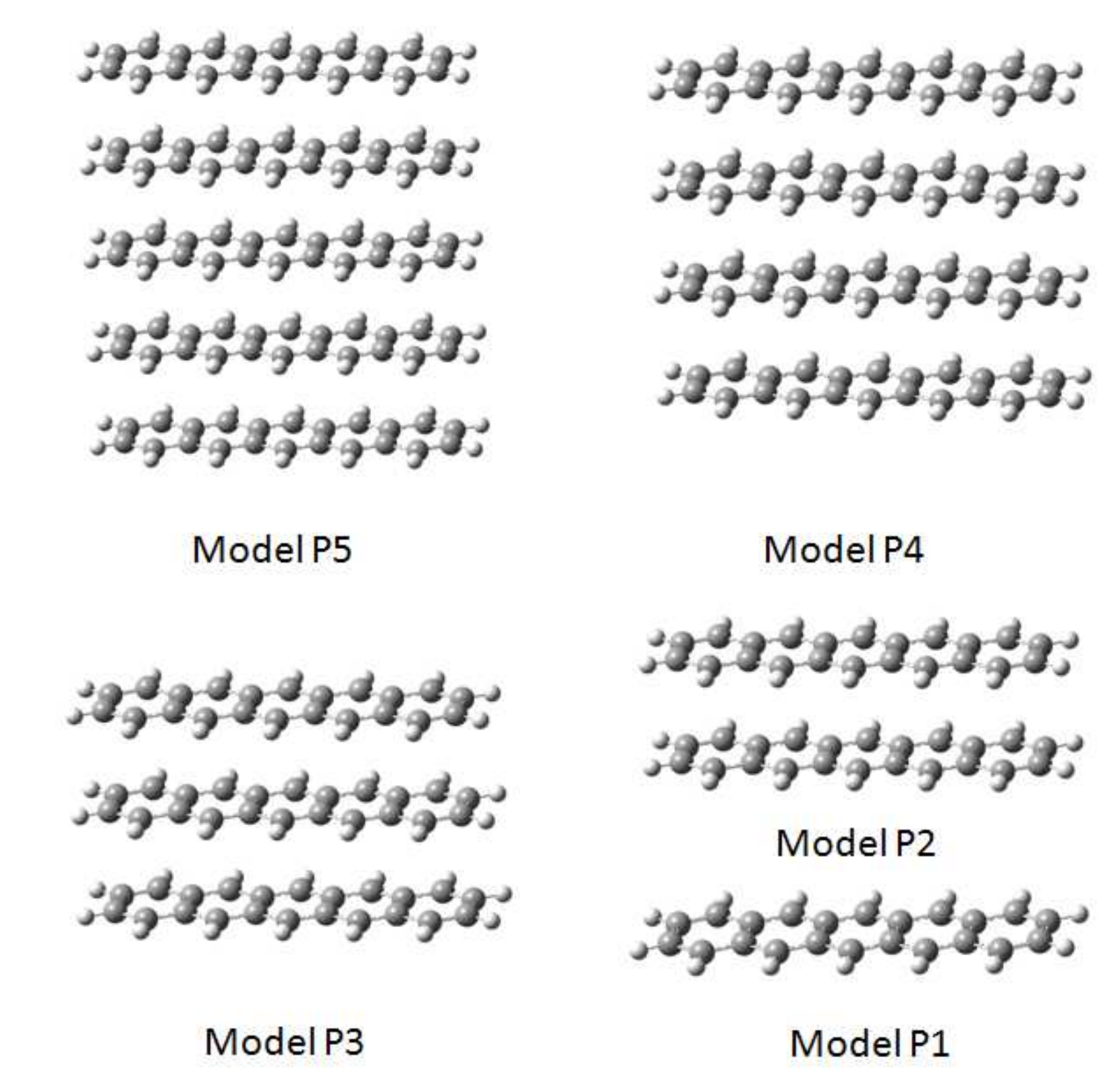}
\caption{
\label{fig:Pn}
The configurations of the five models of parallel stacked pentacene (P$n$ stands for
$n$ parallel stacked pentacenes).
}
\end{figure}

% -------------------------------------------------------
Our structural calculations started with initial x-ray crystallography
geometries taken from the Crystallography Open Database (COD) \cite{DM94,SHDN07}.
We will use the standard numbering of pentacene carbon shown in Fig.~\ref{fig:numbering}.
We first optimized the {\em monomer} geometry and calculated its absorption spectrum at each
level.  Vibrational frequencies were calculated to make sure that the optimized structures
were true minima.  We then went on to study an ideal {\em parallel stacked} model in which
B3LYP/6-31G(d,p) optimized monomers were $\pi$-stacked vertically face-to-face with a fixed
distance $R$ between them (Fig.~\ref{fig:Pn}).  The distance $R$ was optimized for the tetramer 
using different methods and then this distance was used in studying stacks of different sizes.  
Finally we studied calculated absorption spectra for cluster models cut out of the experimental 
{\em herringbone} structure without any geometry optimization.
% ------------------------------------------------------
% -------------------------------------------------------

%%%%%%
% EOF
%%%%%%
% -----------------------------------------------
\section{Results}
\label{sec:results}
% \input{results.tex}
% \begin{verbatim}
% ====================================
% File: results.tex
% Last modified: 10 February 2018
% ====================================
% \end{verbatim}

Our goal in this section is to evaluate state-of-the-art (TD-)DFTB calculations
of excitons in pentacene aggregates with
state-of-the-art (TD-)DFT calculations on the same systems. 
We would also like to get a feeling for the relative
importance of ET versus CT excitons.  This involves three levels of calculation on three
classes of systems. 
The three levels of calculation are first high-quality
(TD-)DFT/6-31G(d,p) calculations aimed at obtaining good quality reference
calculations which can be compared to experiment as a reality check.  The
second type of calculation consists of minimum basis set \\ (TD-)DFT/STO-3G
calculations as our ultimate goal is to evaluate the third method, namely
the minimal basis set semi-empirical (TD-)DFTB method.  The systems
considered are first an isolated gas-phase pentacene molecule, second
a series of parallel-stacked pentacene molecules as these parallel the theory
already presented in Sec.~\ref{sec:theory}, and lastly a subunit of the
known structure of crystalline pentacene.

% ========================================================
\subsection{Monomer}
% ========================================================

% ------------------------------------------------------
\begin{figure}
\includegraphics[width=0.4\textwidth]{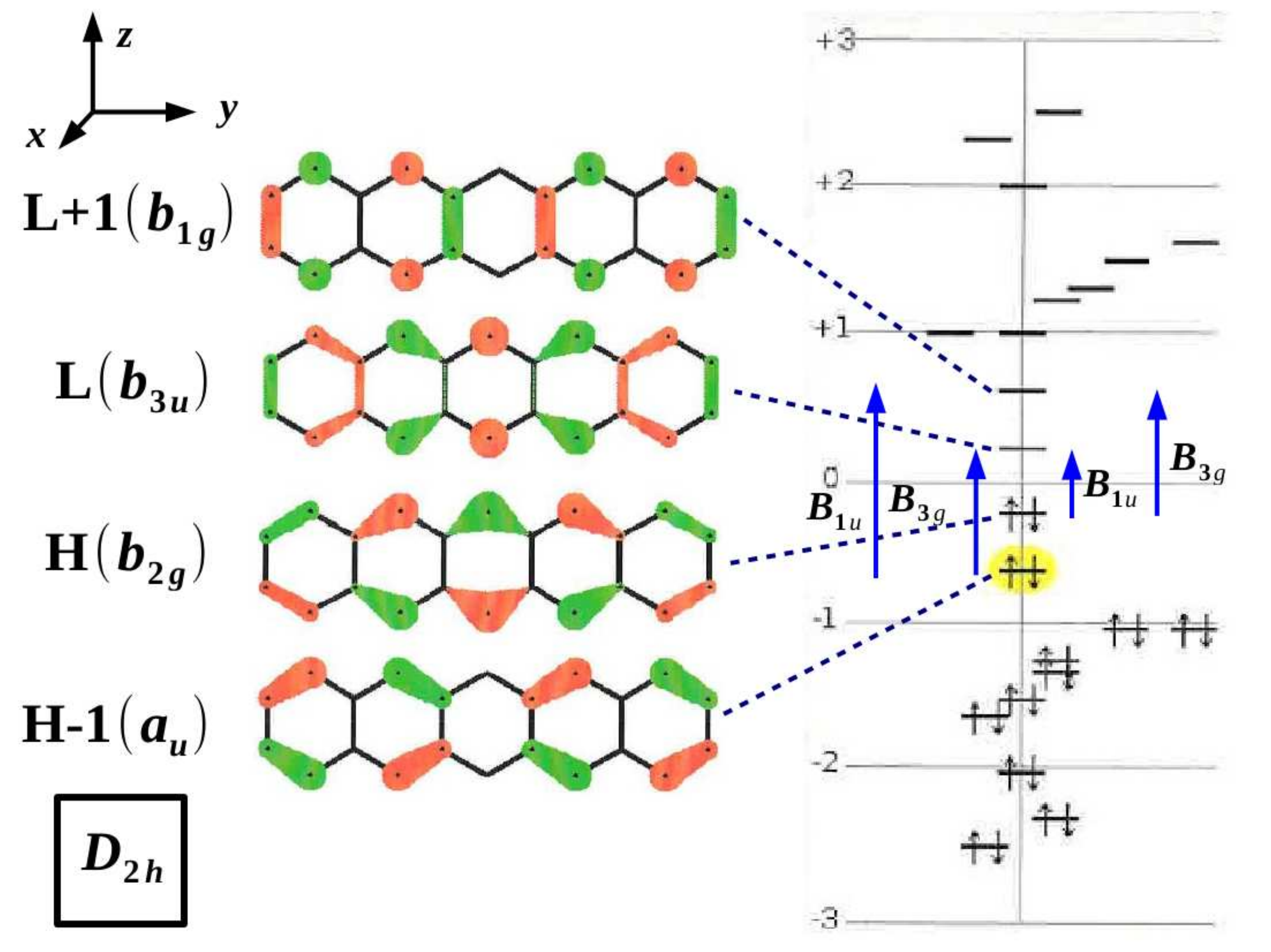}
\caption{
\label{fig:SHMO}
Simple H\"uckel molecular orbital theory results for the pentacene monomer.
}
\end{figure}
% -------------------------------------------------------
Although our primary interest here is in the absorption spectrum of the 
monomer, it is useful to begin with a review of the molecular orbitals (MOs).
Figure~\ref{fig:SHMO} shows the result of a simple H\"uckel MO calculation
with the {\sc SHMO} calculator \cite{SHMO}.  MO symmetries have been assigned 
following the recommended IUPAC nomenclature \cite{M55,M56} and the symmetry
of the expected lowest energy excitations have been assigned.  

\begin{figure}
\includegraphics[width=0.4\textwidth]{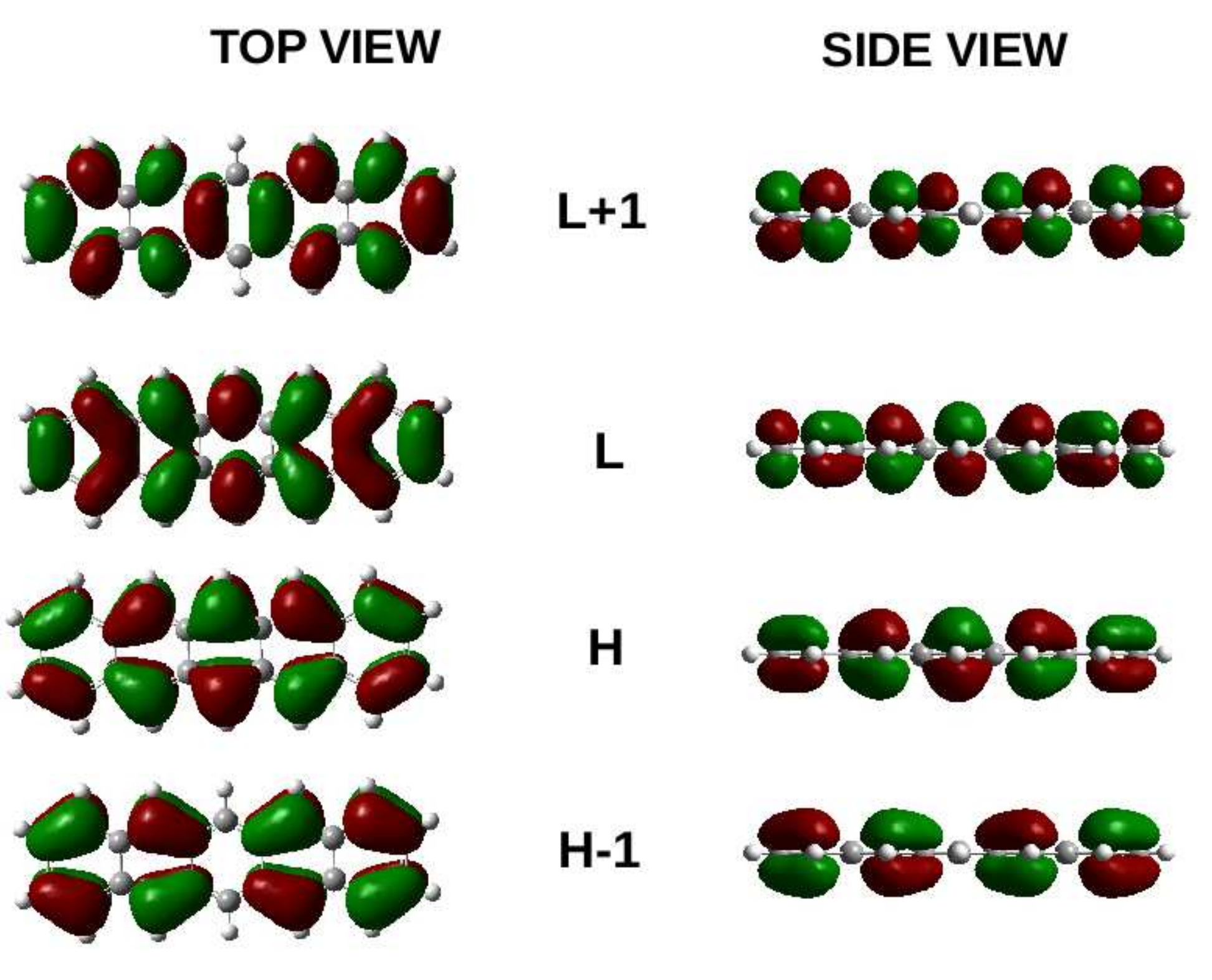}
\caption{
\label{fig:B3LYP_MOs}
Pentacene monomer B3LYP/6-31G(d,p) MOs.
}
\end{figure}

The monomer geometry
has been optimized at the LDA/STO-3G, LDA/6-31G(d,p), B3LYP/STO-3G, B3LYP/6-31G(d,p),
CAM-B3LYP/STO-3G, CAM-B3LYP/6-31G(d,p), DFTB, and lc-DFTB levels of theory.
The orbitals at the resultant optimized gometries have been visualized 
(e.g., Fig.~\ref{fig:B3LYP_MOs}) and are
found to be qualitatively similar to those obtained from simple H\"uckel MO
theory.  This is important as it is then relatively easy to make a connection 
between the results of stacked pentacene molecules and the theoretical discussion
of Sec.~\ref{sec:theory} for stacked ethylene molecules.
% ------------------------------------------------------
% -------------------------------------------------------

% ------------------------------------------------------
\begin{figure}
\begin{tabular}{cc}
a) & \\
% b) & \includegraphics[width=0.45\textwidth]{./graphics/MONOMER/MONOMER_LDA.eps} \\
b) & \includegraphics[width=0.45\textwidth]{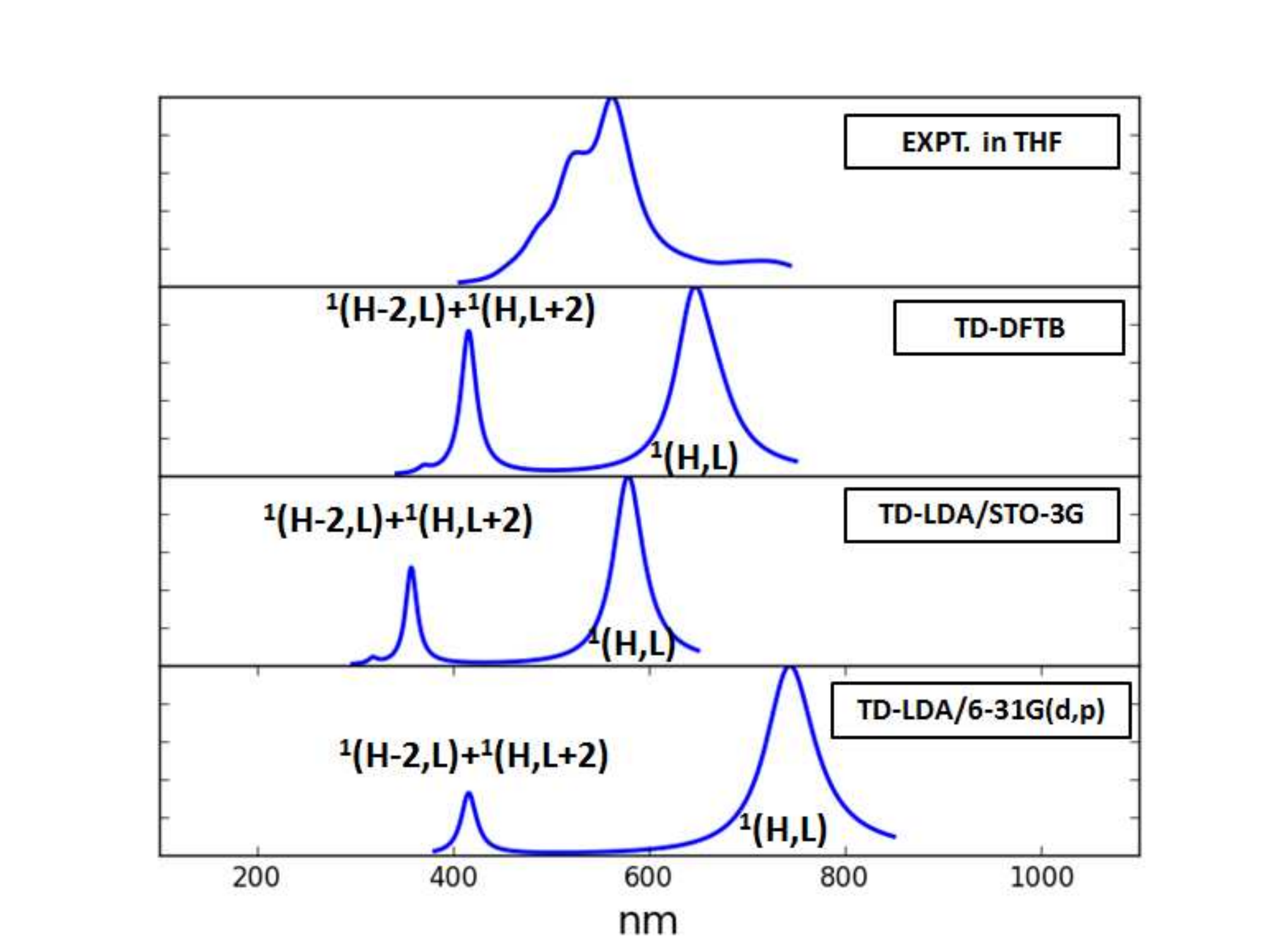} \\
% c) & \includegraphics[width=0.45\textwidth]{./graphics/MONOMER/MONOMER_B3LYP.eps} \\
c) & \includegraphics[width=0.45\textwidth]{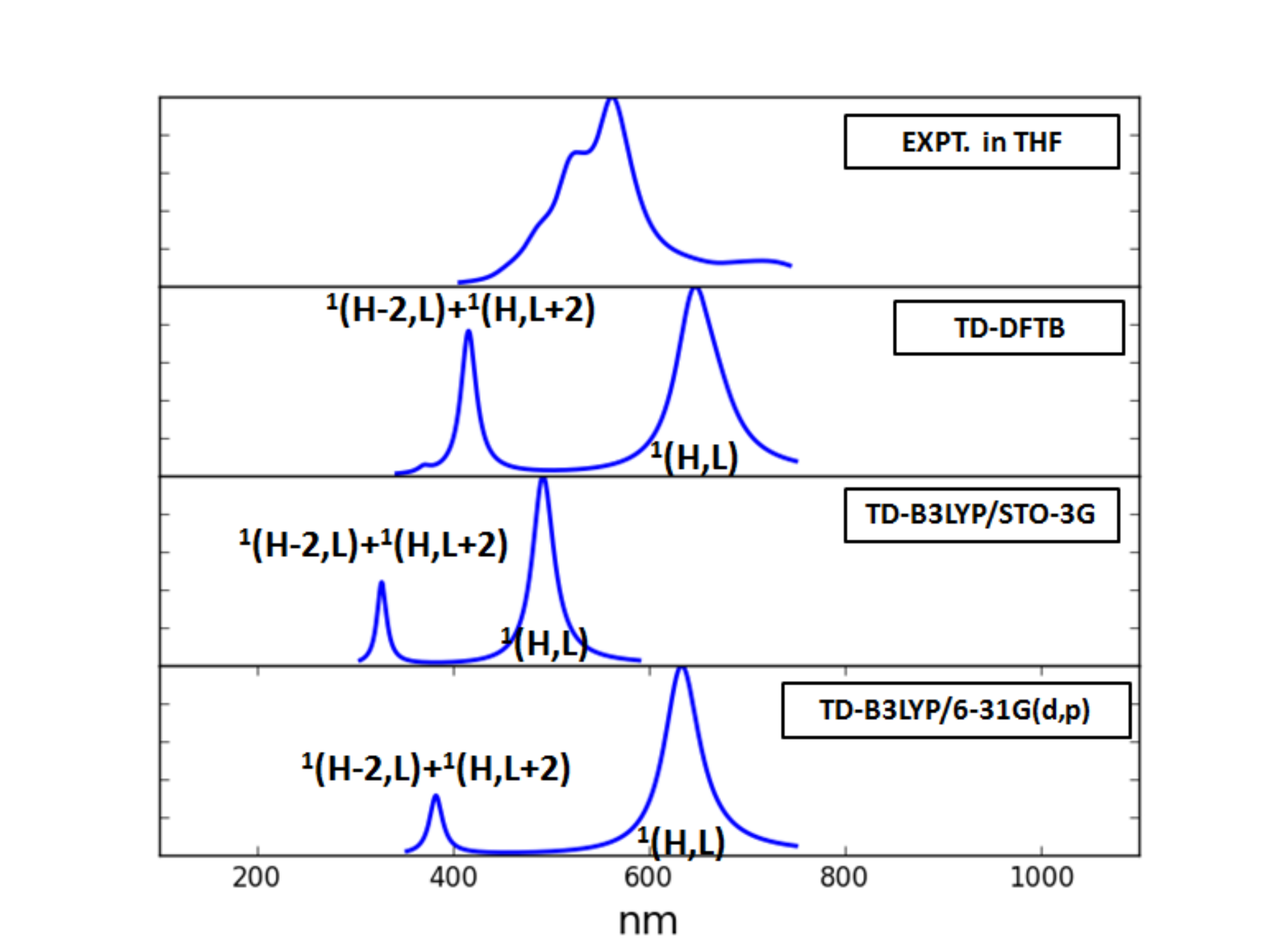} \\
%   & \includegraphics[width=0.45\textwidth]{./graphics/MONOMER/MONOMER_CAM_B3LYP.eps}
   & \includegraphics[width=0.45\textwidth]{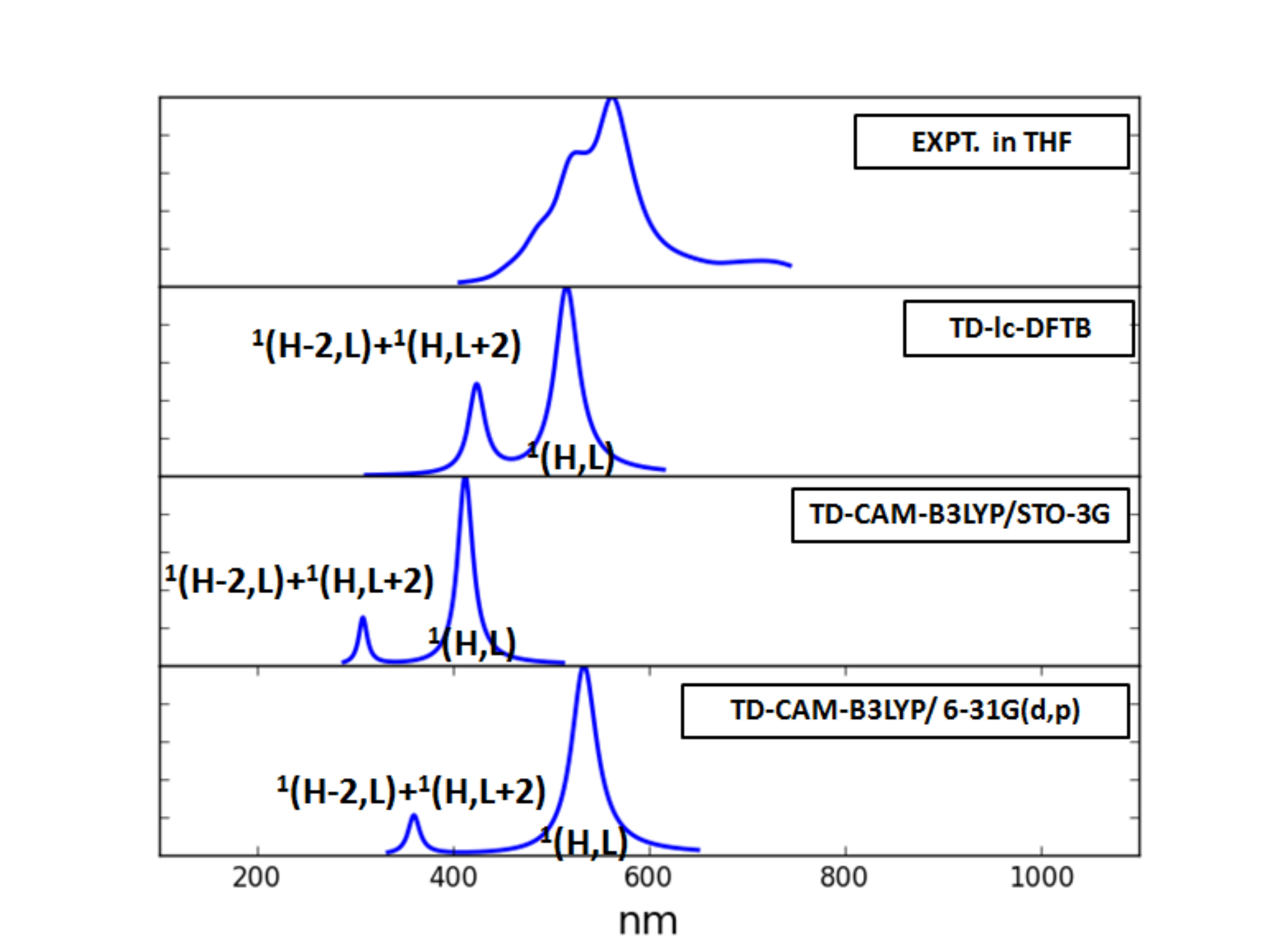}
\end{tabular}
\caption{
\label{fig:monomerspectra}
Pentacene monomer absorption spectra: (a) TD-LDA, TD-DFTB, and experiment;
(b) TD-B3LYP, TD-DFTB, and experiment; (c) TD-CAM-B3LYP, TD-lc-DFTB, and experiment.
The experimental curve is a spectrum measured in tetrahydrofuran.  Intensities
are in arbitrary units.
}
\end{figure}
% -------------------------------------------------------
Figure~\ref{fig:monomerspectra} compares the calculated monomer absorption spectra
with the experimental spectrum measured in tetrahydrofuran (data obtained by
plot digitization \cite{PlotDigitizer} of Fig.\ 1 of Ref.~\onlinecite{MRK+04}).
Note that TD-DFT and TD-DFTB (and TD-lc-DFTB) calculations give qualitatively 
similar spectra in terms of the number and spacing of peaks, though not all
peaks are shown in Fig.~\ref{fig:monomerspectra}.  Our concern is primarily
with the lowest energy (i.e., longest wavelength) transitions.

Let us first look at the TD-DFT calculations with the 6-31G(d,p) basis set
using different functionals.  The TD-LDA/6-31G(d,p) spectrum is red-shifted
with respect to the experimental spectrum.  The TD-B3LYP/6-31G(d,p) spectrum, which
includes some HF exchange via a global hybrid, brings us closer to the 
experimental spectrum.  Finally the TD-CAM-B3LYP/6-31G(d,p) spectrum, which
includes even more HF exchange to decribe the long-range part of the electron-electron
repulsion, matches the experimental spectrum very well.  Of course, this should
be taken with a certain amount of scepticism because the experimental spectrum
is measured in solution while the theoretical calculations are for the gas phase
and neglect any vibrational contributions.

Let us now turn to the TD-DFTB and TD-lc-DFTB calculations.  Since these are
semi-empirical calculations, they are expected to be similar to TD-DFT/STO-3G
calculations in that DFTB calculations are parameterized assuming a minimal basis
set.  This might be expected to show up in the number of underlying degrees of
freedom and hence in the complexity of the calculated absorption spectra.  Indeed
this does seem to be the case in that the longest wavelength TD-CAM-B3YLP/6-31G(d,p)
peak shows more complexity than does the longest wavelength TD-lc-DFTB or TD-DFTB
peak.  (This difference is {\em not} visible in Fig.~\ref{fig:monomerspectra}, but 
rather in the underlying stick spectra.) 
However the TD-lc-DFTB and TD-DFTB spectra are red-shifted compared to
the correspondingly TD-DFT/STO-3G spectra.  This brings the TD-DFTB spectrum
in remarkably good correspondance with the TD-B3LYP/6-31G(d,p) spectrum and the
TD-lc-DFTB spectrum in remarkably good correspondance with the TD-CAM-B3LYP/6-31G(d,p)
spectrum.  

% \input{./tables/CTETP1.tex}
% =======================================
% File: CTETP1.tex
% Last modified: 22 August 2017
% =======================================

\begin{table}
\squeezetable
\caption{Monomer lowest energy peak $^1$(H,L) calculated with various 
methods.
\label{tab:CTETP1}}
\begin{tabular}{cccc}
\hline \hline
\multicolumn{4}{c}{Method} \\
State & $f$ (unitless) & $\lambda$ (nm) & $\Delta E$ (eV) \\
\hline
\multicolumn{4}{c}{TD-LDA/6-31G(d,p)//LDA/6-31G(d,p)}\\
$1\, ^1B_{1u}$ & 0.0234 & 744 & 1.67 \\
\multicolumn{4}{c}{TD-LDA/STO-3G//LDA/STO-3G} \\                
$1\, ^1B_{1u}$ & 0.0325 & 579 & 2.14 \\
\multicolumn{4}{c}{TD-B3LYP/6-31G(d,p)//B3LYP/6-31G(d,p)}\\     
$1\, ^1B_{1u}$ & 0.0415 & 633 & 1.96 \\
\multicolumn{4}{c}{TD-B3LYP/STO-3G//B3LYP/STO-3G}\\
$1\, ^1B_{1u}$ & 0.0596 & 492 & 2.52 \\
\multicolumn{4}{c}{TD-DFTB//DFTB} \\
$1\, ^1B_{1u}$ & 0.1594 & 646 & 1.92 \\ 
\multicolumn{4}{c}{TD-CAM-B3LYP/6-31G(d,p)//B3LYP/6-31G(d,p)}\\
$1\, ^1B_{1u}$ & 0.0750 & 534 & 2.32 \\
\multicolumn{4}{c}{TD-CAM-B3LYP/STO-3G//CAM-B3LYP/STO-3G}\\      
$1\, ^1B_{1u}$ & 0.1070 & 412 & 3.01 \\
\multicolumn{4}{c}{TD-lc-DFTB//lc-DFTB}\\                       
$1\, ^1B_{1u}$ & 0.3212 & 515 & 2.40 \\
\multicolumn{4}{c}{HF/6-31G(d,p)//B3LYP/6-31G(d,p)}\\
$1\, ^1B_{1u}$ & 0.1436 & 491 & 2.53 \\
\multicolumn{4}{c}{HF/STO-3G//B3LYP/6-31G(d,p)}\\      
$1\, ^1B_{1u}$ & 0.2279 & 357 & 3.47 \\
\hline \hline
\end{tabular}
\end{table}

%%%%%%%
% EOF
%%%%%%%
Some rough assignments are given, based upon MO contributions to the TD-DFT
and TD-DFTB (or TD-lc-DFTB) coefficients.  The lowest energy peaks 
(Table~\ref{tab:CTETP1}) are singlet HOMO $\rightarrow$ LUMO transitions 
[$^1$(H,L)].  The $^1$(H,L) TD-CAM-B3YLP/6-31G(d,p)
peak is at 534 nm, which may be compared with the corresponding experimental
value of about 540 nm from gas phase spectroscopy
\cite{HHMH98} and spectroscopy of isolated pentacene molecules in rare gas
matrices \cite{HHS+00}.  The next lowest energy peaks have mixed $^1$(H-2,L)
and $^1$(H,L+2) character as we have mentioned (Sec.~\ref{sec:theory}) 
often occurs in the excitation spectra of $\pi$-conjugated  molecules.

% ------------------------------------------------------
\begin{figure}
\includegraphics[width=0.4\textwidth]{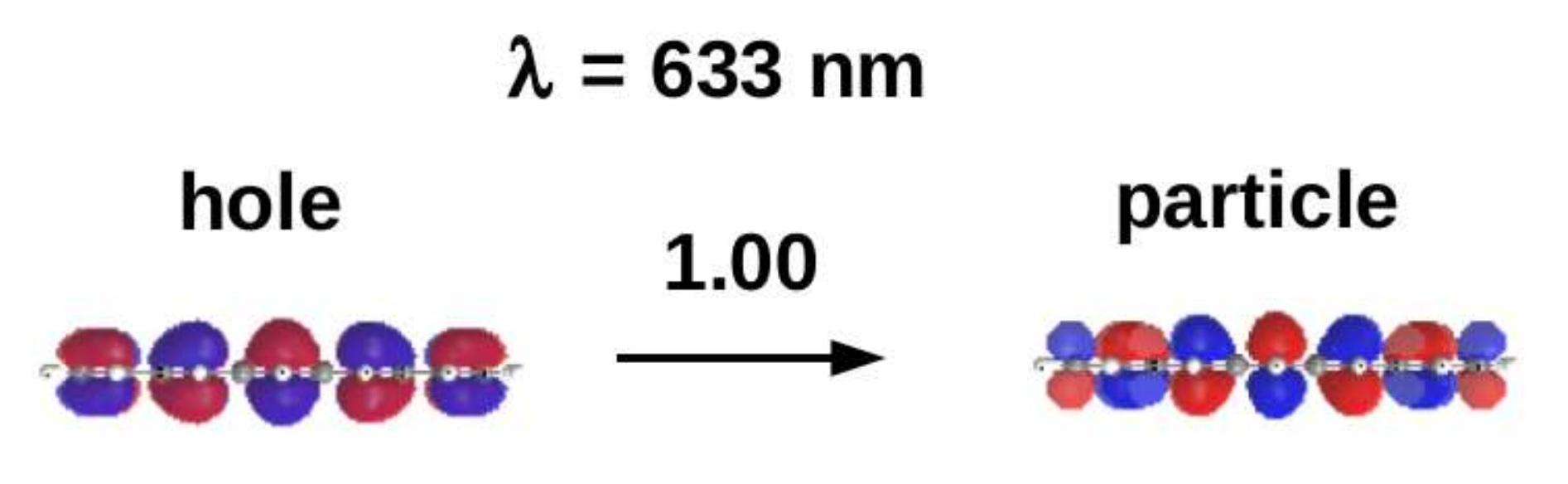}
\caption{
\label{fig:P1_NTOs}
Monomer NTOs and renormalized coefficient:
TD-B3LYP/6-31G(d,p)//B3LYP/6-31G(d,p) 633 nm. % 621 nm.
}
\end{figure}
% -------------------------------------------------------
It is especially important to confirm our $^1(H,L)$ peak assignment.  
Figure~\ref{fig:P1_NTOs} shows the natural transition orbitals (NTOs) 
associated with the lower energy peak in the spectrum.  Comparison with 
the nodal structure in Figs.~\ref{fig:SHMO} and \ref{fig:B3LYP_MOs} 
confirms that this is indeed the $^1(H,L)$ transition.

Notice that there is a close analogy between the TOTEM model in ethylene
and that of pentacene.  In particular, the part of the H and L MOs on 
carbons 6 and 13 in pentacene (see Fig.~\ref{fig:numbering}) 
corresponds to a $\pi^* \rightarrow \pi$.  
This is sufficiently analogous to the $\pi \rightarrow \pi^*$ transition 
in  ethylene that essentially the same theoretical analysis
goes through for pentacene as for ethylene and we will make great use of
this observation in the next subsection.

% ========================================================
\subsection{Stacking}
% ========================================================

We are concerned with the model of equally-spaced stacked pentacenes shown
in Fig.~\ref{fig:Pn}.  This model, though far from the observed herringbone
structure of solid pentacene, is interesting because of its obvious analogy to
graphite and because it may be readily compared with the model of 
equally-spaced stacked ethylenes discussed in the previous section.

% -----------------------------------------
\subsubsection{Intermolecular forces}
% -----------------------------------------

Equally-spaced parallel stacked pentacenes were prepared by optimizing 
the intermolecular distance for stacked tetramers without reoptimizing the 
individual molecules.  
The tetramer stacked structure is expected to be bound together by van der Waals 
forces at a distance similar to that in graphene, namely about 3 {\AA} \cite{ALS11}.  

% ------------------------------------------------------
\begin{figure}
\begin{tabular}{cc}
a) & \\
% b) & \includegraphics[width=0.45\textwidth]{./graphics/PES_LDA.eps} \\
b) & \includegraphics[width=0.45\textwidth]{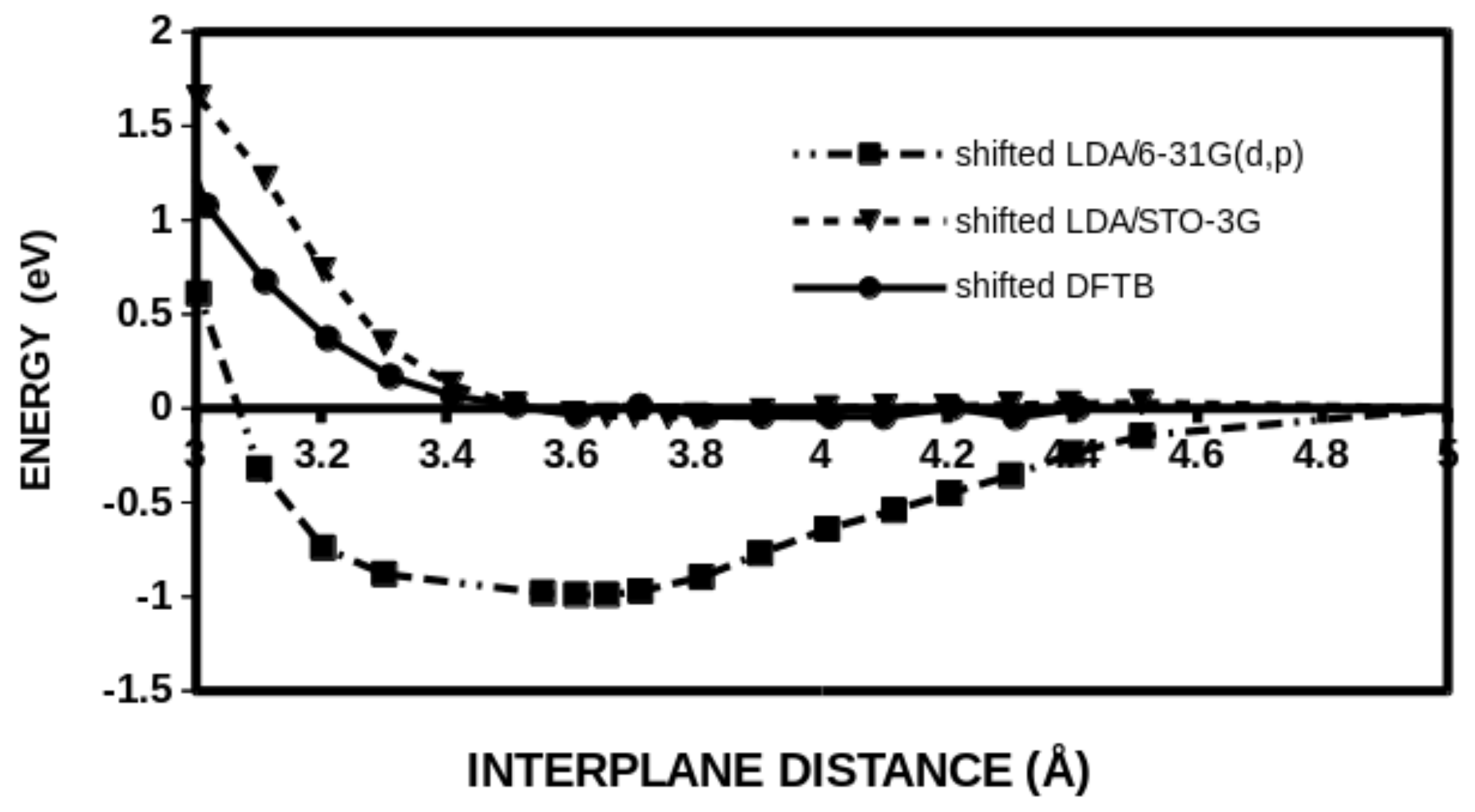} \\
% c) & \includegraphics[width=0.45\textwidth]{./graphics/PES_B3LYP.eps} \\
c) & \includegraphics[width=0.45\textwidth]{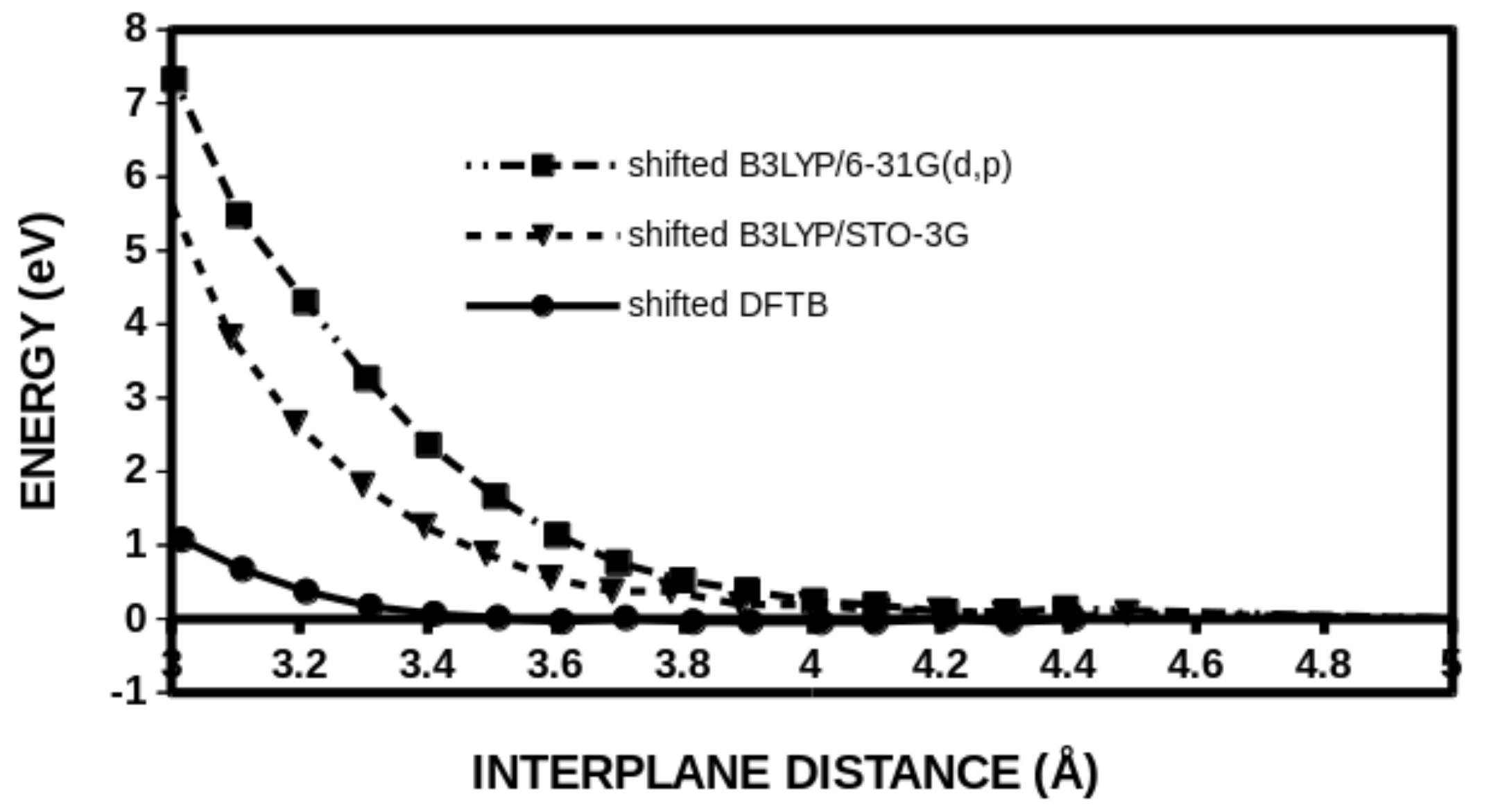} \\
%   & \includegraphics[width=0.45\textwidth]{./graphics/PES_CAM_B3LYP.eps}
   & \includegraphics[width=0.45\textwidth]{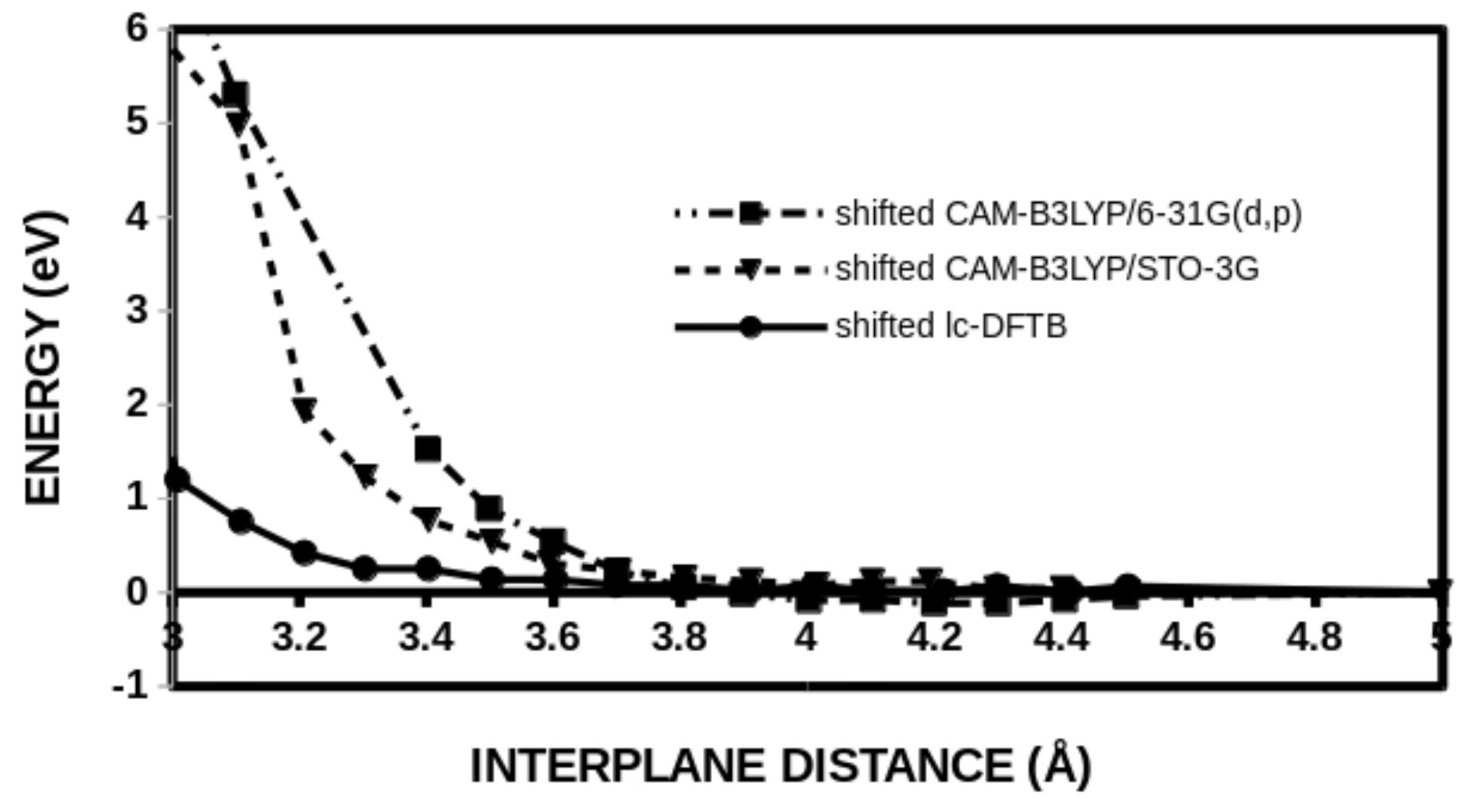}
\end{tabular}
\caption{
\label{fig:tetramerPES}
Pentacene tetramer potential energy surfaces without dispersion correction: 
(a) LDA and DFTB;
(b) B3LYP and DFTB; 
(c) CAM-B3LYP, lc-DFTB.
}
\end{figure}
% -------------------------------------------------------
Figure~\ref{fig:tetramerPES} shows the resultant
PES for molecules without dispersion correction.  As a general rule, 
DFT can only describe forces between atoms in regions of space where the
electron density is significant.  Uncorrected DFT is usually unable to describe
van der Waals binding as such binding takes place at intermolecular distances
where the molecular densities do not overlap significantly.  As seen in 
Fig.~\ref{fig:tetramerPES}, LDA/6-31G(d,p) shows an accidental minimum at about 
3.65 {\AA} but LDA/STO-3G does not bind.  The other functionals do not bind whichever 
basis is used.  DFTB also does not bind, but it is less repulsive than the 
other calculations shown here.

% ------------------------------------------------------
\begin{figure}
\begin{tabular}{cc}
a) & \\
% b) & \includegraphics[width=0.45\textwidth]{./graphics/PES_B3LYP_D3.eps} \\
b) & \includegraphics[width=0.45\textwidth]{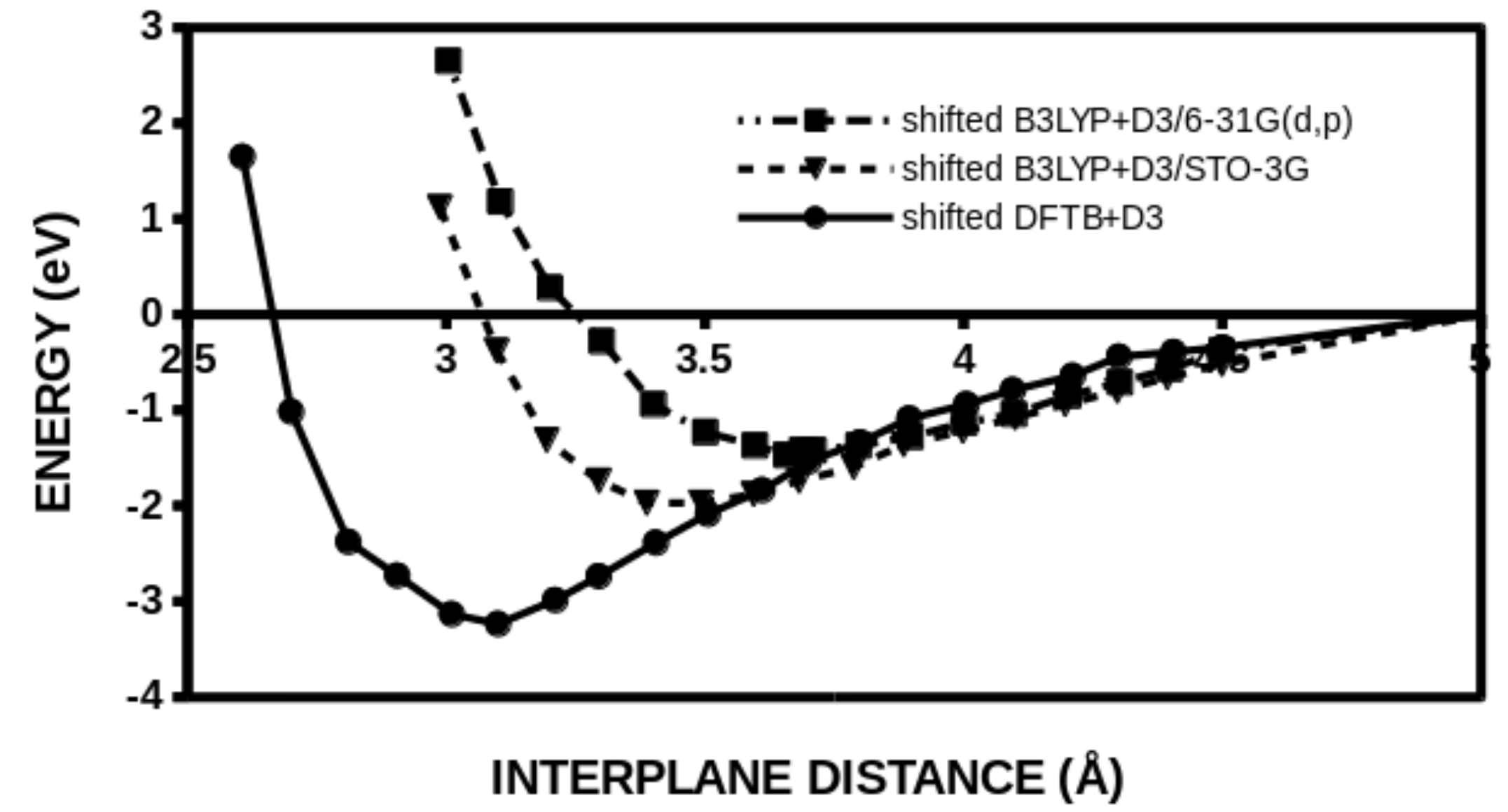} \\
%   & \includegraphics[width=0.45\textwidth]{./graphics/PES_CAM_B3LYP_D3.eps}
   & \includegraphics[width=0.45\textwidth]{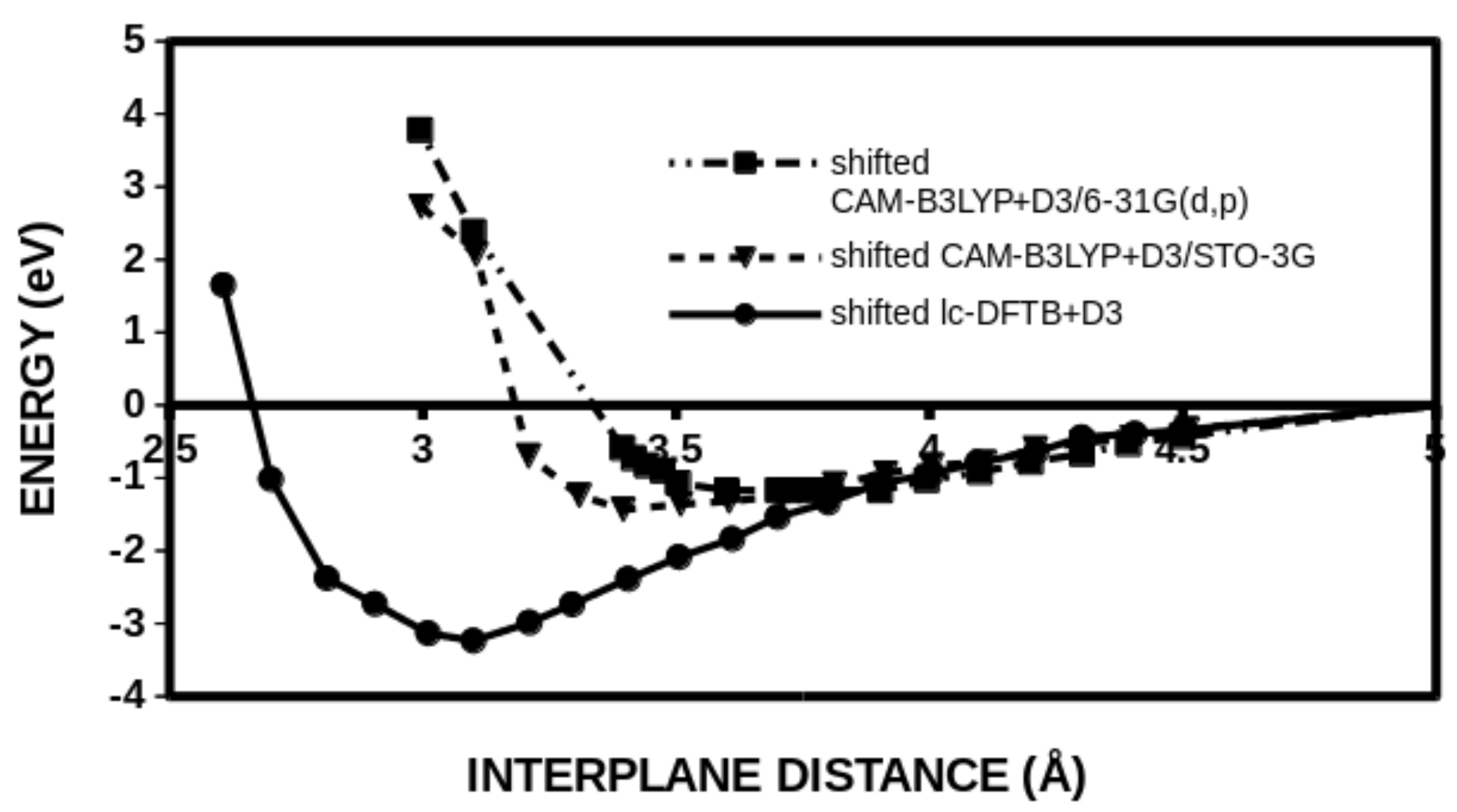}
\end{tabular}
\caption{
\label{fig:tetramerPESD3}
Pentacene tetramer potential energy surfaces with dispersion correction: 
(a) B3LYP+D3 and DFTB+D3; 
(b) CAM-B3LYP+D3, lc-DFTB+D3.
}
\end{figure}
% -------------------------------------------------------
Figure~\ref{fig:tetramerPESD3} shows the improved curves obtained using Grimme's
D3 dispersion correction.  The minima are located at at about 3.7 {\AA} for B3LYP+D3, 
at about 
3.72 {\AA} % 3.8 {\AA} 
for CAM-B3LYP+D3, and at about 3.1 {\AA} for DFTB+D3.
% \input{./tables/minima.tex}
% =======================================
% File: minima.tex
% Last modified: 19 November 2017
% =======================================
\begin{table}
\caption{Intermolecular distances obtained for the tetramer of parallel stacked
pentacene molecules.  \label{tab:minima}}
\begin{tabular}{cc}
\hline \hline
Method & Distance (\AA) \\
\hline
LDA/STO-3G & 3.7 \\
LDA/6-31G(d,p) & 3.6 \\
B3LYP+D3/STO-3G & 3.5 \\
B3LYP+D3/6-31G(d,p) & 3.68 \\
CAM-B3LYP+D3/STO-3G & 3.5 \\
CAM-B3LYP+D3/6-31G(d,p) & 3.72 \\
DFTB+D3 & 3.1 \\
lc-DFTB+D3 & 3.1 \\
\hline \hline
\end{tabular}
\end{table}
%%%%%%%
% EOF
%%%%%%%

Optimized intermolecular distances are summarized in Table~\ref{tab:minima}.

% -------------------------------------------
\subsubsection{Energy versus charge transfer}
% -------------------------------------------

Now that we have determined an optimal stacking distance (namely 3.71 {\AA}),
we can use the analogy to stacked ethylenes developed in the previous section to estimate the relative
contributions of CT versus ET in the exciton model.  This is possible by concentrating
on the central part (carbons 6 and 13 of Fig.~\ref{fig:numbering})
of the H($b_{2g}$) and L($b_{3u}$) MOs in Fig.~\ref{fig:SHMO}.
This part of the pentacene  H($b_{2g}$) MO resembles the ethylene $\pi^*$ MO while the 
pentacene L($b_{3u}$) MO resembles the ethylene $\pi$ MO (Fig.~\ref{fig:ethylene_MO}).
Figure~\ref{fig:P1_NTOs} shows a side view of the pentacene H $\rightarrow$ L transition.
Looked at this way, the only important difference between the MOs for stacked pentacenes
and the MOs for stacked ethylenes is that one MO diagram is the inverse of the other
(i.e., bonding and antibonding orbitals have been interchanged).  

% \input{tables/CTETP2.tex}
% =======================================
% File: CTETP2.tex
% Last modified: 16 January 2018
% =======================================
\begin{table}
\squeezetable
\caption{Relative percentages of CT and ET excitonic transitions to the principle
transition for two parallel stacked pentacenes as a function of method.
See Eq.~(\ref{eq:theory.4.2}). DS is the Davydov splitting between the CT and ET
excitonic transitions.  
\label{tab:CTETP2}}
\begin{tabular}{cccccc}
\hline \hline
\multicolumn{6}{c}{Method} \\
State & $f$ (unitless) & $\lambda$ (nm) & $\Delta E$ (eV) & ET\footnotemark[1] & CT\footnotemark[2] \\
\hline
\multicolumn{6}{c}{TD-LDA/6-31G(d,p)//LDA/6-31G(d,p)}\\
$1\, ^1B_{1u}$ & 0.0002 & 1033 & 1.20 & 0.3\%   &  99.\%  \\
$2\, ^1B_{1u}$ & 0.0322 & 733  & 1.69 & 99.\%   & 0.3\% \\
          &     &      & DS$_2$ = 0.49 eV & &         \\
\multicolumn{6}{c}{TD-LDA/STO-3G//LDA/STO-3G} \\                
$1\, ^1B_{1u}$ & 0.0000 & 818  & 1.52 & 0.005\% & 99.\% \\
$2\, ^1B_{1u}$ & 0.0499 & 572  & 2.17 & 99.\%   & 0.005\% \\
          &    &      & DS$_2$ = 0.65 eV & &         \\
\multicolumn{6}{c}{TD-B3LYP/6-31G(d,p)//B3LYP+D3/6-31G(d,p)}\\     
$1\, ^1B_{1u}$ & 0.0005 & 735  & 1.69 & 2.9\%   & 98.\% \\
$2\, ^1B_{1u}$ & 0.0576 & 621  & 2.00 & 98.\%   & 2.9\% \\
          &    &      & DS$_2$ = 0.31 eV & &         \\
\multicolumn{6}{c}{TD-B3LYP/STO-3G//B3LYP+D3/STO-3G}\\
$1\, ^1B_{1u}$ & 0.0000 & 604  & 2.05 & 0.02\%  & 100.\% \\
$2\, ^1B_{1u}$ & 0.0897 & 484  & 2.56 & 99.\%   & 2.9\% \\
        &      &      & DS$_2$ = 0.51 eV & &         \\
\multicolumn{6}{c}{TD-DFTB//DFTB+D3} \\
$1\, ^1B_{1u}$ & 0.0001 & 872  & 1.42 & 0.05\%  & 100.\% \\
$2\, ^1B_{1u}$ & 0.2831 & 634  & 1.96 & 100.\%   & 0.06\% \\
        &      &      & DS$_2$ = 0.54 eV & &         \\
\multicolumn{6}{c}{TD-CAM-B3LYP/6-31G(d,p)//B3LYP+D3/6-31G(d,p)}\\
$1\, ^1B_{1u}$ & 0.1036 & 523  & 2.37 & 97.\%    & 3.\%  \\
$2\, ^1B_{1u}$ & 0.0033 & 489  & 2.54 & 3.\%    & 97.\%  \\
        &      &      & DS$_2$ = -0.17 eV & &         \\
\multicolumn{6}{c}{TD-CAM-B3LYP/STO-3G//CAM-B3LYP+D3/STO-3G}\\      
$1\, ^1B_{1u}$ & 0.0004 & 423  & 2.93 & 0.2\%   & 99.\% \\
$2\, ^1B_{1u}$  & 0.1628 & 404  & 3.07 & 99.\%   & 0.2\% \\
        &      &      & DS$_2$ = 0.14 eV & &         \\
\multicolumn{6}{c}{TD-lc-DFTB//lc-DFTB+D3}\\                       
$1\, ^1B_{1u}$ & 0.5782 & 495  & 2.50 & 99.8\%   & 0.24\% \\
$2\, ^1B_{1u}$ & 0.0013 & 451  & 2.75 & 0.21\%   & 99.8\% \\
        &      &      & DS$_2$ = -0.25 eV & &         \\
\multicolumn{6}{c}{TD-HF/6-31G(d,p)//B3LYP+D3/6-31G(d,p)}\\
$1\, ^1B_{1u}$ & 0.2087 & 474 & 2.62 & 100.\%   & 0.07\% \\
$2\, ^1B_{1u}$ & 0.0002 & 332 & 3.73 & 0.06\%  & 100.\% \\
        &      &      & DS$_2$ = -1.11 eV & &         \\
\multicolumn{6}{c}{TD-HF/STO-3G//B3LYP+D3/6-31G(d,p)}\\      
$1\, ^1B_{1u}$ & 0.3557 & 347 & 3.58 & 100.\%   & 0.01\% \\
$2\, ^1B_{1u}$ & 0.0001 & 282 & 4.40 & 0.01\%   & 100.\% \\
        &      &      & DS$_2$ = -0.82 eV & &         \\
\hline \hline
\end{tabular}
\footnotetext[1]{$(c_1+c_2)^2/2$.}
\footnotetext[2]{$(c_1-c_2)^2/2$.}
\end{table}
%%%%%%%
% EOF
%%%%%%%
% ------------------------------------------------------
\begin{figure}
\begin{tabular}{ll}
a) & \\
% b) & \includegraphics[width=0.4\textwidth]{./graphics/P2_NTOs_621nm.eps} \\
b) & \includegraphics[width=0.4\textwidth]{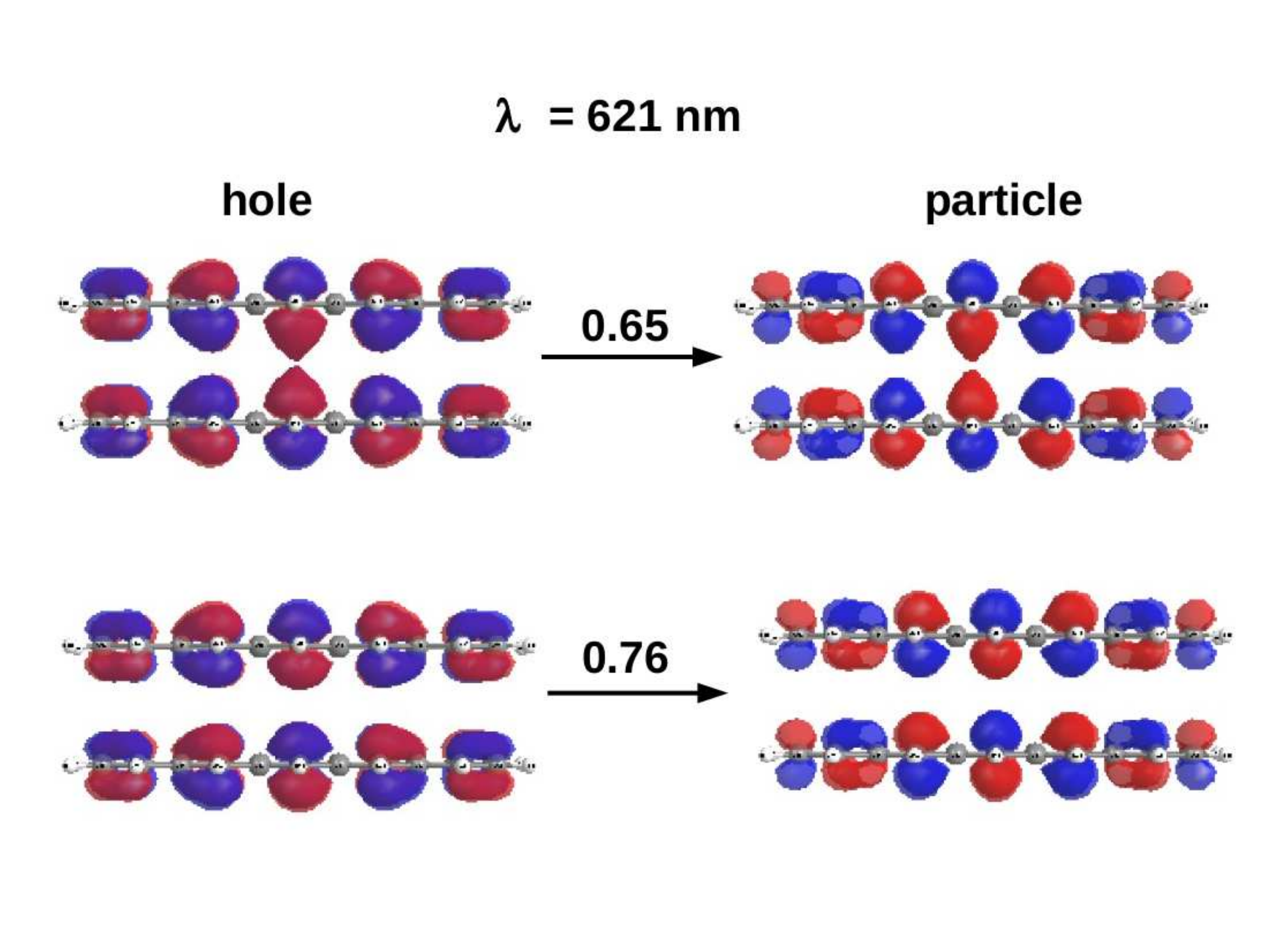} \\
%   & \includegraphics[width=0.4\textwidth]{./graphics/P2_NTOs_735nm.eps} 
   & \includegraphics[width=0.4\textwidth]{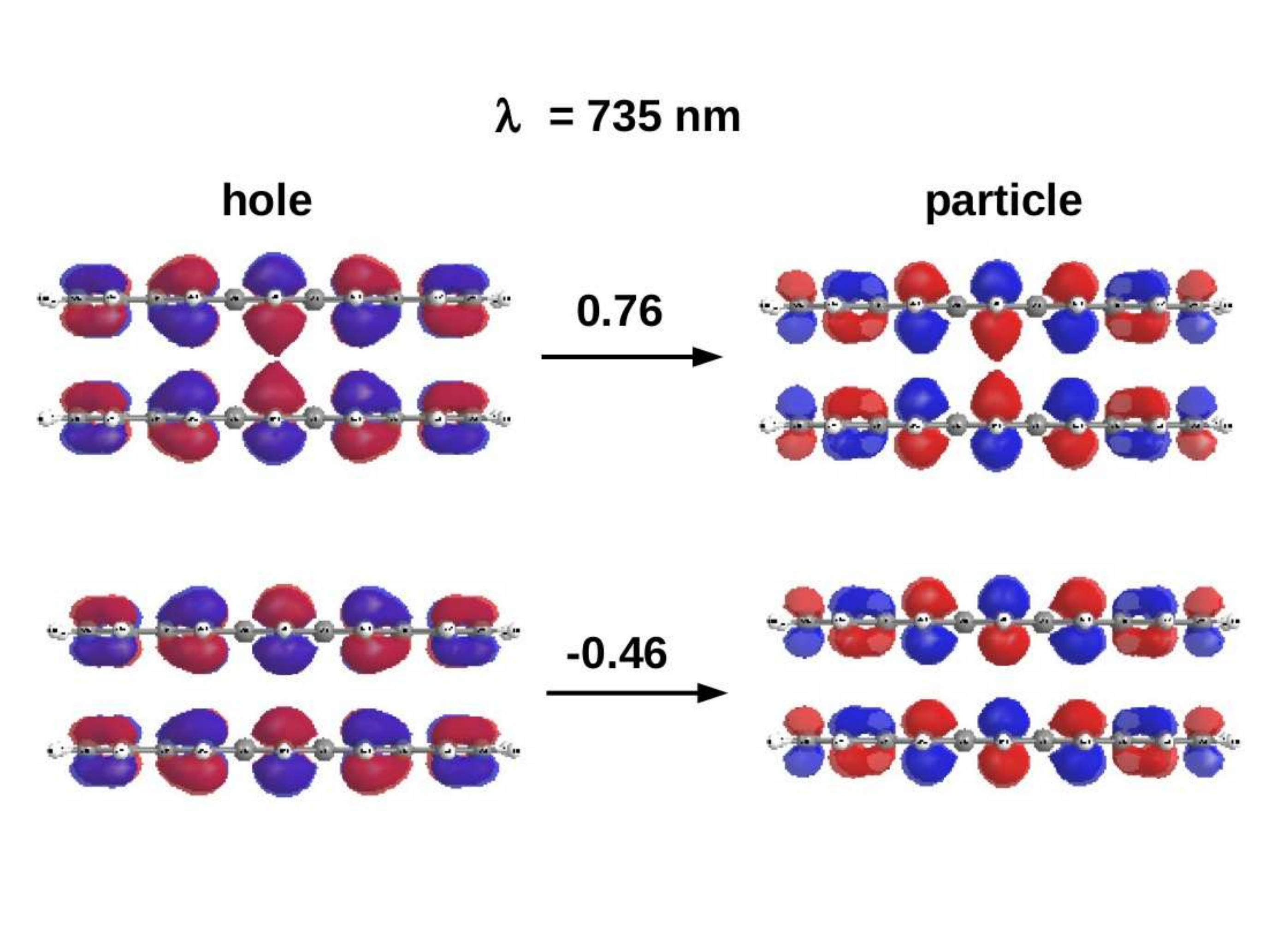} 
\end{tabular}
\caption{
\label{fig:P2_NTOs}
Dimer NTOs and renormalized coefficients:
(a) TD-B3LYP/6-31G(d,p)//D3-B3LYP/6-31G(d,p) 621 nm,
(b) TD-B3LYP/6-31G(d,p)//D3-B3LYP/6-31G(d,p) 735 nm.
}
\end{figure}
% -------------------------------------------------------
Figure~\ref{fig:P2_NTOs} shows the stacked pentacene dimer MOs which may be compared
with the stacked ethylene dimer MOs (Fig.~\ref{fig:stacked_MOs}). It is easy to identify
the coefficient $c_1$ for the $^1(H,L+1)$ configuration and the coefficient $c_2$ for
the $^1(H-1,L)$ configuration. Table~\ref{tab:CTETP2} shows how the excitations
split into a bright ET exciton and a much darker CT exciton.  The energy splitting
DS = ET - CT is the Davydov splitting.  Our exciton model (Fig.~\ref{fig:Kasha})
% ------------------------------------------------------
\begin{figure}
\includegraphics[width=0.5\textwidth]{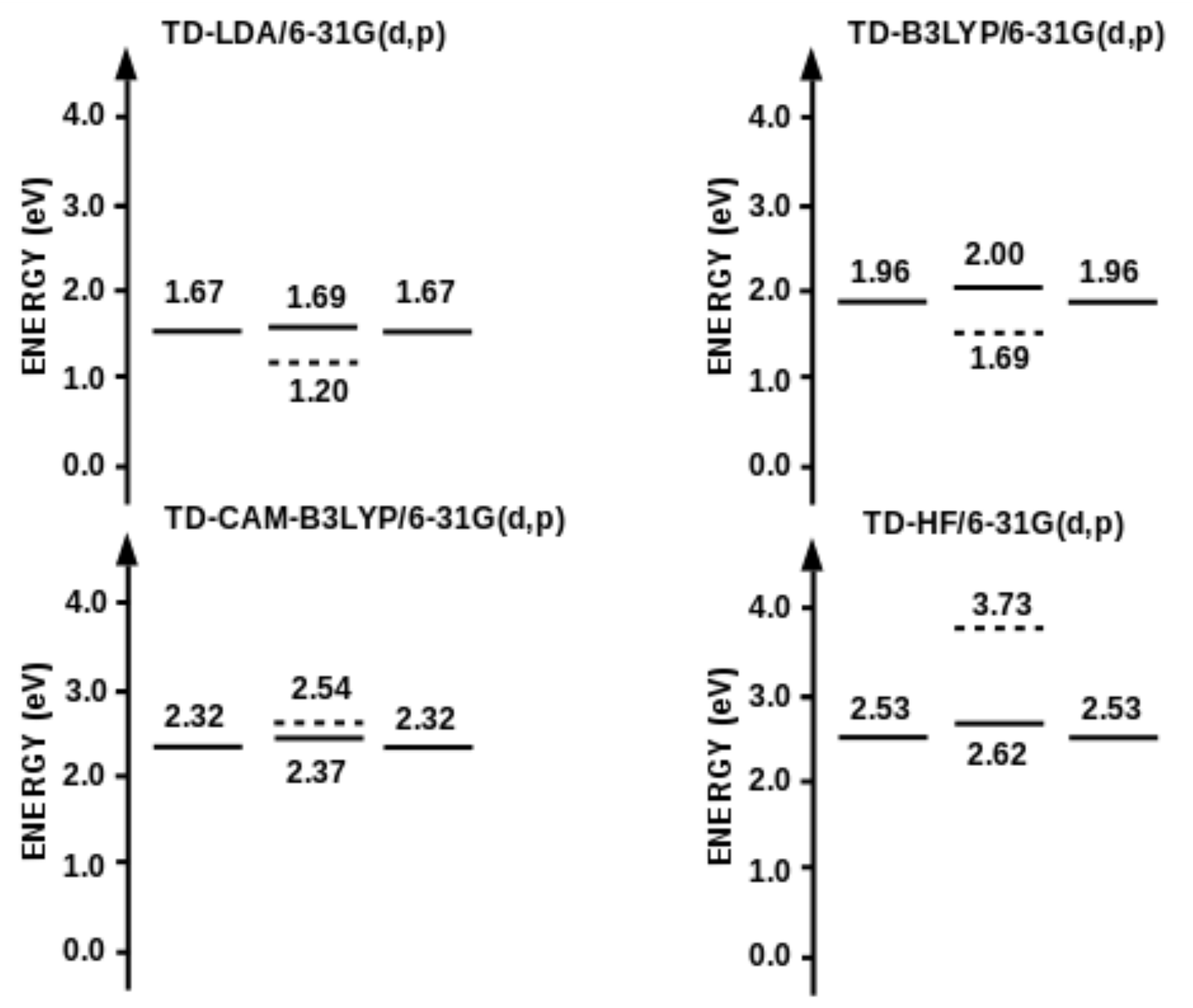}
\caption{
\label{fig:Kasha_reality_check}
Comparison of exciton diagrams for different functionals using the data from
Tables~\ref{tab:CTETP1} and \ref{tab:CTETP2}.
}
\end{figure}
% -------------------------------------------------------
predicts a positive DS and this is exactly what is seen in our TD-LDA/6-31G(d,p), 
TD-LDA/STO-3G, TD-B3LYP/6-31G(d,p), TD-B3LYP/STO-3G, TD-CAM-B3LYP/STO-3G, and 
TD-DFTB calculations.  However, improving the description of charge transfer by 
adding more HF exchange leads to negative values of DS in the TD-CAM-B3LYP/6-31G(d,p),
TD-HF/6-31G(d,p), TD-HF/STO-3G, and TD-lc-DFTB models and a very different 
picture of exciton coupling (Fig.~\ref{fig:Kasha_reality_check}) 
than that seen in Kasha's original exciton model.  
Careful rereading of the classic exciton theory article of Kasha, Rawls, 
and El-Bayoumi \cite{KRE65} reveals that they took into account only 
dipole-dipole interactions but not charge-transfer effects.  As these charge
transfer effects are implicit in our calculations, we may
explain the observation that Hartree-Fock exchange leads to CT excitonic states of 
higher energy than ET states by the large amount of energy needed to separate
charges.
Interestingly the DS obtained from TD-DFTB//DFTB+D3 resembles most closely that 
obtained with the TD-B3LYP/STO-3G//B3LYP+D3/STO-3G or 
TD-LDA/STO-3G//LDA+D3/STO-3G,
consistent with the idea that the DS is primarily determined by the overlap which 
is too small when a minimal basis set is used.  The situation changes
markedly in going to the long-range corrected functionals.  Here the TD-lc-DFTB
DS is closer to the TD-CAM-DFTB/6-31G(d,p) DS than to the TD-CAM-DFTB/STO-3G DS.

% \input{tables/CTETP3.tex}
% =======================================
% File: CTETP3.tex
% Last modified: 5 November 2017
% =======================================
\begin{table}
\squeezetable
\caption{Relative percentages of CT and ET excitonic transitions to the principle
transition for three parallel stacked pentacenes as a function of method.
See Eq.~(\ref{eq:theory.5}). 
DS is the Davydov splitting between the lowest energy CT and the highest energy ET excitonic transitions.
\label{tab:CTETP3a}}
\begin{tabular}{ccccc}
\hline \hline
\multicolumn{4}{c}{Method} \\
State & $f$ (unitless) & $\lambda$ (nm) & $\Delta E$ (eV) \\
\hline
\multicolumn{4}{c}{TD-LDA/6-31G(d,p)//LDA/6-31G(d,p)}\\
$1\, ^1B_{1u}$ & 0.0000 & 2444 & 0.51  \\  % 1st state
      &                & \multicolumn{2}{c}{Analysis} \\
\cline{3-4} 
        &    & 100\% & $^1$(H,L) \\  % C1**2
        &    & 0\% \footnotemark[1] & (ET$_{12}$+ET$_{23}$)/$\sqrt{3}$ \\ % (3*C2+2*C3+3*C4)**2/24
        &    & 0\% \footnotemark[2] & ($\sqrt{3}/2$)[ET$_{13}$-(1/3)(ET$_{12}$+ET$_{23}$)] \\ % (2/3)*C3**2
        &    & 0\% \footnotemark[3] & (CT$_{12}$+CT$_{23}$)/$\sqrt{2}$ \\    % (C2-C4)**2/2
        &    & 0\% \footnotemark[4] & CT$_{13}$                       \\    % (C2-2*C3+C4)**2/8
        &    & 0\% & $^1$(H-2,L+2) \\  % C5**2
$2\, ^1B_{1u}$ & 0.0002 & 1040 & 1.20  \\ % 4th state
      &                & \multicolumn{2}{c}{Analysis} \\
\cline{3-4} 
        &    & 0\% & $^1$(H,L) \\  % C1**2
        &    & 8.9\% \footnotemark[1] & (ET$_{12}$+ET$_{23}$)/$\sqrt{3}$ \\ % (3*C2+2*C3+3*C4)**2/24
        &    & 2.9\% \footnotemark[2] & ($\sqrt{3}/2$)[ET$_{13}$-(1/3)(ET$_{12}$+ET$_{23}$)] \\ % (2/3)*C3**2
        &    & 81.9\% \footnotemark[3] & {\bf (CT$_{12}$+CT$_{23}$)/$\sqrt{2}$} \\    % (C2-C4)**2/2
        &    & 6.3\% \footnotemark[4] & CT$_{13}$                       \\    % (C2-2*C3+C4)**2/8
        &    & 0\% & $^1$(H-2,L+2) \\  % C5**2
$3\, ^1B_{1u}$ & 0.0003 & 950 & 1.30 \\ % 5th state
      &                & \multicolumn{2}{c}{Analysis} \\
\cline{3-4} 
        &    & 0\% & $^1$(H,L) \\  % C1**2
        &    & 1.2\% \footnotemark[1] & (ET$_{12}$+ET$_{23}$)/$\sqrt{3}$ \\ % (3*C2+2*C3+3*C4)**2/24
        &    & 34.2\% \footnotemark[2] & ($\sqrt{3}/2$)[ET$_{13}$-(1/3)(ET$_{12}$+ET$_{23}$)] \\ % (2/3)*C3**2
        &    & 7.6\% \footnotemark[3] & (CT$_{12}$+CT$_{23}$)/$\sqrt{2}$ \\    % (C2-C4)**2/2
        &    & 54.6\% \footnotemark[4] & CT$_{13}$                       \\    % (C2-2*C3+C4)**2/8
        &    & 2.3\% & $^1$(H-2,L+2) \\  % C5**2
$4\, ^1B_{1u}$ & 0.0405 & 730. & 1.70 \\  % 6th state
      &                & \multicolumn{2}{c}{Analysis} \\
\cline{3-4} 
        &     & 0\% & $^1$(H,L) \\  % C1**2
        &    & 80.6\% \footnotemark[1] & {\bf (ET$_{12}$+ET$_{23}$)/$\sqrt{3}$} \\ % (3*C2+2*C3+3*C4)**2/24
        &    & 19.1\% \footnotemark[2] & ($\sqrt{3}/2$)[ET$_{13}$-(1/3)(ET$_{12}$+ET$_{23}$)] \\ % (2/3)*C3**2
        &    & 0.3\% \footnotemark[3] & (CT$_{12}$+CT$_{23}$)/$\sqrt{2}$ \\    % (C2-C4)**2/2
        &    & 0.02\% \footnotemark[4] & CT$_{13}$                       \\    % (C2-2*C3+C4)**2/8
        &    & 0\% & $^1$(H-2,L+2) \\  % C5**2
\multicolumn{4}{c}{DS$_3$ = 0.50 eV} \\  % DS$_2$ = 0.49 eV
\hline
\multicolumn{4}{c}{TD-LDA/STO-3G//LDA/STO-3G} \\                
$1\, ^1B _{1u}$ & 0.0000 & 1024 & 1.21 \\  % 1st state
      &                & \multicolumn{2}{c}{Analysis} \\
\cline{3-4} 
      &    & 88.4\% & $^1$(H,L) \\  % C1**2
      &    & 0.1\% \footnotemark[1] & (ET$_{12}$+ET$_{23}$)/$\sqrt{3}$ \\ % (3*C2+2*C3+3*C4)**2/24
      &    & 3.5\% \footnotemark[2] & ($\sqrt{3}/2$)[ET$_{13}$-(1/3)(ET$_{12}$+ET$_{23}$)] \\ % (2/3)*C3**2
      &    & 2.3\% \footnotemark[3] & (CT$_{12}$+CT$_{23}$)/$\sqrt{2}$ \\    % (C2-C4)**2/2
      &    & 5.7\% \footnotemark[4] & CT$_{13}$                       \\    % (C2-2*C3+C4)**2/8
      &    & 0\% & $^1$(H-2,L+2) \\  % C5**2
$2\, ^1B _{1u}$ & 0.0000 & 816 & 1.52 \\  % 4th state
      &                & \multicolumn{2}{c}{Analysis} \\
\cline{3-4} 
      &    & 0\% & $^1$(H,L) \\  % C1**2
      &    & 0.3\% \footnotemark[1] & (ET$_{12}$+ET$_{23}$)/$\sqrt{3}$ \\ % (3*C2+2*C3+3*C4)**2/24
      &    & 7.3\% \footnotemark[2] & ($\sqrt{3}/2$)[ET$_{13}$-(1/3)(ET$_{12}$+ET$_{23}$)] \\ % (2/3)*C3**2
      &    & 79.7\% \footnotemark[3] & {\bf (CT$_{12}$+CT$_{23}$)/$\sqrt{2}$} \\    % (C2-C4)**2/2
      &    & 12.7\% \footnotemark[4] & CT$_{13}$                       \\    % (C2-2*C3+C4)**2/8
      &    & 0\% & $^1$(H-2,L+2) \\  % C5**2
$3\, ^1B _{1u}$ & 0.0000 & 754 & 1.64 \\  % 5th state
      &                & \multicolumn{2}{c}{Analysis} \\
\cline{3-4} 
      &    &  7.2\% & $^1$(H,L) \\  % C1**2
      &    &  0.1\% \footnotemark[1] & (ET$_{12}$+ET$_{23}$)/$\sqrt{3}$ \\ % (3*C2+2*C3+3*C4)**2/24
      &    & 39.2\% \footnotemark[2] & ($\sqrt{3}/2$)[ET$_{13}$-(1/3)(ET$_{12}$+ET$_{23}$)] \\ % (2/3)*C3**2
      &    & 22.8\% \footnotemark[3] & (CT$_{12}$+CT$_{23}$)/$\sqrt{2}$ \\    % (C2-C4)**2/2
      &    &  4.1\% \footnotemark[4] & CT$_{13}$                       \\    % (C2-2*C3+C4)**2/8
      &    & 26.7\% & $^1$(H-2,L+2) \\  % C5**2
$4\, ^1B _{1u}$ & 0.0664 & 571  & 2.17 \\ % 7th state
      &                & \multicolumn{2}{c}{Analysis} \\
\cline{3-4} 
      &     & 0\% & $^1$(H,L) \\  % C1**2
      &    & 78.5\% \footnotemark[1] & {\bf (ET$_{12}$+ET$_{23}$)/$\sqrt{3}$} \\ % (3*C2+2*C3+3*C4)**2/24
      &    & 21.4\% \footnotemark[2] & ($\sqrt{3}/2$)[ET$_{13}$-(1/3)(ET$_{12}$+ET$_{23}$)] \\ % (2/3)*C3**2
      &    &  0.05\% \footnotemark[3] & (CT$_{12}$+CT$_{23}$)/$\sqrt{2}$ \\    % (C2-C4)**2/2
      &    &  0.05\% \footnotemark[4] & CT$_{13}$                       \\    % (C2-2*C3+C4)**2/8
      &    & 0\% & $^1$(H-2,L+2) \\  % C5**2
% $5\, ^1B _{1u}$ & 0.0001 & 457  & 2.27 \\ % 9th state
%       &                & \multicolumn{2}{c}{Analysis} \\
% \cline{3-4} 
%       &    & 3.246 26.2\% & $^1$(H,L) \\  % C1**2
%       &    & 1.821 14.7\% \footnotemark[1] & (ET$_{12}$+ET$_{23}$)/$\sqrt{3}$ \\ % (3*C2+2*C3+3*C4)**2/24
%       &    & 1.671 13.5\% \footnotemark[2] & ($\sqrt{3}/2$)[ET$_{13}$-(1/3)(ET$_{12}$+ET$_{23}$)] \\ % (2/3)*C3**2
%       &    & 2.200 17.7\% \footnotemark[3] & (CT$_{12}$+CT$_{23}$)/$\sqrt{2}$ \\    % (C2-C4)**2/2
%       &    & 3.464 27.9\% \footnotemark[4] & CT$_{13}$                       \\    % (C2-2*C3+C4)**2/8
%       &    & 0\% & $^1$(H-2,L+2) \\  % C5**2
\multicolumn{4}{c}{DS$_3$ = 0.65 eV} \\  % DS$_2$ = 0.65 eV
\hline \hline
\end{tabular}
\footnotetext[1]{$(3c_2+2c_3+3c_4)^2/24$.}
\footnotetext[2]{$2c_3^2/3$.}
\footnotetext[3]{$(c_2-c_4)^2/2$.}
\footnotetext[4]{$(c_2-2c_3+c_4)^2/8$.}
\end{table}
% ------------------------------------------------------------------
\begin{table}
\squeezetable
\caption{Relative percentages of CT and ET excitonic transitions to the principle
transition for three parallel stacked pentacenes as a function of method.
See Eq.~(\ref{eq:theory.5}). 
DS is the Davydov splitting between the lowest energy CT and the highest energy ET excitonic transitions.
\label{tab:CTETP3b}}
\begin{tabular}{cccccc}
\hline \hline
\multicolumn{4}{c}{Method} \\
State & $f$ (unitless) & $\lambda$ (nm) & $\Delta E$ (eV)  \\
\hline
\multicolumn{4}{c}{TD-B3LYP/6-31G(d,p)//B3LYP+D3/6-31G(d,p)}\\     
$1\, ^1B _{1u}$ & 0.0001 & 1220  & 1.02 \\  % 1st state
      &                & \multicolumn{2}{c}{Analysis} \\
\cline{3-4} 
        &     & 100.0\% & $^1$(H,L) \\  % C1**2
        &    & 0\% \footnotemark[1] & (ET$_{12}$+ET$_{23}$)/$\sqrt{3}$ \\ % (3*C2+2*C3+3*C4)**2/24
        &    & 0\% \footnotemark[2] & ($\sqrt{3}/2$)[ET$_{13}$-(1/3)(ET$_{12}$+ET$_{23}$)] \\ % (2/3)*C3**2
        &    & 0\% \footnotemark[3] & (CT$_{12}$+CT$_{23}$)/$\sqrt{2}$ \\    % (C2-C4)**2/2
        &    & 0\% \footnotemark[4] & CT$_{13}$                       \\    % (C2-2*C3+C4)**2/8
        &    & 0\% & $^1$(H-2,L+2) \\  % C5**2
$2\, ^1B _{1u}$ & 0.0006 & 734  & 1.69  \\ % 4th state
      &                & \multicolumn{2}{c}{Analysis} \\
\cline{3-4} 
        &    &  0\% & $^1$(H,L) \\  % C1**2
        &    &  0.8\% \footnotemark[1] & (ET$_{12}$+ET$_{23}$)/$\sqrt{3}$ \\ % (3*C2+2*C3+3*C4)**2/24
        &    &  0\% \footnotemark[2] & ($\sqrt{3}/2$)[ET$_{13}$-(1/3)(ET$_{12}$+ET$_{23}$)] \\ % (2/3)*C3**2
        &    & 98.8\% \footnotemark[3] & {\bf (CT$_{12}$+CT$_{23}$)/$\sqrt{2}$} \\    % (C2-C4)**2/2
        &    &  0.3\% \footnotemark[4] & CT$_{13}$                       \\    % (C2-2*C3+C4)**2/8
        &    &  0\% & $^1$(H-2,L+2) \\  % C5**2
$3\, ^1B _{1u}$ & 0.0012 & 657  & 1.89  \\ % 5th state
      &                & \multicolumn{2}{c}{Analysis} \\
\cline{3-4} 
        &    &  0\% & $^1$(H,L) \\  % C1**2
        &    &  0.6\% \footnotemark[1] & (ET$_{12}$+ET$_{23}$)/$\sqrt{3}$ \\ % (3*C2+2*C3+3*C4)**2/24
        &    & 39.7\% \footnotemark[2] & ($\sqrt{3}/2$)[ET$_{13}$-(1/3)(ET$_{12}$+ET$_{23}$)] \\ % (2/3)*C3**2
        &    &  0.2\% \footnotemark[3] & (CT$_{12}$+CT$_{23}$)/$\sqrt{2}$ \\    % (C2-C4)**2/2
        &    & 59.5\% \footnotemark[4] & CT$_{13}$                       \\    % (C2-2*C3+C4)**2/8
        &    &  0\% & $^1$(H-2,L+2) \\  % C5**2
$4\, ^1B _{1u}$ & 0.0723 & 618  & 2.01 \\ % 6th state
      &                & \multicolumn{2}{c}{Analysis} \\
\cline{3-4} 
        &    &  0\% & $^1$(H,L) \\  % C1**2
        &    & 82.9\% \footnotemark[1] & {\bf (ET$_{12}$+ET$_{23}$)/$\sqrt{3}$} \\ % (3*C2+2*C3+3*C4)**2/24
        &    & 15.8\% \footnotemark[2] & ($\sqrt{3}/2$)[ET$_{13}$-(1/3)(ET$_{12}$+ET$_{23}$)] \\ % (2/3)*C3**2
        &    &  0.9\% \footnotemark[3] & (CT$_{12}$+CT$_{23}$)/$\sqrt{2}$ \\    % (C2-C4)**2/2
        &    &  0.4\% \footnotemark[4] & CT$_{13}$                       \\    % (C2-2*C3+C4)**2/8
        &    &  0\% & $^1$(H-2,L+2) \\  % C5**2
\multicolumn{4}{c}{DS$_3$ = 0.32 eV} \\  % DSE$_2$ = 0.31 eV
\hline
\multicolumn{4}{c}{TD-B3LYP/STO-3G//B3LYP+D3/STO-3G}\\
$1\, ^1B _{1u}$ & 0.0001 & 801  & 1.55 \\  % 1st state
      &                & \multicolumn{2}{c}{Analysis} \\
\cline{3-4} 
        &     & 100\% & $^1$(H,L) \\  % C1**2
        &    & 0\% \footnotemark[1] & (ET$_{12}$+ET$_{23}$)/$\sqrt{3}$ \\ % (3*C2+2*C3+3*C4)**2/24
        &    & 0\% \footnotemark[2] & ($\sqrt{3}/2$)[ET$_{13}$-(1/3)(ET$_{12}$+ET$_{23}$)] \\ % (2/3)*C3**2
        &    & 0\% \footnotemark[3] & (CT$_{12}$+CT$_{23}$)/$\sqrt{2}$ \\    % (C2-C4)**2/2
        &    & 0\% \footnotemark[4] & CT$_{13}$                       \\    % (C2-2*C3+C4)**2/8
        &    & 0\% & $^1$(H-2,L+2) \\  % C5**2
$2\, ^1B _{1u}$ & 0.0000 & 602  & 2.06 \\ % 4th state
      &                & \multicolumn{2}{c}{Analysis} \\
\cline{3-4} 
        &    & 0\% & $^1$(H,L) \\  % C1**2
        &    & 0.0\% \footnotemark[1] & (ET$_{12}$+ET$_{23}$)/$\sqrt{3}$ \\ % (3*C2+2*C3+3*C4)**2/24
        &    & 0\% \footnotemark[2] & ($\sqrt{3}/2$)[ET$_{13}$-(1/3)(ET$_{12}$+ET$_{23}$)] \\ % (2/3)*C3**2
        &    & 100.0\% \footnotemark[3] & {\bf (CT$_{12}$+CT$_{23}$)/$\sqrt{2}$} \\    % (C2-C4)**2/2
        &    & 0.0\% \footnotemark[4] & CT$_{13}$                       \\    % (C2-2*C3+C4)**2/8
        &    & 0\% & $^1$(H-2,L+2) \\  % C5**2
$3\, ^1B _{1u}$ & 0.0003 & 539  & 2.30 \\  % 5th state
      &                & \multicolumn{2}{c}{Analysis} \\
\cline{3-4} 
        &    &  0\% & $^1$(H,L) \\  % C1**2
        &    &  7.0\% \footnotemark[1] & (ET$_{12}$+ET$_{23}$)/$\sqrt{3}$ \\ % (3*C2+2*C3+3*C4)**2/24
        &    & 79.6\% \footnotemark[2] & ($\sqrt{3}/2$)[ET$_{13}$-(1/3)(ET$_{12}$+ET$_{23}$)] \\ % (2/3)*C3**2
        &    &  0.0\% \footnotemark[3] & (CT$_{12}$+CT$_{23}$)/$\sqrt{2}$ \\    % (C2-C4)**2/2
        &    & 13.4\% \footnotemark[4] & CT$_{13}$                       \\    % (C2-2*C3+C4)**2/8
        &    &  0\% & $^1$(H-2,L+2) \\  % C5**2
$4\, ^1B _{1u}$ & 0.1180 & 482  & 2.57 \\ % 7th state
      &                & \multicolumn{2}{c}{Analysis} \\
\cline{3-4} 
        &    &  0\% & $^1$(H,L) \\  % C1**2
        &    & 82.9\% \footnotemark[1] & {\bf (ET$_{12}$+ET$_{23}$)/$\sqrt{3}$} \\ % (3*C2+2*C3+3*C4)**2/24
        &    & 17.1\% \footnotemark[2] & ($\sqrt{3}/2$)[ET$_{13}$-(1/3)(ET$_{12}$+ET$_{23}$)] \\ % (2/3)*C3**2
        &    &  0.0\% \footnotemark[3] & (CT$_{12}$+CT$_{23}$)/$\sqrt{2}$ \\    % (C2-C4)**2/2
        &    &  0.0\% \footnotemark[4] & CT$_{13}$                       \\    % (C2-2*C3+C4)**2/8
        &    &  0\% & $^1$(H-2,L+2) \\  % C5**2
\multicolumn{4}{c}{DS$_3$ = 0.51 eV} \\  % DS$_2$ = 0.51 eV
\hline \hline
\end{tabular}
\footnotetext[1]{$(3c_2+2c_3+3c_4)^2/24$.}
\footnotetext[2]{$2c_3^2/3$.}
\footnotetext[3]{$(c_2-c_4)^2/2$.}
\footnotetext[4]{$(c_2-2c_3+c_4)^2/8$.}
\end{table}
% ------------------------------------------------------------------
\begin{table}
\squeezetable
\caption{Relative percentages of CT and ET excitonic transitions to the principle
transition for three parallel stacked pentacenes as a function of method.
See Eq.~(\ref{eq:theory.5}). 
DS is the Davydov splitting between the lowest energy CT and the highest energy ET excitonic transitions.
\label{tab:CTETP3c}}
\begin{tabular}{cccccc}
\hline \hline
\multicolumn{4}{c}{Method} \\
State & $f$ (unitless) & $\lambda$ (nm) & $\Delta E$ (eV)  \\
\hline
\multicolumn{4}{c}{TD-DFTB//DFTB+D3} \\
$1\, ^1B _{1u}$ & 0.0000 & 1606 & 0.772 \\ % 1st state
      &                & \multicolumn{2}{c}{Analysis} \\
\cline{3-4} 
        &    & 86.9\% & $^1$(H,L) \\  % C1**2
        &    &  1.6\% \footnotemark[1] & (ET$_{12}$+ET$_{23}$)/$\sqrt{3}$ \\ % (3*C2+2*C3+3*C4)**2/24
        &    &  6.6\% \footnotemark[2] & ($\sqrt{3}/2$)[ET$_{13}$-(1/3)(ET$_{12}$+ET$_{23}$)] \\ % (2/3)*C3**2
        &    & 0\% \footnotemark[3] & (CT$_{12}$+CT$_{23}$)/$\sqrt{2}$ \\    % (C2-C4)**2/2
        &    &  4.9\% \footnotemark[4] & CT$_{13}$                       \\    % (C2-2*C3+C4)**2/8
        &    & 0\% & $^1$(H-2,L+2) \\  % C5**2
$2\, ^1B _{1u}$ & 0.0002 & 872 & 1.42 \\ % 4th state
      &                & \multicolumn{2}{c}{Analysis} \\
\cline{3-4} 
        &    & 0\% & $^1$(H,L) \\  % C1**2
        &    & 0.01\% \footnotemark[1] & (ET$_{12}$+ET$_{23}$)/$\sqrt{3}$ \\ % (3*C2+2*C3+3*C4)**2/24
        &    & 0\% \footnotemark[2] & ($\sqrt{3}/2$)[ET$_{13}$-(1/3)(ET$_{12}$+ET$_{23}$)] \\ % (2/3)*C3**2
        &    & 99.9\% \footnotemark[3] & {\bf (CT$_{12}$+CT$_{23}$)/$\sqrt{2}$ } \\    % (C2-C4)**2/2
        &    & 0.03\% \footnotemark[4] & CT$_{13}$                       \\    % (C2-2*C3+C4)**2/8
        &    & 0\% & $^1$(H-2,L+2) \\  % C5**2
$3\, ^1B _{1u}$ & 0.0004 & 819 & 1.51 \\ % 5th state
      &                & \multicolumn{2}{c}{Analysis} \\
\cline{3-4} 
        &    &  0\% & $^1$(H,L) \\  % C1**2
        &    &  6.0\% \footnotemark[1] & (ET$_{12}$+ET$_{23}$)/$\sqrt{3}$ \\ % (3*C2+2*C3+3*C4)**2/24
        &    & 39.8\% \footnotemark[2] & ($\sqrt{3}/2$)[ET$_{13}$-(1/3)(ET$_{12}$+ET$_{23}$)] \\ % (2/3)*C3**2
        &    &  7.5\% \footnotemark[3] & (CT$_{12}$+CT$_{23}$)/$\sqrt{2}$ \\    % (C2-C4)**2/2
        &    & 46.7\% \footnotemark[4] & CT$_{13}$                       \\    % (C2-2*C3+C4)**2/8
        &    &  0\% & $^1$(H-2,L+2) \\  % C5**2
$4\, ^1B _{1u}$ & 0.4122 & 624 & 1.99 \\ % 10th state
      &                & \multicolumn{2}{c}{Analysis} \\
\cline{3-4} 
        &     & 0\% & $^1$(H,L) \\  % C1**2
        &    &  4.2\% \footnotemark[1] & (ET$_{12}$+ET$_{23}$)/$\sqrt{3}$ \\ % (3*C2+2*C3+3*C4)**2/24
        &    & 47.6\% \footnotemark[2] & {\bf ($\sqrt{3}/2$)[ET$_{13}$-(1/3)(ET$_{12}$+ET$_{23}$)]} \\ % (2/3)*C3**2
        &    & 40.3\% \footnotemark[3] & (CT$_{12}$+CT$_{23}$)/$\sqrt{2}$ \\    % (C2-C4)**2/2
        &    &  7.9\% \footnotemark[4] & CT$_{13}$                       \\    % (C2-2*C3+C4)**2/8
        &    &  0\% & $^1$(H-2,L+2) \\  % C5**2
\multicolumn{4}{c}{DS$_3$ = 0.57 eV} \\ % DS$_2$ = 0.60 eV
\hline
\multicolumn{4}{c}{TD-CAM-B3LYP/6-31G(d,p)//B3LYP+D3/6-31G(d,p)}\\
$1\, ^1B _{1u}$ & 0.0005 & 764  & 1.62 \\ % 1st state
      &                & \multicolumn{2}{c}{Analysis} \\
\cline{3-4} 
        &    & 94.5\% & $^1$(H,L) \\  % C1**2
        &    &  0.7\% \footnotemark[1] & (ET$_{12}$+ET$_{23}$)/$\sqrt{3}$ \\ % (3*C2+2*C3+3*C4)**2/24
        &    &  2.7\% \footnotemark[2] & ($\sqrt{3}/2$)[ET$_{13}$-(1/3)(ET$_{12}$+ET$_{23}$)] \\ % (2/3)*C3**2
        &    &  0\% \footnotemark[3] & (CT$_{12}$+CT$_{23}$)/$\sqrt{2}$ \\    % (C2-C4)**2/2
        &    &  2.0\% \footnotemark[4] & CT$_{13}$                       \\    % (C2-2*C3+C4)**2/8
        &    &  0\% & $^1$(H-2,L+2) \\  % C5**2
$2\, ^1B _{1u}$ & 0.1301 & 520  & 2.38 \\ % 4th state
      &                & \multicolumn{2}{c}{Analysis} \\
\cline{3-4} 
        &    &  0\% & $^1$(H,L) \\  % C1**2
        &    & 74.4\% \footnotemark[1] & {\bf (ET$_{12}$+ET$_{23}$)/$\sqrt{3}$} \\ % (3*C2+2*C3+3*C4)**2/24
        &    & 20.6\% \footnotemark[2] & ($\sqrt{3}/2$)[ET$_{13}$-(1/3)(ET$_{12}$+ET$_{23}$)] \\ % (2/3)*C3**2
        &    &  4.9\% \footnotemark[3] & (CT$_{12}$+CT$_{23}$)/$\sqrt{2}$ \\    % (C2-C4)**2/2
        &    &  0.06\% \footnotemark[4] & CT$_{13}$                       \\    % (C2-2*C3+C4)**2/8
        &    & 0\% & $^1$(H-2,L+2) \\  % C5**2
$3\, ^1B _{1u}$ & 0.0074 & 489  & 2.53 \\ % 5th state
      &                & \multicolumn{2}{c}{Analysis} \\
\cline{3-4} 
        &    &  0\% & $^1$(H,L) \\  % C1**2
        &    &  4.5\% \footnotemark[1] & (ET$_{12}$+ET$_{23}$)/$\sqrt{3}$ \\ % (3*C2+2*C3+3*C4)**2/24
        &    &  0\% \footnotemark[2] & ($\sqrt{3}/2$)[ET$_{13}$-(1/3)(ET$_{12}$+ET$_{23}$)] \\ % (2/3)*C3**2
        &    & 94.0\% \footnotemark[3] & {\bf (CT$_{12}$+CT$_{23}$)/$\sqrt{2}$} \\    % (C2-C4)**2/2
        &    &  1.5\% \footnotemark[4] & CT$_{13}$                       \\    % (C2-2*C3+C4)**2/8
        &    &  0\% & $^1$(H-2,L+2) \\  % C5**2
$4\, ^1B _{1u}$ & 0.0001 & 435  & 2.85 \\ % 7th state
      &                & \multicolumn{2}{c}{Analysis} \\
\cline{3-4} 
        &    &  4.2\% & $^1$(H,L) \\  % C1**2
        &    &  2.0\% \footnotemark[1] & (ET$_{12}$+ET$_{23}$)/$\sqrt{3}$ \\ % (3*C2+2*C3+3*C4)**2/24
        &    & 22.9\% \footnotemark[2] & ($\sqrt{3}/2$)[ET$_{13}$-(1/3)(ET$_{12}$+ET$_{23}$)] \\ % (2/3)*C3**2
        &    &  7.9\% \footnotemark[3] & (CT$_{12}$+CT$_{23}$)/$\sqrt{2}$ \\    % (C2-C4)**2/2
        &    & 40.3\% \footnotemark[4] & CT$_{13}$                       \\    % (C2-2*C3+C4)**2/8
        &    & 22.7\% & $^1$(H-2,L+2) \\  % C5**2
\multicolumn{4}{c}{DS$_3$ = -0.15 eV} \\  % DS$_2$ = 0.17 eV
\hline \hline
\end{tabular}
\footnotetext[1]{$(3c_2+2c_3+3c_4)^2/24$.}
\footnotetext[2]{$2c_3^2/3$.}
\footnotetext[3]{$(c_2-c_4)^2/2$.}
\footnotetext[4]{$(c_2-2c_3+c_4)^2/8$.}
\end{table}
% ------------------------------------------------------------------
\begin{table}
\squeezetable
\caption{Relative percentages of CT and ET excitonic transitions to the principle
transition for three parallel stacked pentacenes as a function of method.
See Eq.~(\ref{eq:theory.5}). 
DS is the Davydov splitting between the lowest energy CT and the highest energy ET excitonic transitions.
\label{tab:CTETP3d}}
\begin{tabular}{cccccc}
\hline \hline
\multicolumn{4}{c}{Method} \\
State & $f$ (unitless) & $\lambda$ (nm) & $\Delta E$ (eV)  \\
\hline
\multicolumn{4}{c}{TD-CAM-B3LYP/STO-3G//CAM-B3LYP+D3/STO-3G}\\      
$1\, ^1B _{1u}$ & 0.0006 & 544  & 2.28 \\ % 1st step
      &                & \multicolumn{2}{c}{Analysis} \\
\cline{3-4} 
        &    & 93.0\% & $^1$(H,L) \\  % C1**2
        &    &  4.1\% \footnotemark[1] & (ET$_{12}$+ET$_{23}$)/$\sqrt{3}$ \\ % (3*C2+2*C3+3*C4)**2/24
        &    &  1.7\% \footnotemark[2] & ($\sqrt{3}/2$)[ET$_{13}$-(1/3)(ET$_{12}$+ET$_{23}$)] \\ % (2/3)*C3**2
        &    &  0\% \footnotemark[3] & (CT$_{12}$+CT$_{23}$)/$\sqrt{2}$ \\    % (C2-C4)**2/2
        &    &  1.2\% \footnotemark[4] & CT$_{13}$                       \\    % (C2-2*C3+C4)**2/8
        &    &  0\% & $^1$(H-2,L+2) \\  % C5**2
$2\, ^1B _{1u}$ & 0.0000 & 423  & 2.93 \\ % 4th step
      &                & \multicolumn{2}{c}{Analysis} \\
\cline{3-4} 
        &    &  0\% & $^1$(H,L) \\  % C1**2
        &    &  4.4\% \footnotemark[1] & (ET$_{12}$+ET$_{23}$)/$\sqrt{3}$ \\ % (3*C2+2*C3+3*C4)**2/24
        &    &  0\% \footnotemark[2] & ($\sqrt{3}/2$)[ET$_{13}$-(1/3)(ET$_{12}$+ET$_{23}$)] \\ % (2/3)*C3**2
        &    & 89.7\% \footnotemark[3] & {\bf (CT$_{12}$+CT$_{23}$)/$\sqrt{2}$} \\    % (C2-C4)**2/2
        &    &  5.9\% \footnotemark[4] & CT$_{13}$                       \\    % (C2-2*C3+C4)**2/8
        &    &  0\% & $^1$(H-2,L+2) \\  % C5**2
$3\, ^1B _{1u}$ & 0.2165 & 402  & 3.08 \\ % 5th step
      &                & \multicolumn{2}{c}{Analysis} \\
\cline{3-4} 
        &    & 0\% & $^1$(H,L) \\  % C1**2
        &    & 79.6\% \footnotemark[1] & {\bf (ET$_{12}$+ET$_{23}$)/$\sqrt{3}$} \\ % (3*C2+2*C3+3*C4)**2/24
        &    & 20.3\% \footnotemark[2] & ($\sqrt{3}/2$)[ET$_{13}$-(1/3)(ET$_{12}$+ET$_{23}$)] \\ % (2/3)*C3**2
        &    &  0.03\% \footnotemark[3] & (CT$_{12}$+CT$_{23}$)/$\sqrt{2}$ \\    % (C2-C4)**2/2
        &    &  0.1\% \footnotemark[4] & CT$_{13}$                       \\    % (C2-2*C3+C4)**2/8
        &    &  0\% & $^1$(H-2,L+2) \\  % C5**2
$4\, ^1B _{1u}$ & 0.0001 & 371  & 3.34 \\ % 7th state
      &                & \multicolumn{2}{c}{Analysis} \\
\cline{3-4} 
        &    &  0\% & $^1$(H,L) \\  % C1**2
        &    &  3.4\% \footnotemark[1] & (ET$_{12}$+ET$_{23}$)/$\sqrt{3}$ \\ % (3*C2+2*C3+3*C4)**2/24
        &    & 23.9\% \footnotemark[2] & ($\sqrt{3}/2$)[ET$_{13}$-(1/3)(ET$_{12}$+ET$_{23}$)] \\ % (2/3)*C3**2
        &    &  0.7\% \footnotemark[3] & (CT$_{12}$+CT$_{23}$)/$\sqrt{2}$ \\    % (C2-C4)**2/2
        &    & 45.1\% \footnotemark[4] & CT$_{13}$                       \\    % (C2-2*C3+C4)**2/8
        &    & 26.8\% & $^1$(H-2,L+2) \\  % C5**2
\multicolumn{4}{c}{DS$_3$ = 0.15 eV} \\ % DS$_2$ = 0.14 eV
\hline
\multicolumn{4}{c}{TD-lc-DFTB//lc-DFTB+D3}\\                       
$1\, ^1B _{1u}$ & 0.0028 & 705  & 1.76 \\ % 1st state
      &                & \multicolumn{2}{c}{Analysis} \\
\cline{3-4} 
        &    & 88.4\% & $^1$(H,L) \\  % C1**2
        &    &  1.5\% \footnotemark[1] & (ET$_{12}$+ET$_{23}$)/$\sqrt{3}$ \\ % (3*C2+2*C3+3*C4)**2/24
        &    &  5.8\% \footnotemark[2] & ($\sqrt{3}/2$)[ET$_{13}$-(1/3)(ET$_{12}$+ET$_{23}$)] \\ % (2/3)*C3**2
        &    &  0\% \footnotemark[3] & (CT$_{12}$+CT$_{23}$)/$\sqrt{2}$ \\    % (C2-C4)**2/2
        &    &  4.4\% \footnotemark[4] & CT$_{13}$                       \\    % (C2-2*C3+C4)**2/8
        &    &  0\% & $^1$(H-2,L+2) \\  % C5**2
$2\, ^1B _{1u}$ & 0.8238 & 487  & 2.54 \\ % 3rd state
      &                & \multicolumn{2}{c}{Analysis} \\
\cline{3-4} 
        &    &  0\% & $^1$(H,L) \\  % C1**2
        &    & 29.6\% \footnotemark[1] & {\bf (ET$_{12}$+ET$_{23}$)/$\sqrt{3}$} \\ % (3*C2+2*C3+3*C4)**2/24
        &    & 26.0\% \footnotemark[2] & ($\sqrt{3}/2$)[ET$_{13}$-(1/3)(ET$_{12}$+ET$_{23}$)] \\ % (2/3)*C3**2
        &    & 14.2\% \footnotemark[3] & (CT$_{12}$+CT$_{23}$)/$\sqrt{2}$ \\    % (C2-C4)**2/2
        &    & 30.2\% \footnotemark[4] & CT$_{13}$                       \\    % (C2-2*C3+C4)**2/8
        &    &  0\% & $^1$(H-2,L+2) \\  % C5**2
$3\, ^1B _{1u}$ & 0.0018 & 452  & 2.74 \\  % 7th state
      &                & \multicolumn{2}{c}{Analysis} \\
\cline{3-4} 
        &    &   0\% & $^1$(H,L) \\  % C1**2
        &    &   0.01\% \footnotemark[1] & (ET$_{12}$+ET$_{23}$)/$\sqrt{3}$ \\ % (3*C2+2*C3+3*C4)**2/24
        &    &   0\% \footnotemark[2] & ($\sqrt{3}/2$)[ET$_{13}$-(1/3)(ET$_{12}$+ET$_{23}$)] \\ % (2/3)*C3**2
        &    & 100.0\% \footnotemark[3] & {\bf (CT$_{12}$+CT$_{23}$)/$\sqrt{2}$} \\    % (C2-C4)**2/2
        &    &   0.003\% \footnotemark[4] & CT$_{13}$                       \\    % (C2-2*C3+C4)**2/8
        &    &   0\% & $^1$(H-2,L+2) \\  % C5**2
$4\, ^1B _{1u}$ & 0.0007 & 409  & 3.03 \\ % 11th state
      &                & \multicolumn{2}{c}{Analysis} \\
\cline{3-4} 
        &    &  0\% & $^1$(H,L) \\  % C1**2
        &    &  3.8\% \footnotemark[1] & (ET$_{12}$+ET$_{23}$)/$\sqrt{3}$ \\ % (3*C2+2*C3+3*C4)**2/24
        &    & 15.2\% \footnotemark[2] & ($\sqrt{3}/2$)[ET$_{13}$-(1/3)(ET$_{12}$+ET$_{23}$)] \\ % (2/3)*C3**2
        &    &  0\% \footnotemark[3] & (CT$_{12}$+CT$_{23}$)/$\sqrt{2}$ \\    % (C2-C4)**2/2
        &    & 11.4\% \footnotemark[4] & CT$_{13}$                       \\    % (C2-2*C3+C4)**2/8
        &    & 69.5\% & $^1$(H-2,L+2) \\  % C5**2
\multicolumn{4}{c}{DS$_3$ = -0.20 eV} \\ % DS$_2$ = 0.14 eV
%mec \multicolumn{4}{c}{{\bf STOP}}\\  % ********************************************
\hline \hline
\end{tabular}
\footnotetext[1]{$(3c_2+2c_3+3c_4)^2/24$.}
\footnotetext[2]{$2c_3^2/3$.}
\footnotetext[3]{$(c_2-c_4)^2/2$.}
\footnotetext[4]{$(c_2-2c_3+c_4)^2/8$.}
\end{table}
% ------------------------------------------------------------------
\begin{table}
\squeezetable
\caption{Relative percentages of CT and ET excitonic transitions to the principle
transition for three parallel stacked pentacenes as a function of method.
See Eq.~(\ref{eq:theory.5}). 
DS is the Davydov splitting between the lowest energy CT and the highest energy ET excitonic transitions.
\label{tab:CTETP3e}}
\begin{tabular}{cccccc}
\hline \hline
\multicolumn{4}{c}{Method} \\
State & $f$ (unitless) & $\lambda$ (nm) & $\Delta E$ (eV)  \\
\hline
\multicolumn{4}{c}{TD-HF/STO-3G//B3LYP+D3/6-31G(d,p)}\\      
$1\, ^1B _{1u}$ & 0.0051 & 417  & 2.97 \\ % 1st step
      &                & \multicolumn{2}{c}{Analysis} \\
\cline{3-4} 
        &    & 77.8\% & $^1$(H,L) \\  % C1**2
        &    & 0.2\% \footnotemark[1] & (ET$_{12}$+ET$_{23}$)/$\sqrt{3}$ \\ % (3*C2+2*C3+3*C4)**2/24
        &    & 8.4\% \footnotemark[2] & ($\sqrt{3}/2$)[ET$_{13}$-(1/3)(ET$_{12}$+ET$_{23}$)] \\ % (2/3)*C3**2
        &    & 1.4\% \footnotemark[3] & (CT$_{12}$+CT$_{23}$)/$\sqrt{2}$ \\    % (C2-C4)**2/2
        &    & 9.6\% \footnotemark[4] & CT$_{13}$                       \\    % (C2-2*C3+C4)**2/8
        &    & 6.9\% & $^1$(H-2,L+2) \\  % C5**2
$2\, ^1B _{1u}$ & 0.4703 & 343 & 3.62 \\ % 4th step
      &                & \multicolumn{2}{c}{Analysis} \\
\cline{3-4} 
        &    & 0\% & $^1$(H,L) \\  % C1**2
        &    & 89.4\% \footnotemark[1] & {\bf (ET$_{12}$+ET$_{23}$)/$\sqrt{3}$} \\ % (3*C2+2*C3+3*C4)**2/24
        &    & 21.2\% \footnotemark[2] & ($\sqrt{3}/2$)[ET$_{13}$-(1/3)(ET$_{12}$+ET$_{23}$)] \\ % (2/3)*C3**2
        &    & 0.02\% \footnotemark[3] & (CT$_{12}$+CT$_{23}$)/$\sqrt{2}$ \\    % (C2-C4)**2/2
        &    & 0.02\% \footnotemark[4] & CT$_{13}$                       \\    % (C2-2*C3+C4)**2/8
        &    & 0\% & $^1$(H-2,L+2) \\  % C5**2
$3\, ^1B _{1u}$ & 0.0001 & 283  & 4.38 \\ % 5th step
      &                & \multicolumn{2}{c}{Analysis} \\
\cline{3-4} 
        &    & 4.3\% & $^1$(H,L) \\  % C1**2
        &    & 0.65\% \footnotemark[1] & (ET$_{12}$+ET$_{23}$)/$\sqrt{3}$ \\ % (3*C2+2*C3+3*C4)**2/24
        &    & 0\% \footnotemark[2] & ($\sqrt{3}/2$)[ET$_{13}$-(1/3)(ET$_{12}$+ET$_{23}$)] \\ % (2/3)*C3**2
        &    & 91.1\% \footnotemark[3] & {\bf (CT$_{12}$+CT$_{23}$)/$\sqrt{2}$} \\    % (C2-C4)**2/2
        &    & 0.22\% \footnotemark[4] & CT$_{13}$                       \\    % (C2-2*C3+C4)**2/8
        &    & 3.7\% & $^1$(H-2,L+2) \\  % C5**2
$4\, ^1B _{1u}$ & 0.0000 & 270 & 4.59 \\ % 7th state
      &                & \multicolumn{2}{c}{Analysis} \\
\cline{3-4} 
        &    & 18.3\% & $^1$(H,L) \\  % C1**2
        &    & 1.3\% \footnotemark[1] & (ET$_{12}$+ET$_{23}$)/$\sqrt{3}$ \\ % (3*C2+2*C3+3*C4)**2/24
        &    & 9.5\% \footnotemark[2] & ($\sqrt{3}/2$)[ET$_{13}$-(1/3)(ET$_{12}$+ET$_{23}$)] \\ % (2/3)*C3**2
        &    & 9.5\% \footnotemark[3] & (CT$_{12}$+CT$_{23}$)/$\sqrt{2}$ \\    % (C2-C4)**2/2
        &    & 17.8\% \footnotemark[4] & CT$_{13}$                       \\    % (C2-2*C3+C4)**2/8
        &    & 48.4\% & $^1$(H-2,L+2) \\  % C5**2
\multicolumn{4}{c}{DS$_3$ = -1.06 eV} \\ % DS$_2$ = 0.14 eV
\hline
\multicolumn{4}{c}{TD-HF/6-31G(d,p)//B3LYP+D3/6-31G(d,p)}\\      
$1\, ^1B _{1u}$ & 0.0025 & 662 & 1.87 \\ % 1st state
      &                & \multicolumn{2}{c}{Analysis} \\
\cline{3-4} 
        &    & 81.0\% & $^1$(H,L) \\  % C1**2
        &    & 0.2\% \footnotemark[1] & (ET$_{12}$+ET$_{23}$)/$\sqrt{3}$ \\ % (3*C2+2*C3+3*C4)**2/24
        &    & 7.1\% \footnotemark[2] & ($\sqrt{3}/2$)[ET$_{13}$-(1/3)(ET$_{12}$+ET$_{23}$)] \\ % (2/3)*C3**2
        &    & 3.1\% \footnotemark[3] & (CT$_{12}$+CT$_{23}$)/$\sqrt{2}$ \\    % (C2-C4)**2/2
        &    & 13.9\% \footnotemark[4] & CT$_{13}$                       \\    % (C2-2*C3+C4)**2/8
        &    & 0\% & $^1$(H-2,L+2) \\  % C5**2
$2\, ^1B _{1u}$ & 0.2682 & 468 & 2.65 \\ % 3rd state
      &                & \multicolumn{2}{c}{Analysis} \\
\cline{3-4} 
        &    & 0\% & $^1$(H,L) \\  % C1**2
        &    & 89.0\% \footnotemark[1] & {\bf (ET$_{12}$+ET$_{23}$)/$\sqrt{3}$} \\ % (3*C2+2*C3+3*C4)**2/24
        &    & 21.8\% \footnotemark[2] & ($\sqrt{3}/2$)[ET$_{13}$-(1/3)(ET$_{12}$+ET$_{23}$)] \\ % (2/3)*C3**2
        &    & 0.11\% \footnotemark[3] & (CT$_{12}$+CT$_{23}$)/$\sqrt{2}$ \\    % (C2-C4)**2/2
        &    & 0.003\% \footnotemark[4] & CT$_{13}$                       \\    % (C2-2*C3+C4)**2/8
        &    & 0\% & $^1$(H-2,L+2) \\  % C5**2
$3\, ^1B _{1u}$ & 0.0002 & 334 & 3.71 \\  % 7th state
      &                & \multicolumn{2}{c}{Analysis} \\
\cline{3-4} 
        &    & 4.0\% & $^1$(H,L) \\  % C1**2
        &    & 1.5\% \footnotemark[1] & (ET$_{12}$+ET$_{23}$)/$\sqrt{3}$ \\ % (3*C2+2*C3+3*C4)**2/24
        &    & 0.\% \footnotemark[2] & ($\sqrt{3}/2$)[ET$_{13}$-(1/3)(ET$_{12}$+ET$_{23}$)] \\ % (2/3)*C3**2
        &    & 90.6\% \footnotemark[3] & {\bf (CT$_{12}$+CT$_{23}$)/$\sqrt{2}$} \\    % (C2-C4)**2/2
        &    & 5.1\% \footnotemark[4] & CT$_{13}$                       \\    % (C2-2*C3+C4)**2/8
        &    & 3.4\% & $^1$(H-2,L+2) \\  % C5**2
$4\, ^1B _{1u}$ & 0.0004 & 315  & 3.94 \\ % 11th state
      &                & \multicolumn{2}{c}{Analysis} \\
\cline{3-4} 
        &    & 18.1\% & $^1$(H,L) \\  % C1**2
        &    & 1.1\% \footnotemark[1] & (ET$_{12}$+ET$_{23}$)/$\sqrt{3}$ \\ % (3*C2+2*C3+3*C4)**2/24
        &    & 14.7\% \footnotemark[2] & ($\sqrt{3}/2$)[ET$_{13}$-(1/3)(ET$_{12}$+ET$_{23}$)] \\ % (2/3)*C3**2
        &    & 11.5\% \footnotemark[3] & (CT$_{12}$+CT$_{23}$)/$\sqrt{2}$ \\    % (C2-C4)**2/2
        &    & 25.1\% \footnotemark[4] & CT$_{13}$                       \\    % (C2-2*C3+C4)**2/8
        &    & 36.8\% & $^1$(H-2,L+2) \\  % C5**2
\multicolumn{4}{c}{DS$_3$ = -0.76 eV} \\ % DS$_2$ = 0.14 eV
%mec \multicolumn{4}{c}{{\bf STOP}}\\  % ********************************************
\hline \hline
\end{tabular}
\footnotetext[1]{$(3c_2+2c_3+3c_4)^2/24$.}
\footnotetext[2]{$2c_3^2/3$.}
\footnotetext[3]{$(c_2-c_4)^2/2$.}
\footnotetext[4]{$(c_2-2c_3+c_4)^2/8$.}
\end{table}
%%%%%%%
% EOF
%%%%%%%
Tables~\ref{tab:CTETP3a}, \ref{tab:CTETP3b}, \ref{tab:CTETP3c}, 
\ref{tab:CTETP3d}, and \ref{tab:CTETP3e}
apply the analysis of Sec.~\ref{sec:theory} to the equally-spaced parallel stacked trimer.
As expected, instead of a Davydov pair of ET and CT excitations, we find a Davydov triplet
corresponding to the $2\, ^1B_{1u}$, $3\, ^1B_{1u}$, and $4\, ^1B_{1u}$ states.
When the Davydov pairs can be identified, we have highlighted their assignment in terms
of nearest neighbor interactions in bold face in the tables.  
Cases where the Davydov pairs are clear are:
TD-LDA/6-31G(d,p)//LDA/6-31G(d,p), 
TD-LDA/STO-3G//LDA/STO-3G,
TD-B3LYP/6-31G(d,p)//B3LYP+D3/6-31G(d,p), 
TD-B3LYP/STO-3G//B3LYP+D3/STO-3G,
TD-DFTB//DFTB+D3.
In these cases, the (ET$_{12}$+ET$_{23}$)-dominated state is the highest
energy transition of the Davydov triplet and also has the highest oscillator
strength while the (CT$_{12}$+CT$_{23}$)-dominated state is the lowest energy transition
of the Davydov triplet and has significantly less oscillator strength.  A third
contribution to the Davydov triplet lies between the two other states and also has
only a feeble transition energy.  
This assignment is a bit less clear in the TD-DFTB//DFTB+D3 case because there is
significant mixing between ET$_{12}$+ET$_{23}$ and ET$_{13}$ in the brightest configuration.
Henceforth we shall simply assume that the peak with the highest oscillator strength
is an (ET$_{12}$+ET$_{23}$)-dominated state.
% ------------------------------------------------------
\begin{figure}
\includegraphics[width=0.5\textwidth]{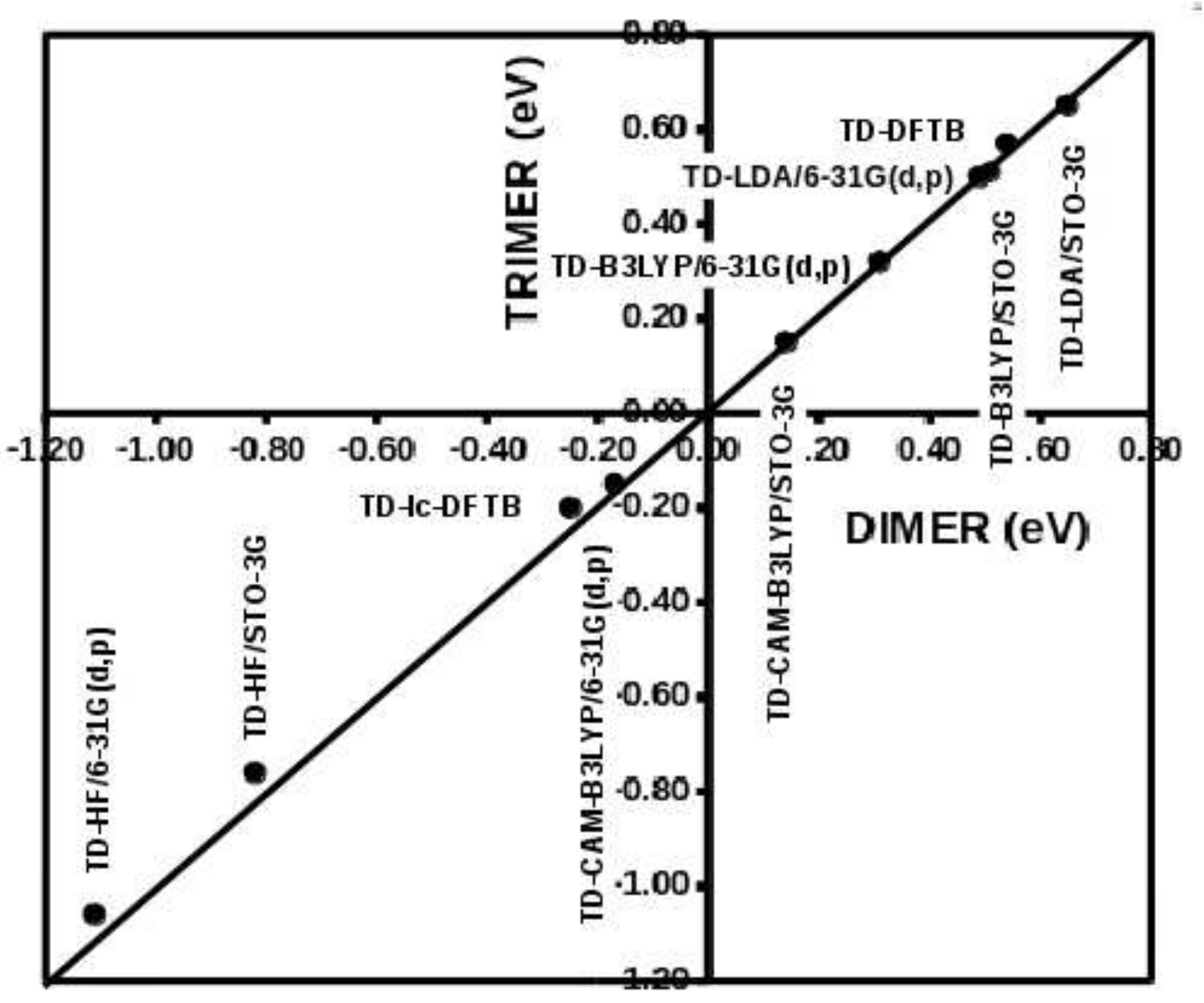} \\
\caption{
\label{fig:DS3vsDS2}
Graph comparing the Davydov splittings, $E$(ET)-$E$(CT), of the trimer and 
dimer.  The 45$^\circ$ line indicates perfect agreement between dimer and trimer Davydov
splittings.
}
\end{figure}
% -------------------------------------------------------
Figure~\ref{fig:DS3vsDS2} shows that calculations with these methods give essentially the
same Davydov splitting for the dimer and for the trimer, and that the dimer (DS$_2$) and
trimer (DS$_3$) splittings are very similar for TD-DFTB and for TD-B3LYP/6-31G(d,p),
as well as being very similar for TD-lc-DFT and for TD-CAM-B3LYP/6-31G(d,p).

% -------------------------------------------
\subsubsection{Spectra}
% -------------------------------------------

% ------------------------------------------------------
\begin{figure}
\begin{tabular}{cc}
a) & \\
% b) & \includegraphics[width=0.43\textwidth]{./graphics/Spectra/TD_LDA_6-31Gdp.eps} \\
b) & \includegraphics[width=0.43\textwidth]{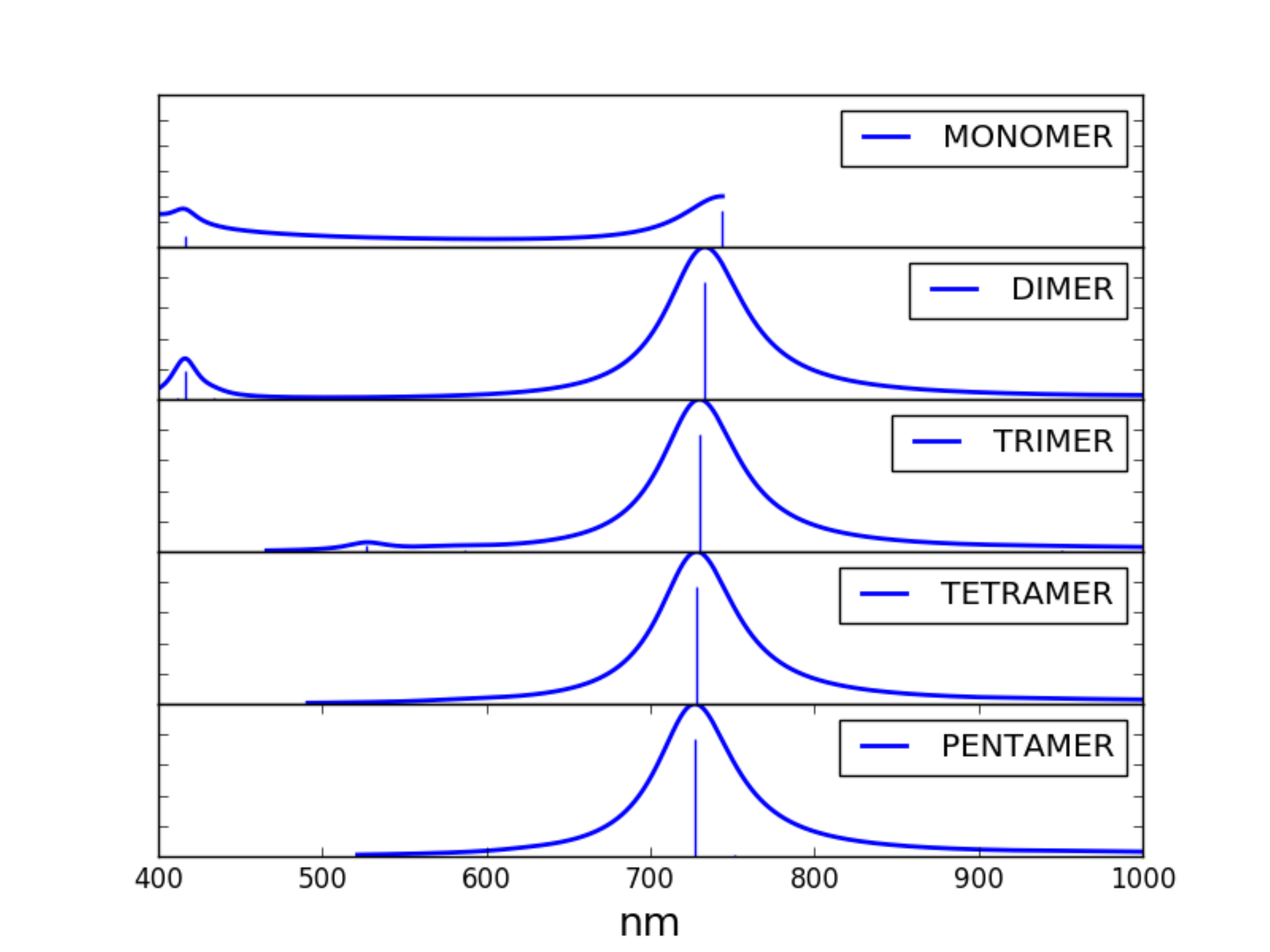} \\
% c) & \includegraphics[width=0.43\textwidth]{./graphics/Spectra/TD_B3LYP_6-31Gdp.eps} \\
c) & \includegraphics[width=0.43\textwidth]{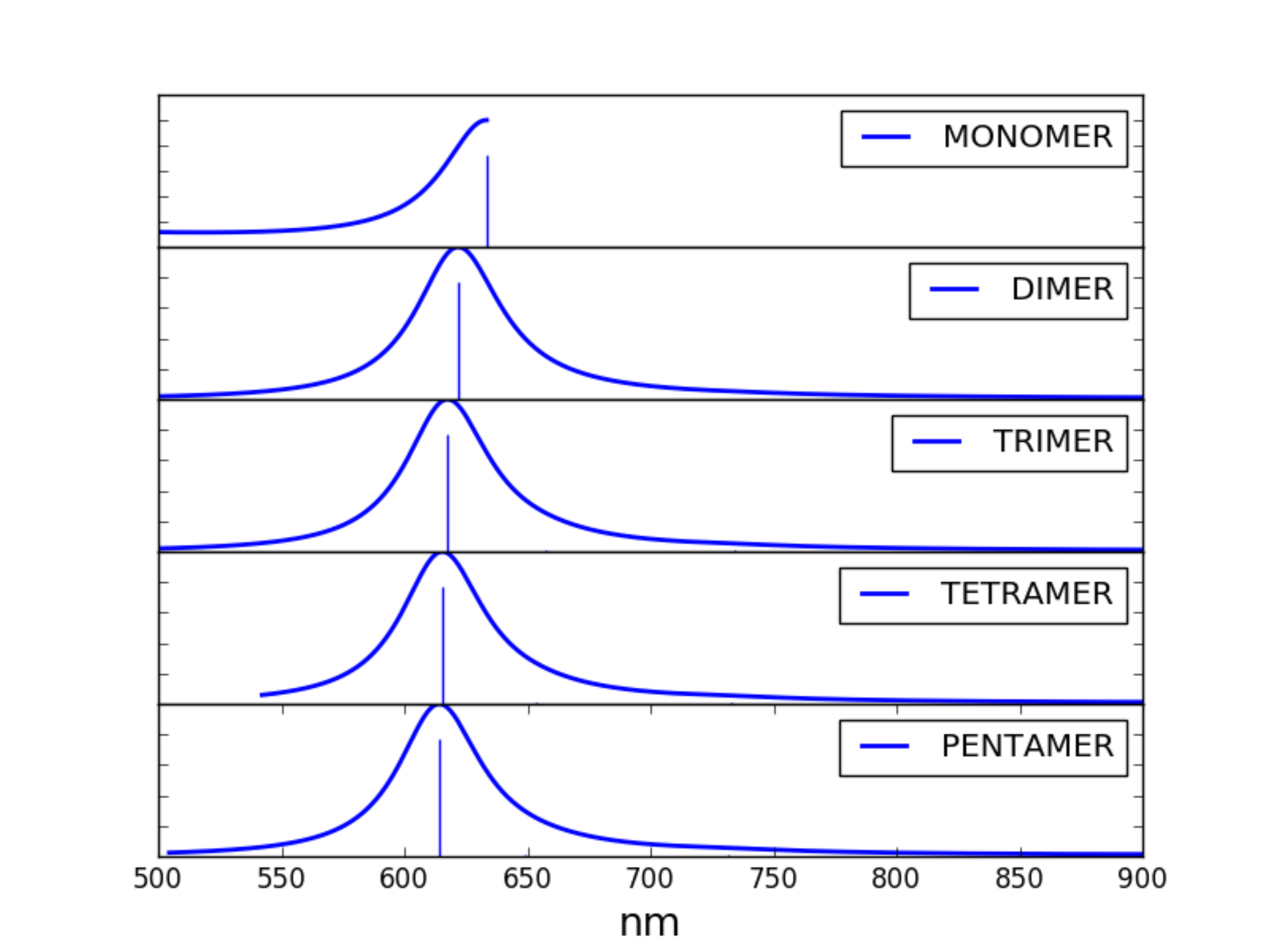} \\
% d) & \includegraphics[width=0.43\textwidth]{./graphics/Spectra/TD-CAM-B3LYP_6-31Gdp.eps} \\
d) & \includegraphics[width=0.43\textwidth]{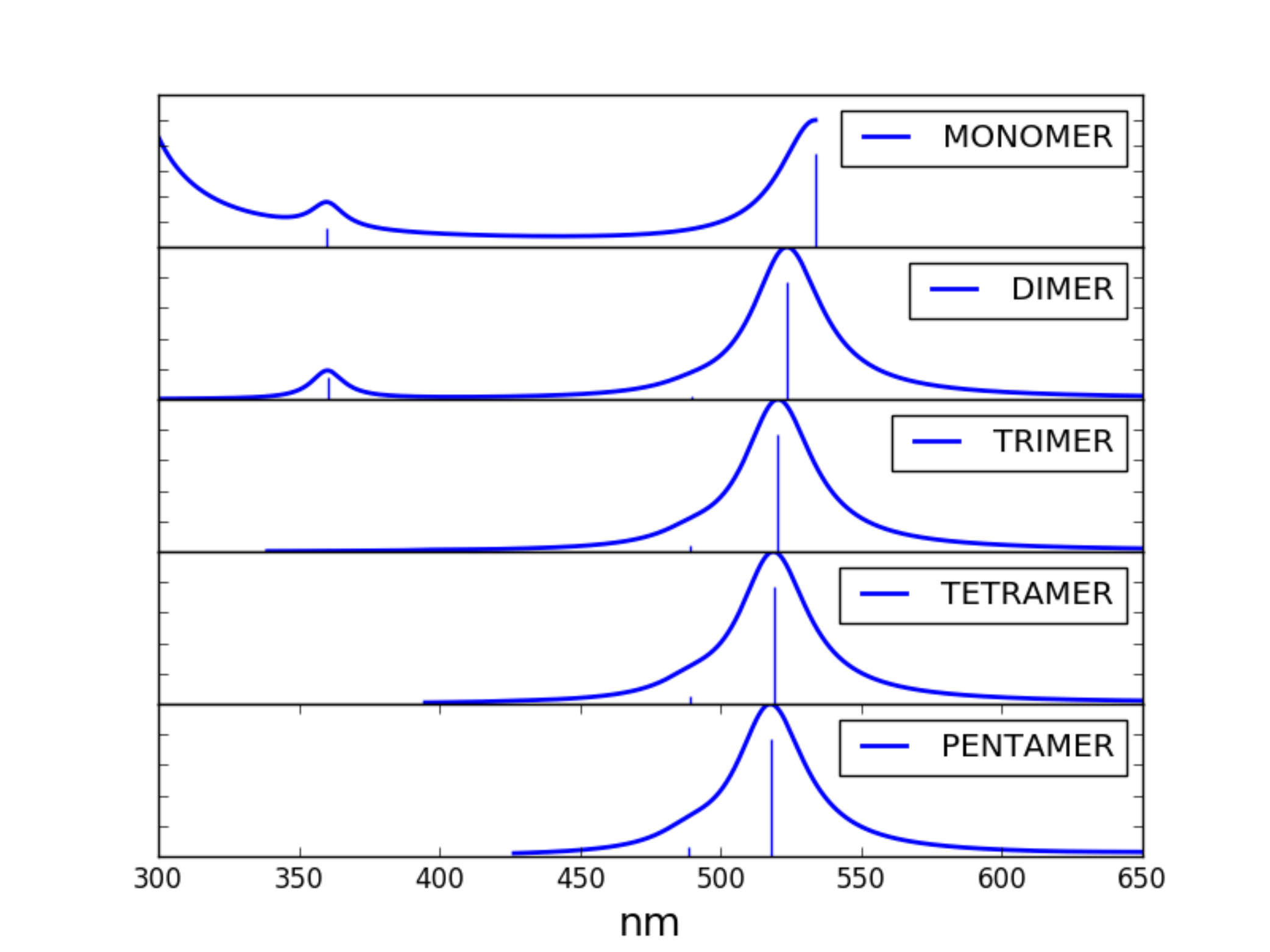} \\
%   & \includegraphics[width=0.43\textwidth]{./graphics/Spectra/TD_HF_6-31Gdp.eps}
   & \includegraphics[width=0.43\textwidth]{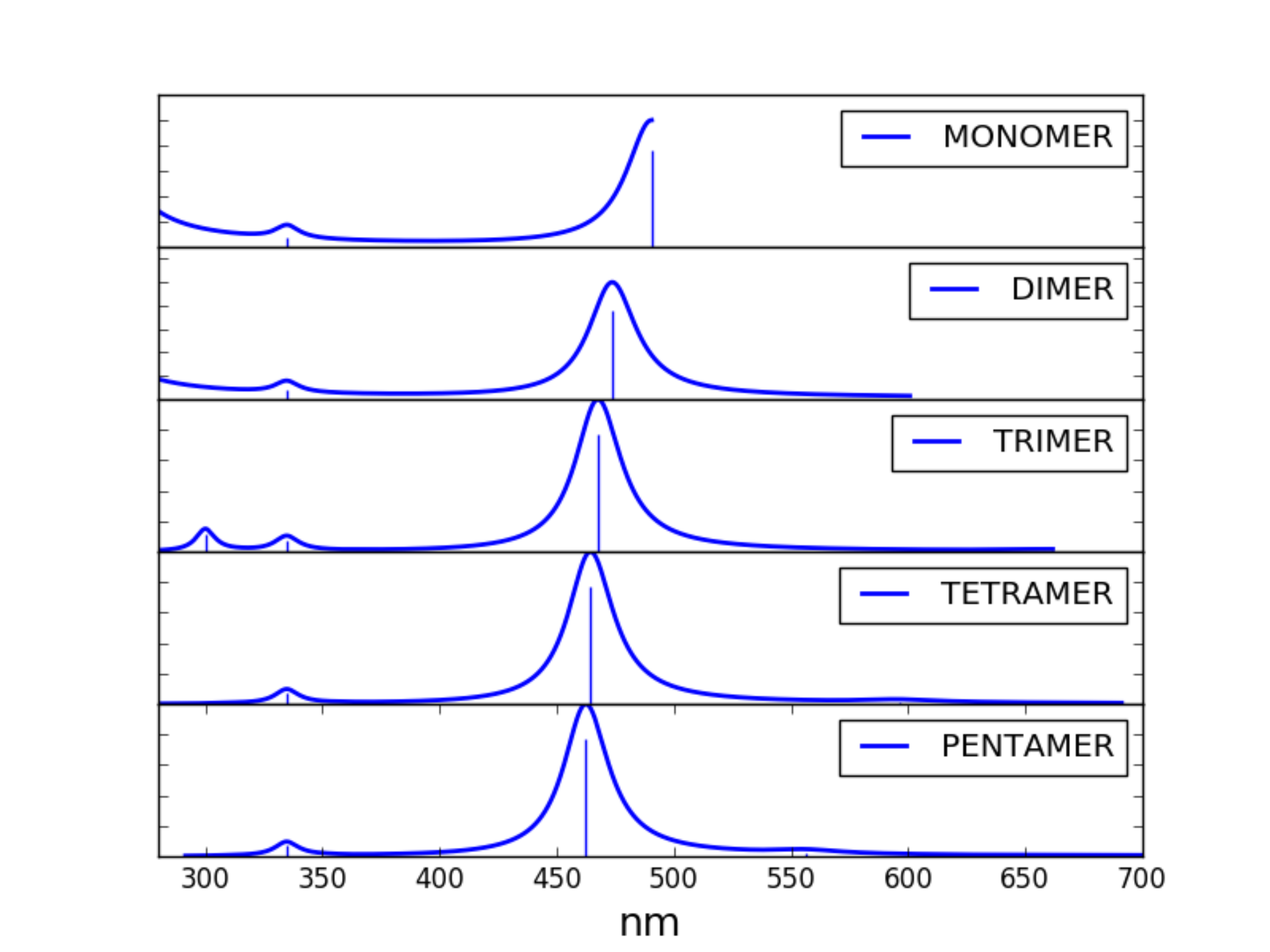}
\end{tabular}
\caption{
\label{fig:stackspectra}
Convergence of spectra as a function of number of pentacene molecules in the stack:
(a) TD-LDA/6-31G(d,p), (b) TD-B3LYP/6-31G(d,p), (c) TD-CAM-B3LYP/6-31G(d,p), and (d) TD-HF/6-31G(d,p).
}
\end{figure}
% -------------------------------------------------------
Calculations beyond the trimer become increasingly complicated to analyze but we may 
compare calculated spectra for increasingly large numbers of parallel stacked pentacene
molecules. The tight-binding calculation in Sec.~\ref{sec:theory} is based upon the hope
that nearest neighbor interactions dominate excitonic effects in spectra.  The comparison
of dimer and trimer DSs seem to at least partially confirm this.  We may make a further check
by seeing how the spectra change as more and more pentacene molecules are stacked.  These
spectra are shown in Fig.~\ref{fig:stackspectra}.  All of the spectra show a main peak 
(i.e., the ET peak) which blue shifts as the pentacene stack grows.  More specifically,
the graphs show a main peak which undergoes the largest shift in going from the monomer 
to the dimer, a smaller shift in going from the dimer to the trimer and then shifts very 
little in going to higher oligomers, consistent with the suppositions behind the tight-binding model.

% ------------------------------------------------------
\begin{figure}
\begin{tabular}{cc}
a) & \\
% b) & \includegraphics[width=0.5\textwidth]{./graphics/Spectra/TD-DFTB_fixed.eps} \\
b) & \includegraphics[width=0.5\textwidth]{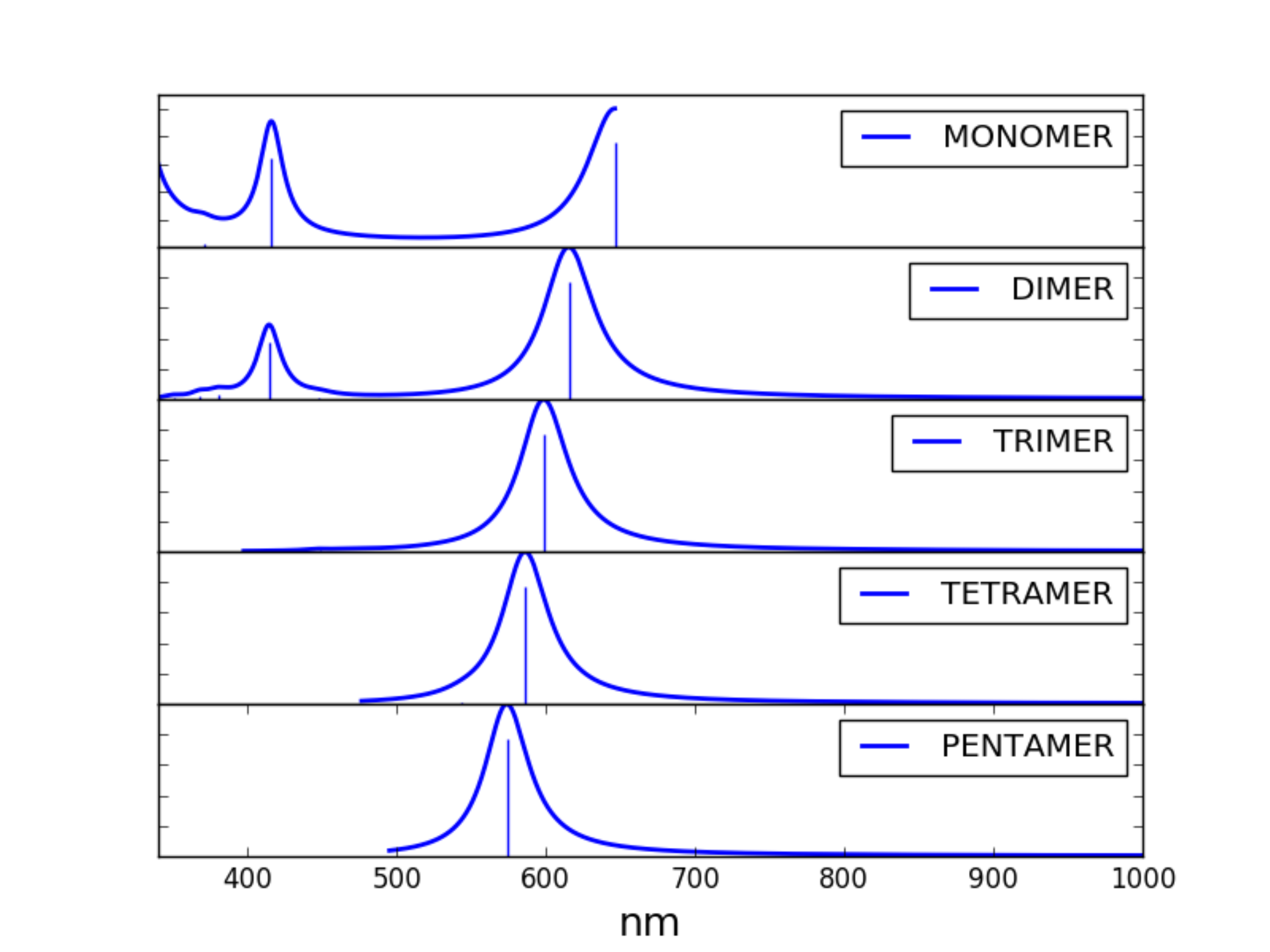} \\
%   & \includegraphics[width=0.5\textwidth]{./graphics/Spectra/TD-DFTB_size_extensive.eps}
   & \includegraphics[width=0.5\textwidth]{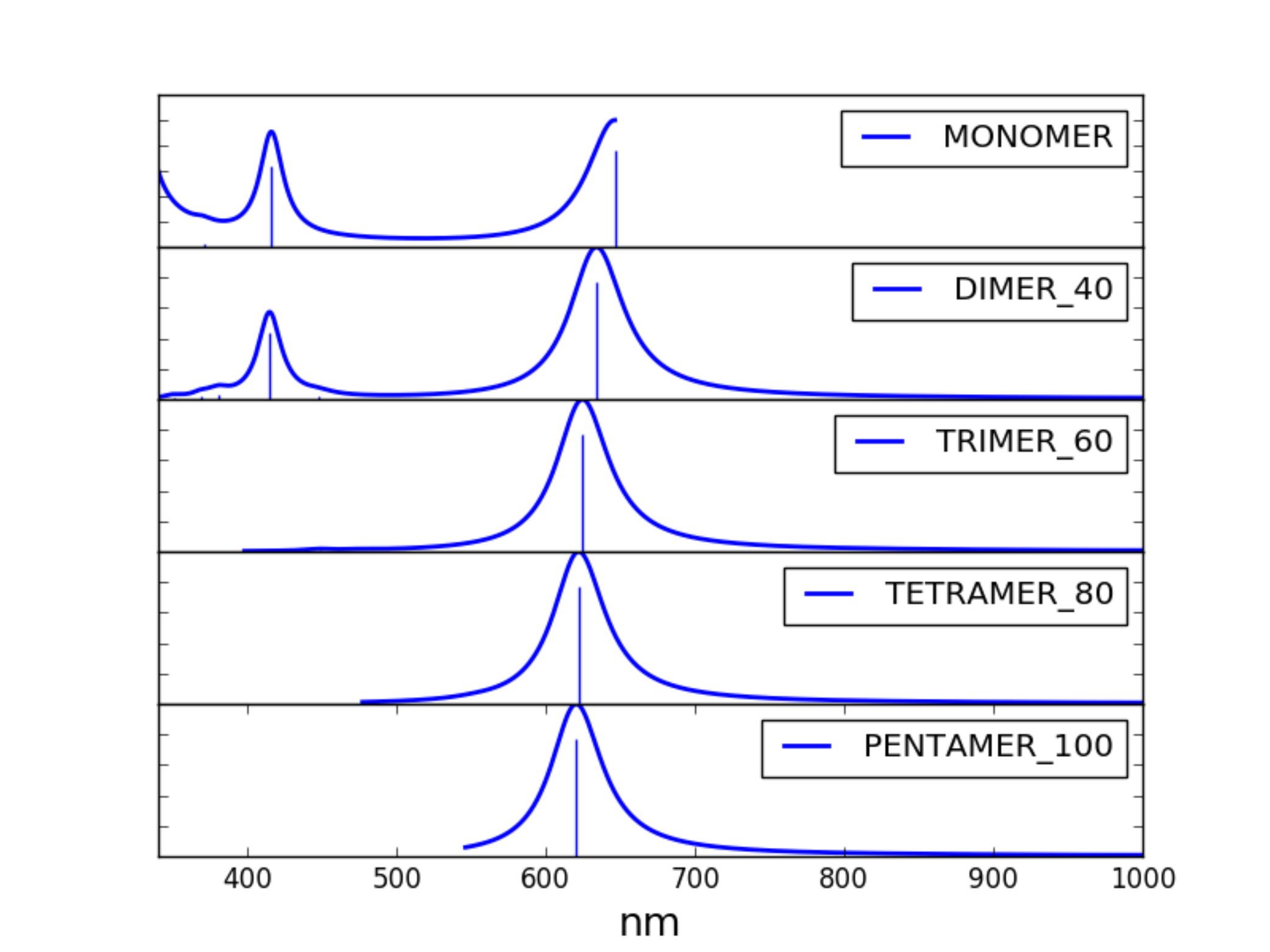}
\end{tabular}
\caption{
\label{fig:TD-DFTBsizeconsistency}
TD-DFTB spectra of stacked pentacene: (a) with a fixed active space, (b) with a size-extensive
active space. 
}
\end{figure}
% -------------------------------------------------------
\begin{figure}
\begin{tabular}{cc}
a) & \\
% b) & \includegraphics[width=0.5\textwidth]{./graphics/Spectra/TD-lc-DFTB_fixed.eps} \\
b) & \includegraphics[width=0.5\textwidth]{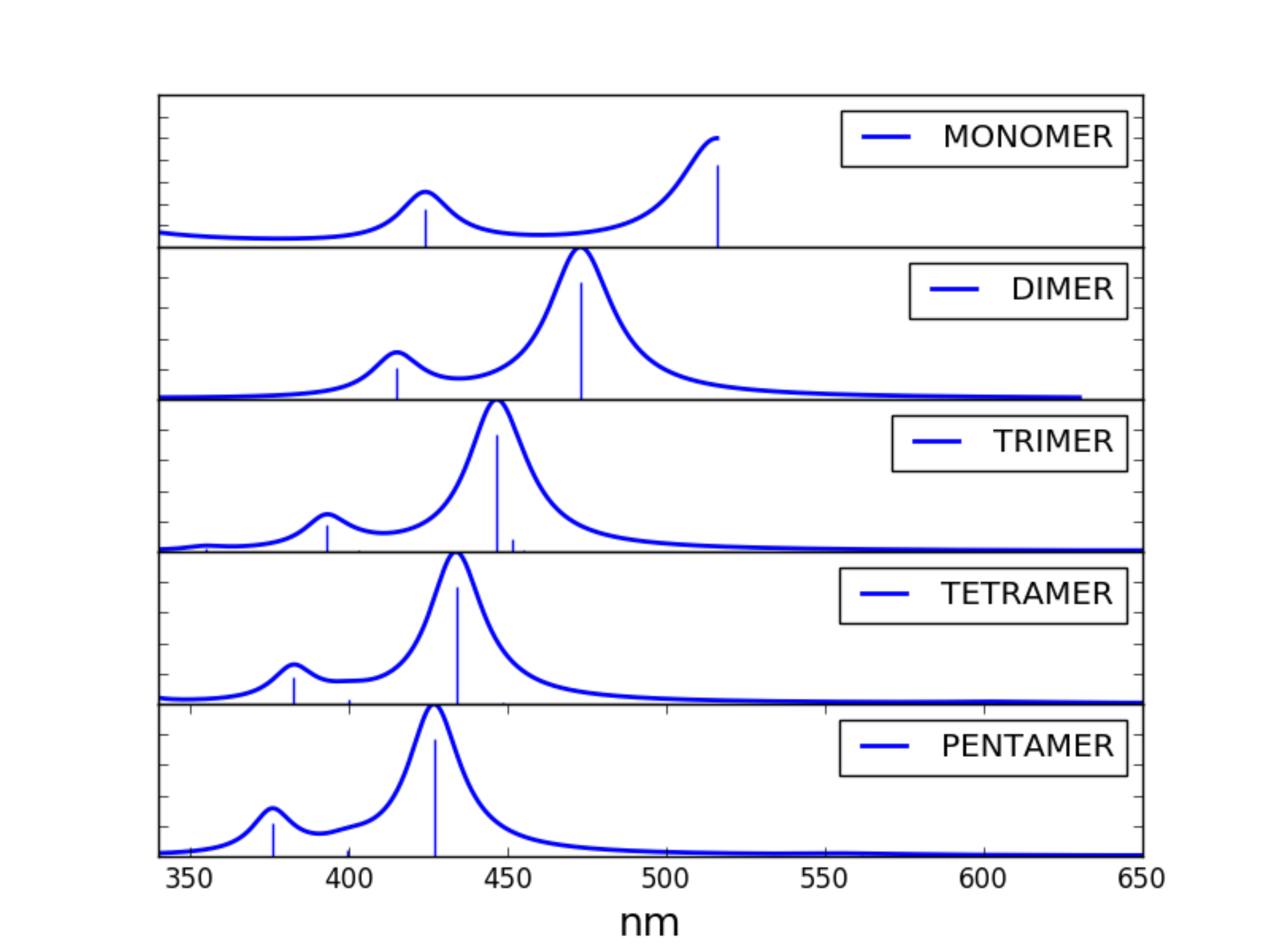} \\
%   & \includegraphics[width=0.5\textwidth]{./graphics/Spectra/TD-lc-DFTB_size_extensive.eps}
   & \includegraphics[width=0.5\textwidth]{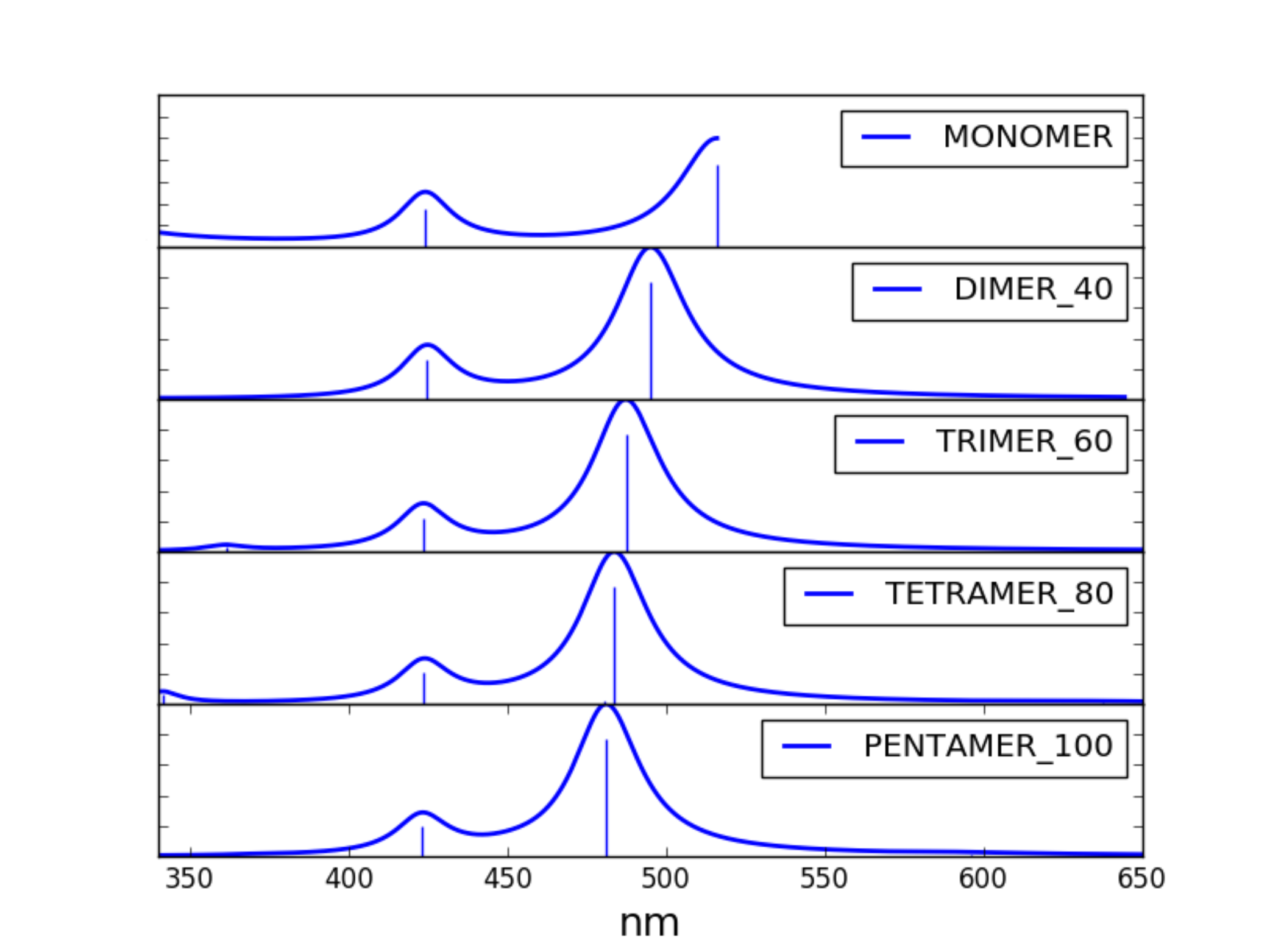}
\end{tabular}
\caption{
\label{fig:TD-lc-DFTBsizeconsistency}
TD-lc-DFTB spectra of stacked pentacene: (a) with a fixed active space, (b) with a size-consistent
active space. 
}
\end{figure}
% -------------------------------------------------------
Our TD-DFT and TD-lc-DFTB calculations led us to become aware of a problem already reported in
Ref.~\onlinecite{HM17}.  It is in the spirit of semi-empirical approaches to make simplifying
approximations which allow the treatment of larger molecules than would otherwise be possible.
This is why {\sc DFTBaby} restricts the space of active orbitals, but it is still up to the user
to decide how to use this option.  One way would be to increase the size of the active space
until converged spectra are achieved.  But this ideal approach is not really practical when going
to larger and larger aggregates of molecules.  Instead, the first idea that comes to mind is to
use the largest active space for which calculations are possible.  In practice, this means using
the same number of occupied and unoccupied orbitals in the active space, independent of the number
of molecules.  We call doing this a calculation with a {\em fixed active space}.  However it has
the important drawback when describing size-dependent trends that fixed active space calculations
have more basis functions per molecule for smaller aggregates than for larger aggregates and so invariably
describe smaller aggregates better than larger aggregates with the introduction of corresponding
systematic errors in the resultant size-dependent properties.  The other approach is to keep the
number of occupied and unoccupied orbitals in the active space proportional to the number of 
molecules.  In this way, we hope to obtain a better description of size-dependent trends, albeit
at some cost of accuracy for smaller aggregats.  We call this doing a calculation with a 
{\em size-consistent active space}.  (There is some confusion in the literature between the
terms ``size-consistent'' and ``size-extensive.''  Both terms are arguably correct here, but 
we shall stick to ``size-consistent.'') 

(Our size-consistent active-space approach resembles other approaches based upon energy and/or
oscillator strength cut-offs \cite{G14,RLL+15}.  Note however that there is an important difference
in philosophy as the latter aim for accurage spectra with large basis sets while our approach aims
at constant accuracy for varying sizes of aggregates.)

\begin{figure}
\begin{tabular}{cc}
a) & \\
% b) & \includegraphics[width=0.4\textwidth]{./graphics/TD-DFTB_TD_B3LYP_6-31Gdp_PENTAMER.eps} \\
b) & \includegraphics[width=0.4\textwidth]{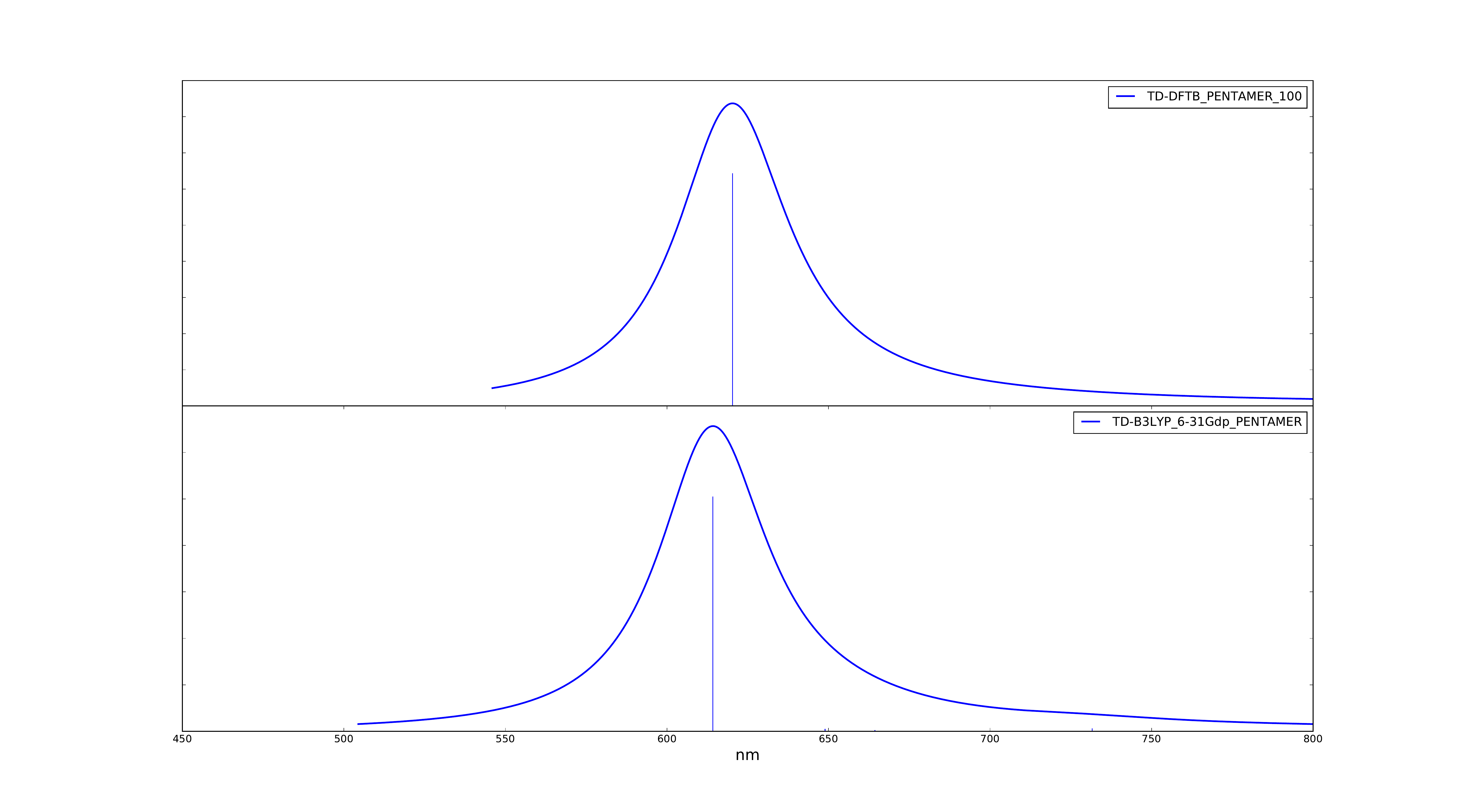} \\
%   & \includegraphics[width=0.4\textwidth]{./graphics/TD-lc-DFTB_TD-CAM-B3LYP_6-31Gdp_PENTAMER.eps} 
   & \includegraphics[width=0.4\textwidth]{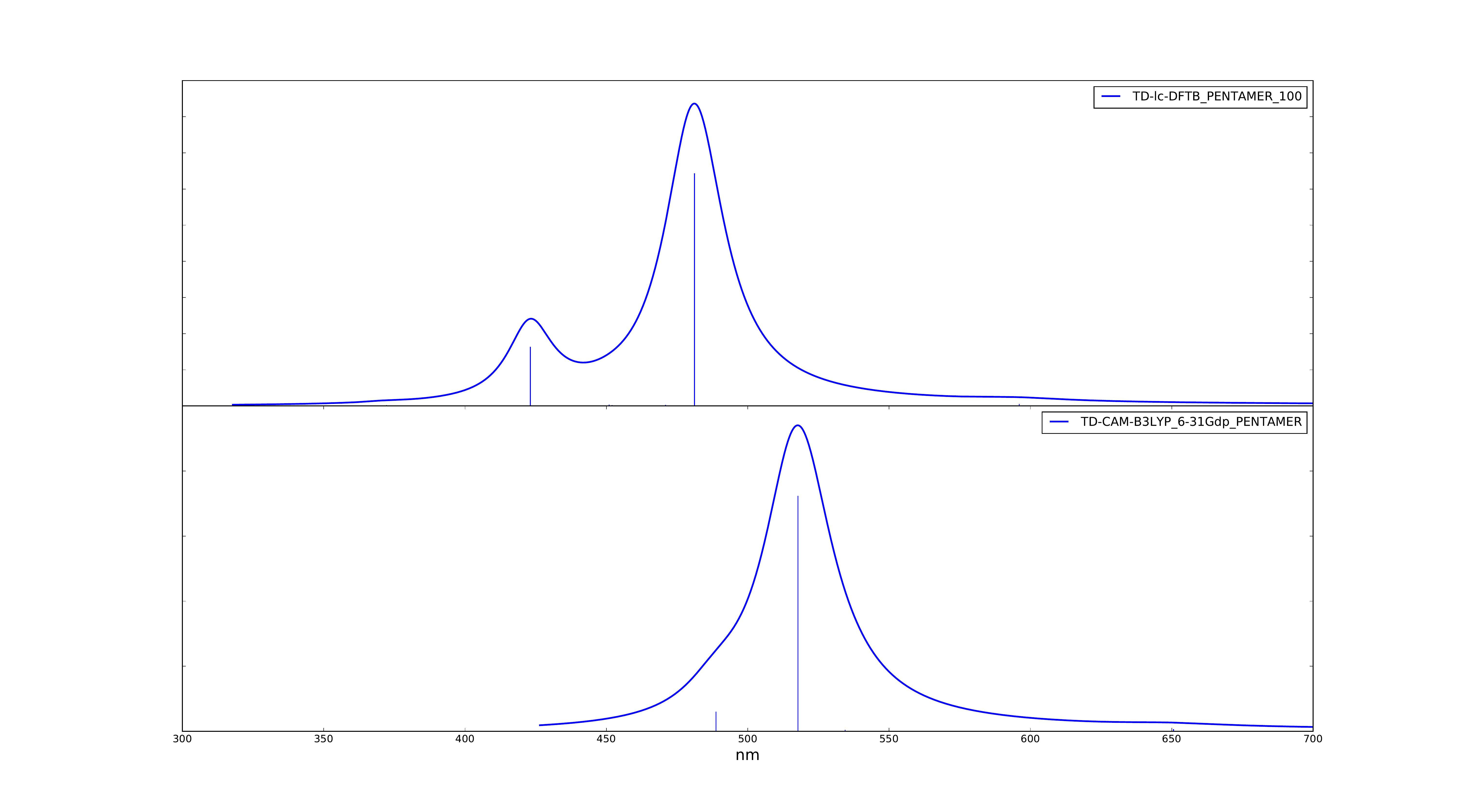} 
\end{tabular}
\caption{
\label{fig:activespace}
Comparison of pentacene spectra: (a) TD-DFTB and TD-B3LYP/6-31G(d,p) and (b) TD-lc-DFTB
and TD-CAM-B3LYP/6-31G(d,p).
}
\end{figure}

Figures~\ref{fig:TD-DFTBsizeconsistency} 
and \ref{fig:TD-lc-DFTBsizeconsistency} compare fixed and size-consistent active space calculations.
The fixed active space calculations use 20 occupied orbitals and 20 unoccupied 
orbitals per aggregate.  The size-consistent active space calculations use 20 occupied and 
20 unoccupied orbitals per pentacene molecule.  
As the figure shows, the calculations with fixed active space
blue shifts much more than do the calculations with self-consistent active space as the number of
pentacene molecules increases.  TD-DFT calculations of excitation energies are variational in
the Tamm-Dancoff approximation and pseudo-variational in the sense that full linear response
calculations often give similar results to using the Tamm-Dancoff approximation.  
For the monomer the fixed active space and size-consistent active space calculations
are identical; however for the aggregates, the size-consistent active space is larger
than the fixed active space calculations, leading to lower excitation energies in the size-consistent
active space calculations.  One would hope that the larger basis set would give better and answers
and that this is the case is shown in Fig.~\ref{fig:activespace} where it 
is seen that TD-DFTB and TD-B3LYP/6-31G(d,p) spectral peak locations differ by only about 10 nm.
Figure~\ref{fig:TD-DFTBsizeconsistency} shows that the difference between the TD-DFTB and TD-B3LYP/6-31G(d,p)
calculations would have been more like 50 nm had the fixed active space been used.
Figure~\ref{fig:activespace} also shows that the differences between TD-lc-DFTB and 
TD-CAM-B3LYP/6-31G(d,p) spectra are larger than for
the TD-DFTB and TD-B3LYP/6-31G(d,p) case when the size-extensive active space is used, with the main
peak in this part of the spectrum having an energy difference of around 30 nm between the two
calculations.  Interestingly both show qualitatively similar Davydov multiplets.
Figure~\ref{fig:TD-lc-DFTBsizeconsistency} shows that the difference between the TD-lc-DFTB and 
TD-CAM-B3LYP(d,p) calculations would have been more like 80 nm had the fixed active space been used.
This is why, except for Figs.~\ref{fig:TD-DFTBsizeconsistency} and \ref{fig:TD-lc-DFTBsizeconsistency},
we have been careful to use a size-consistent active space consisting of 20 occupied and 20 unoccupied
orbitals per molecule in all the TD-DFTB and TD-lc-DFTB reported in this paper.
% ------------------------------------------------------
% -------------------------------------------------------

% ========================================================
\subsection{Herringbone}
% ========================================================

% ------------------------------------------------------
\begin{figure}
\includegraphics[width=0.5\textwidth]{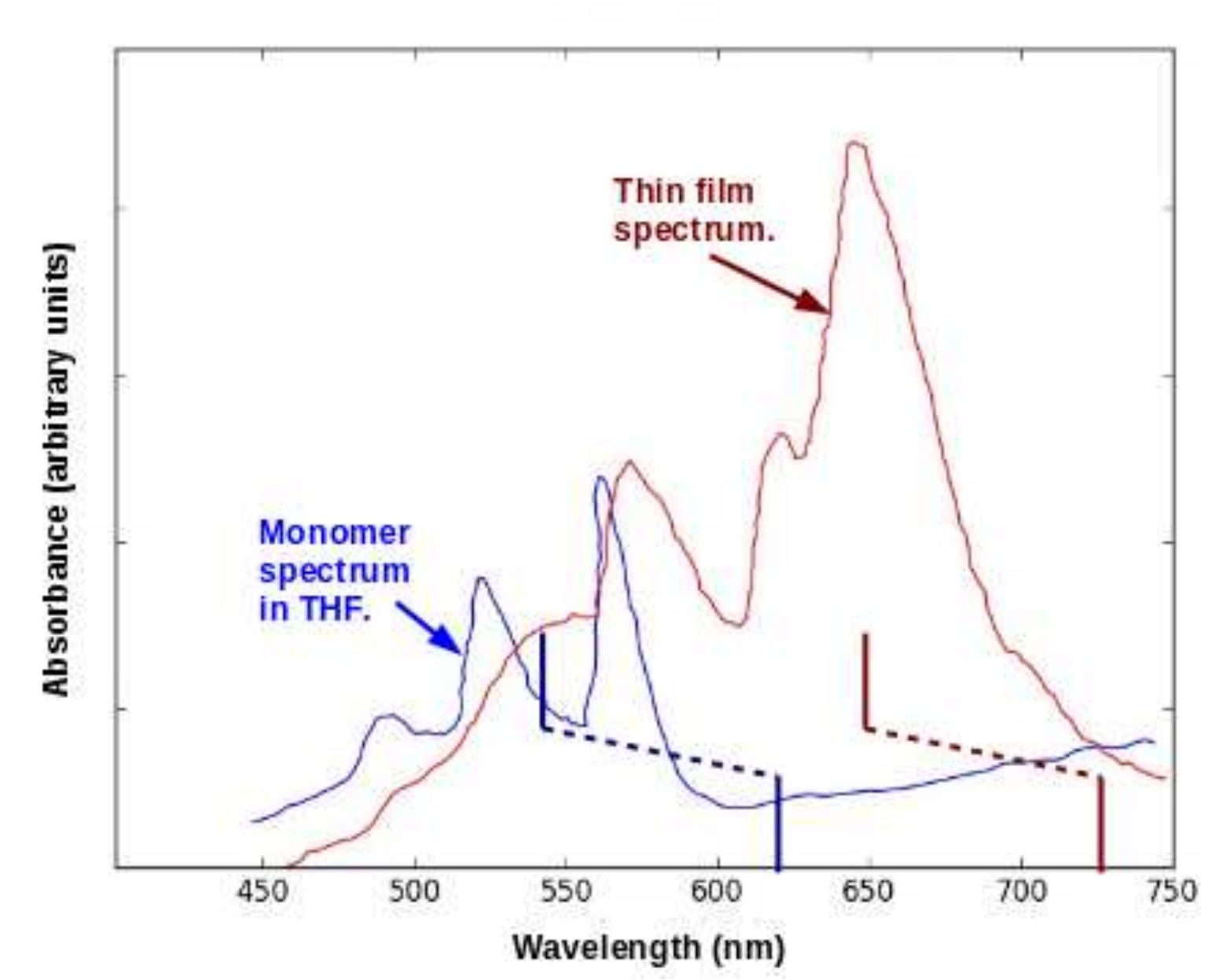}
\caption{
\label{fig:basic_exciton_spectrum}
Excitonic effects on the absorption spectrum of pentacene: curves, pentacene
in tetrahydrofuran (THF) and as a thin film (from Ref.~\onlinecite{MRK+04}); 
stick spectra, lower are the position of unshifted monomer and crystal peaks
calculated using the Bethe-Salpeter equation (BSE) while the upper stick spectra
have been shifted to match the experiment (from Ref.~\onlinecite{CSRG15}.)
See also Ref.~\onlinecite{TNL03}.
}
\end{figure}
% -------------------------------------------------------
The main objective of the present work has been to evaluate the ability of 
TD-DFTB and TD-lc-DFTB to simulate, respectively, the results of 
TD-B3LYP/6-31G(d,p) and TD-CAM-B3LYP/6-31G(d,p) calculations.  While 
this has been largely satisfied by our study of parallel stacked 
pentacene molecules, the case of parallel stacked molecules is too
artificial to allow comparison against experiment (except for the monomer.)
In order to have a reality check, we have also carried out calculations for
cluster models of pentacene crystals.  
The experimental spectrum of the molecule
and of the crystal are available both from experiment \cite{SWB81,MRK+04,HYX+15} and from 
state-of-the-art theoretical calculations \cite{TNL03,ANPM09,CGRS13,SDKN13,CSRG15}.  These are shown in
Fig.~\ref{fig:basic_exciton_spectrum}.  This time excitonic shifts lead to a red
shift, rather than a blue shift.  The structure of the spectrum suggests that both CT and 
ET transitions contribute to the spectrum.  As we shall see, charge transfer is more
important for describing excitonic effects in the absorption spectrum than is
the case for parallel stacked pentamers.

% ------------------------------------------------------
\begin{figure}
\begin{tabular}{cc}
a) & \\
% b) & \includegraphics[width=0.3\textwidth]{./graphics/herringbone5.eps} \\
b) & \includegraphics[width=0.3\textwidth]{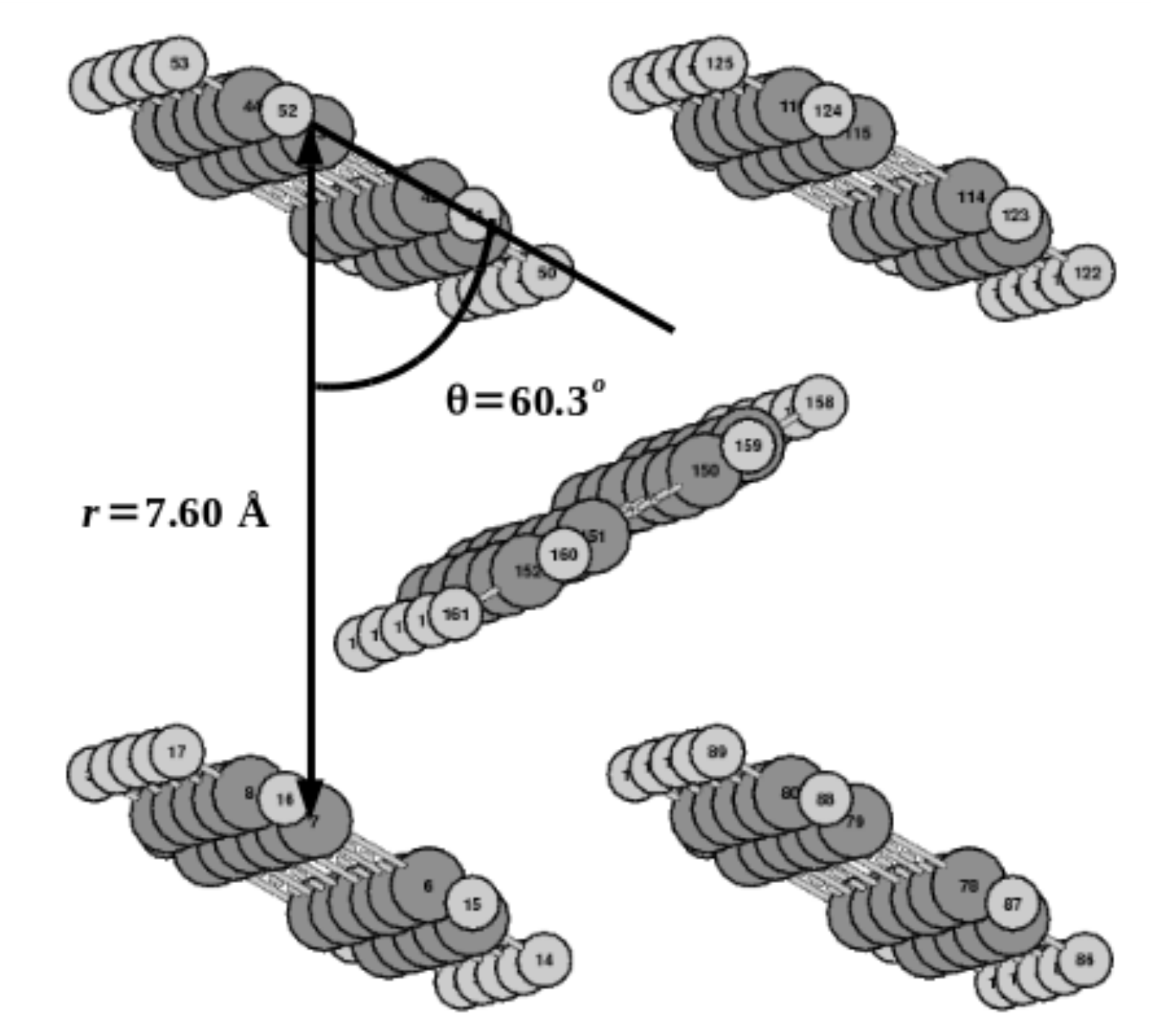} \\
% c)  & \includegraphics[width=0.3\textwidth]{./graphics/herringbone10v.eps} \\
c)  & \includegraphics[width=0.3\textwidth]{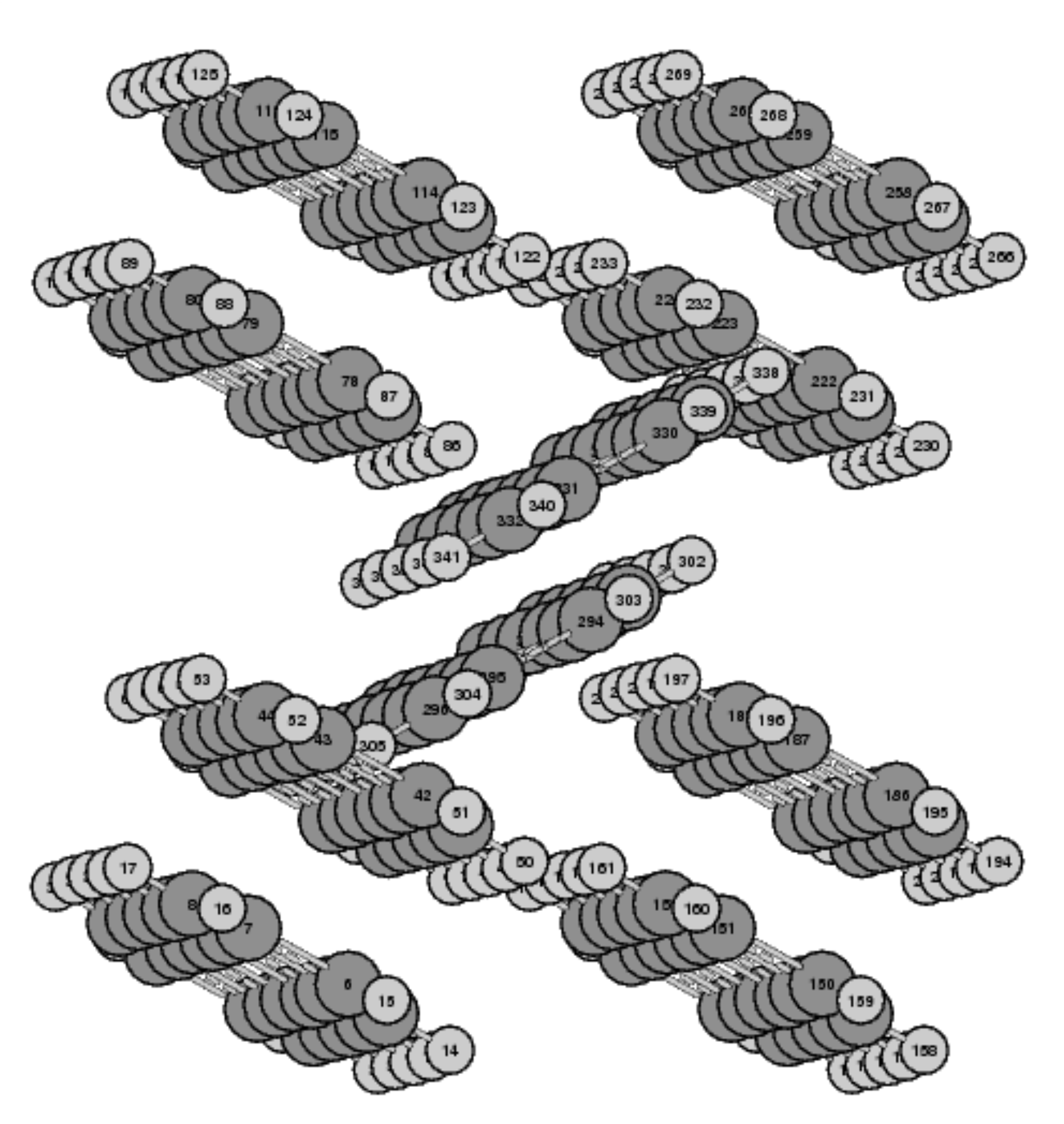} \\
%   & \includegraphics[width=0.3\textwidth]{./graphics/herringbone10h.eps} 
   & \includegraphics[width=0.3\textwidth]{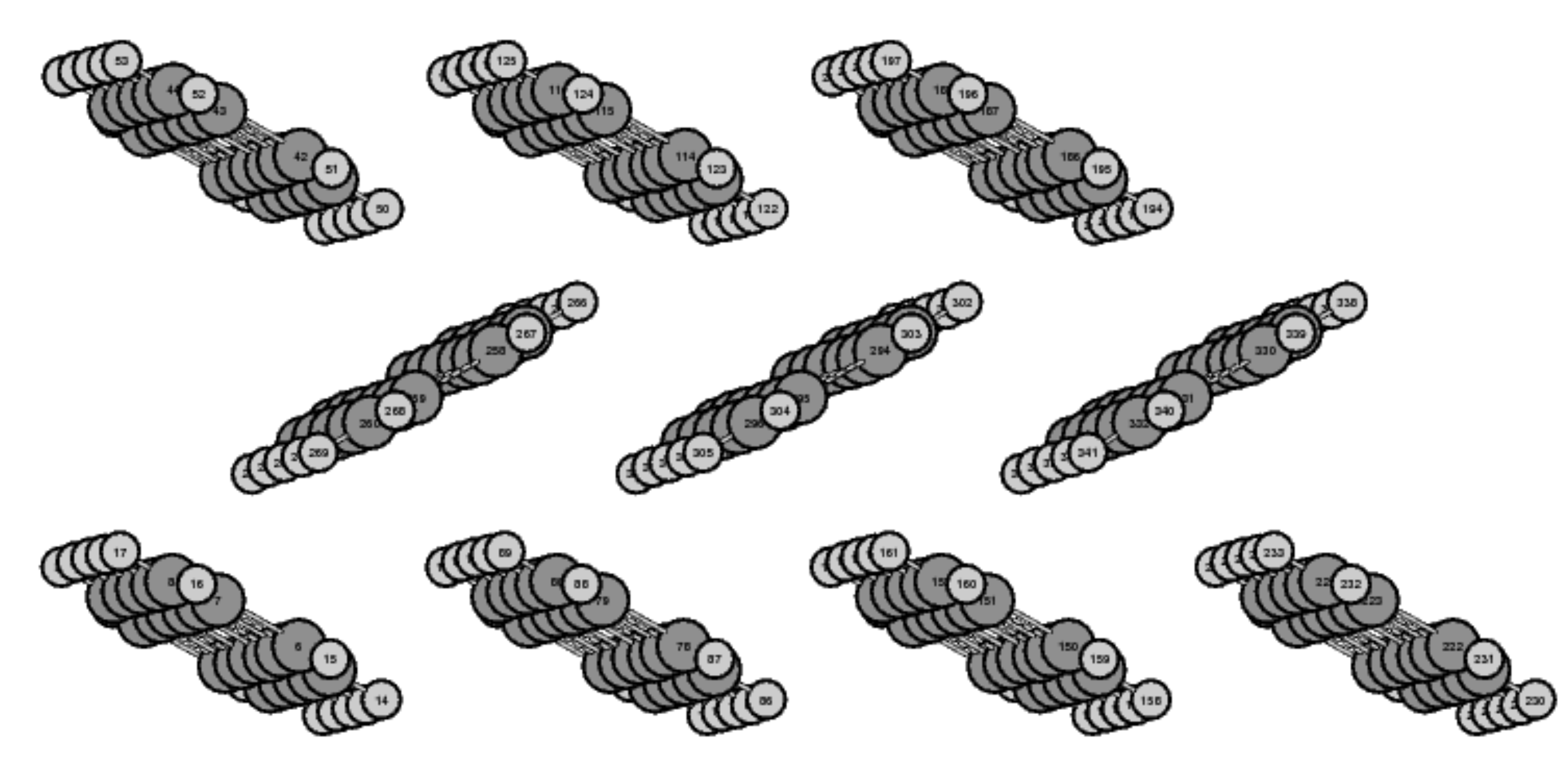} 
\end{tabular}
\caption{
\label{fig:herringbonemodels}
Herringbone cluster models used in this work.  All are 
portions of the x-ray crystal structure: 
(a) pentamer, % {\color{magenta} Ph}, 
(b) ``vertical'' decamer, and % {\color{magenta} Ph5-Ph5}, and 
(c) ``horizontal'' decamer. % {\color{magenta} Ph10}.
}
\end{figure}
% -------------------------------------------------------
We carried out calculations for the cluster models shown in 
Fig.~\ref{fig:herringbonemodels} obtained by cutting out different portions of the
x-ray crystal structure \cite{DM94,SHDN07} without any subsequent relaxation.
Unless otherwise indicated all of the results reported below are for the ``horizontal''
decamer model. The picture of the horizontal model makes it clear that the crystal
is made up of layers of tilted stacks of pentamers whose tilt angles alternate
from layer to layer to provide a herringbone structure.  

% ------------------------------------------------------
\begin{figure}
\includegraphics[width=0.5\textwidth]{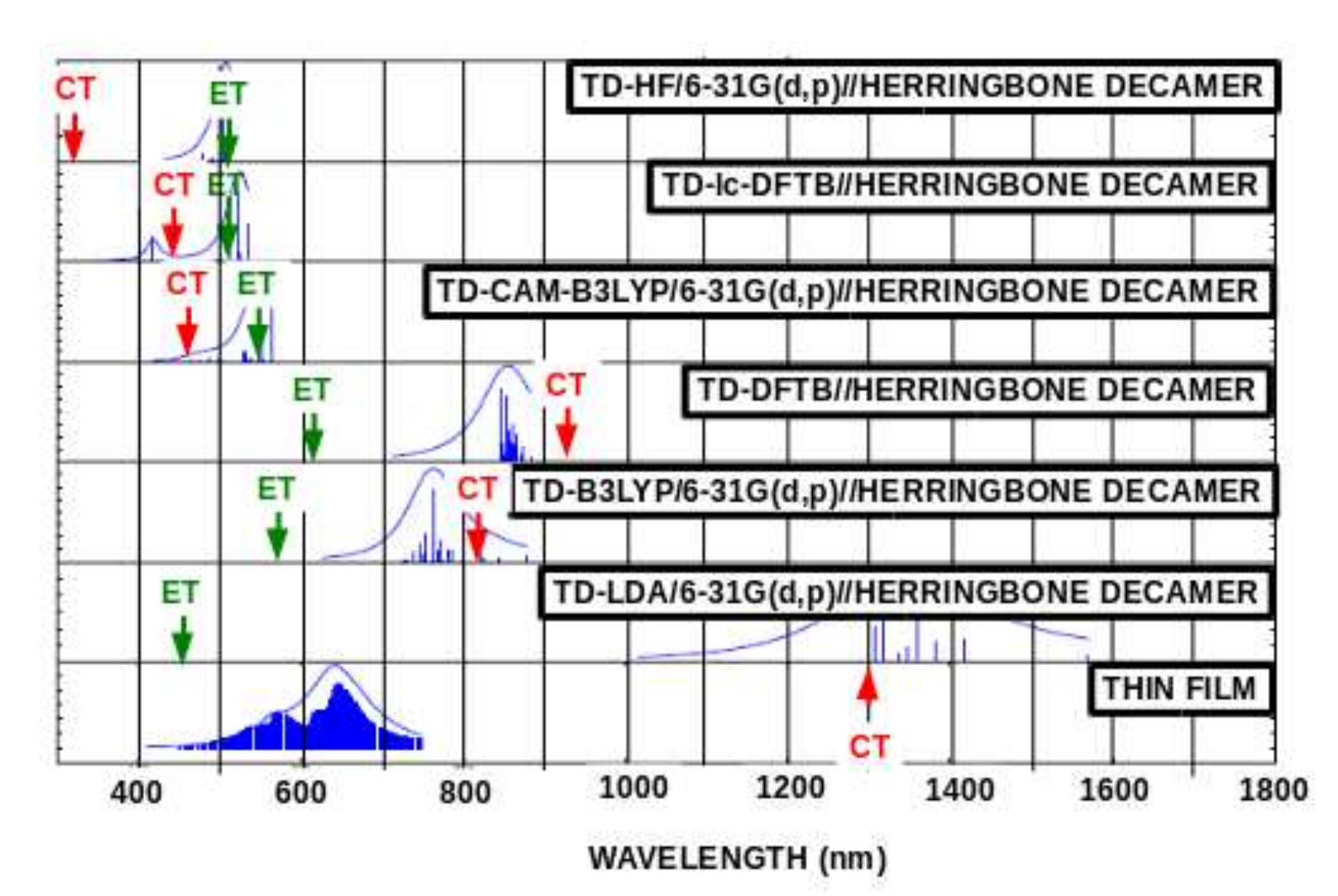}
\caption{
\label{fig:Ph_10}
Comparisons of calculations using various methods with the thin film absorption spectrum
from Ref.~\onlinecite{MRK+04}.  The CT and ET excitation energies were calculated from
Kasha's exciton model using Eq.~(\ref{eq:results.2}).
}
\end{figure}
% -------------------------------------------------------
Figure~\ref{fig:Ph_10} shows the herringbone spectra calculated at various 
levels and compared with the thin film spectrum.
Both the TD-LDA/6-31G(d,p) and the TD-B3LYP/6-31G(d,p) calculations are 
red shifted compared to the thin film experiment with the TD-LDA/6-31G(d,p) 
red shift being quite dramatic.  This is consistent with the idea that the 
TD-LDA/6-31G(d,p) exciton is delocalized over too many molecules while 
the inclusion of some HF exchange in the TD-B3LYP/6-31G(d,p) helps to increase
the excitation energy by localizing the exciton over fewer molecules.  
The TD-DFTB calculation is in semi-quantitative agreement with the 
TD-B3LYP/6-31G(d,p) but are slightly red-shifted.  In contrast, the 
TD-CAM-B3LYP/6-31G(d,p) and TD-HF/6-31G(d,p) calculations are blue shifted
compared to the thin film experiment.  The TD-lc-DFTB calculation is in 
semi-quantitative agreement with the TD-CAM-B3LYP/6-31G(d,p) calculation 
but is slightly blue shifted.  It is difficult to say from this figure 
which of the two calculations --- TD-B3LYP/6-31G(d,p) or 
TD-CAM-B3LYP/6-31G(d,p) --- is a better description of the experiment.

% ------------------------------------------------------
\begin{figure}
\begin{tabular}{cc}
a) & \\
% b) & \includegraphics[width=0.43\textwidth]{./graphics/HERRINGBONE/HERRINGBONE_B3LYP.eps} \\  % check
b) & \includegraphics[width=0.43\textwidth]{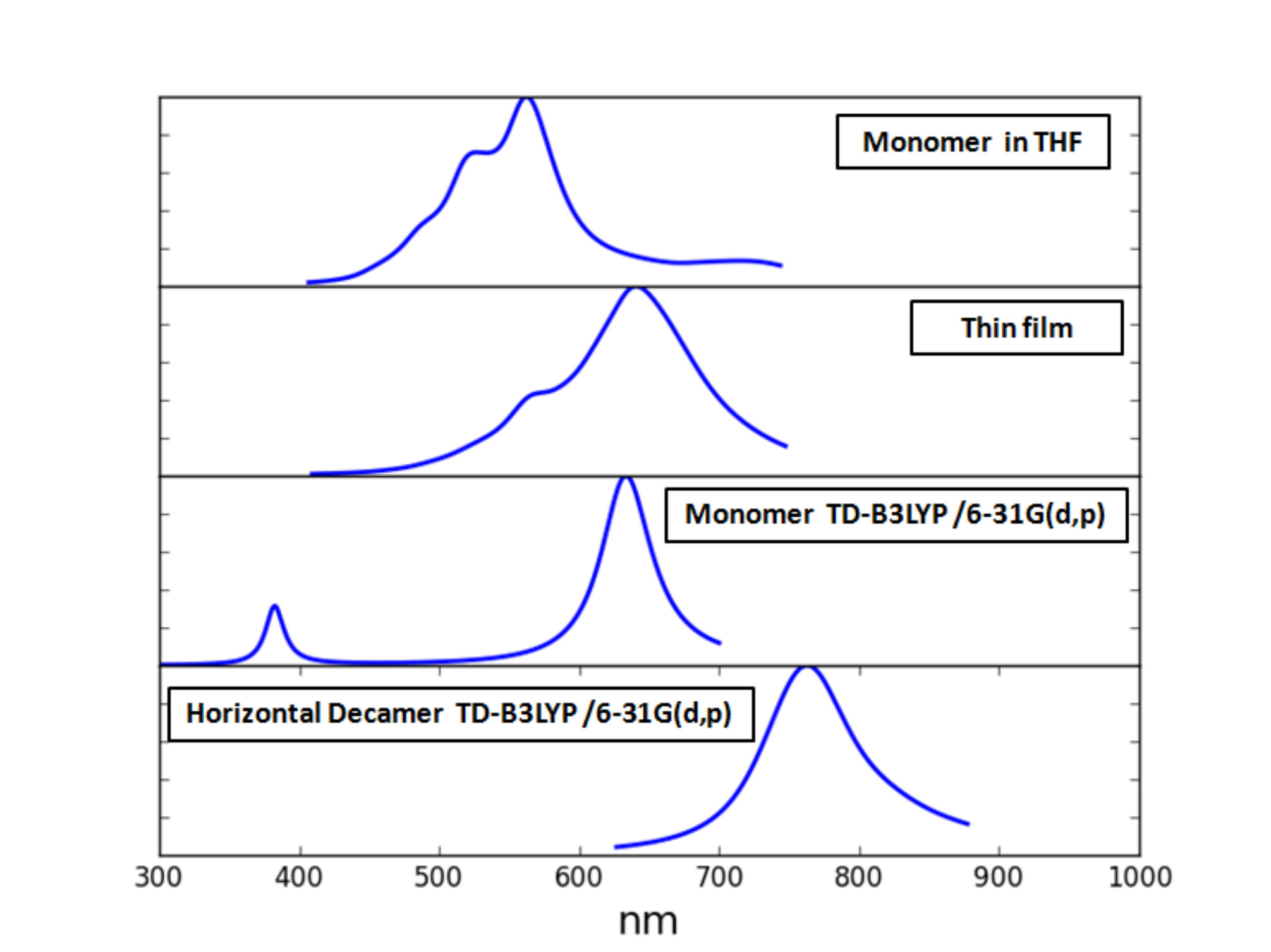} \\  % check
% c) & \includegraphics[width=0.43\textwidth]{./graphics/HERRINGBONE/HERRINGBONE_DFTB.eps} \\
c) & \includegraphics[width=0.43\textwidth]{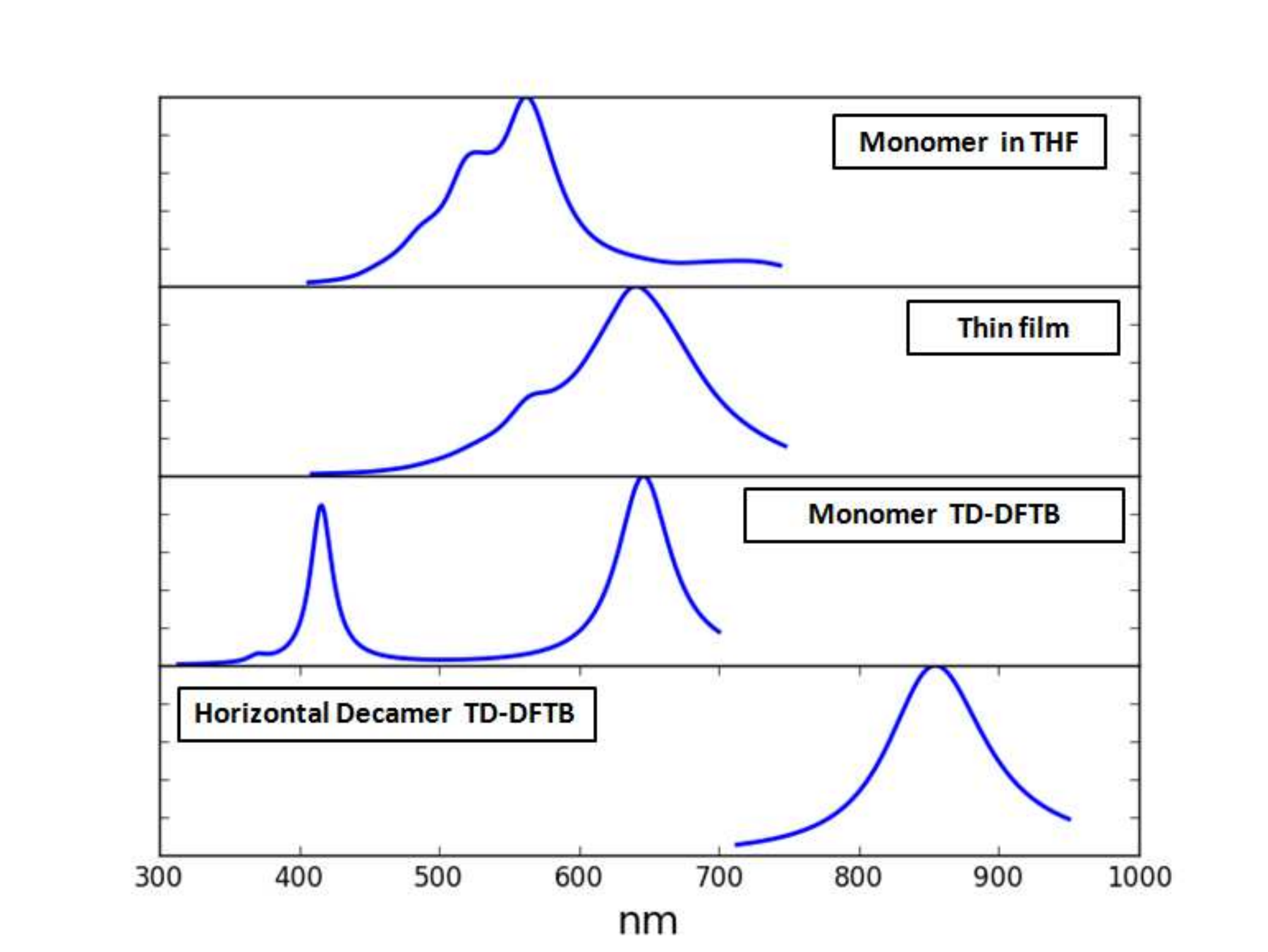} \\
% d) & \includegraphics[width=0.43\textwidth]{./graphics/HERRINGBONE/HERRINGBONE_CAM_B3LYP.eps} \\
d) & \includegraphics[width=0.43\textwidth]{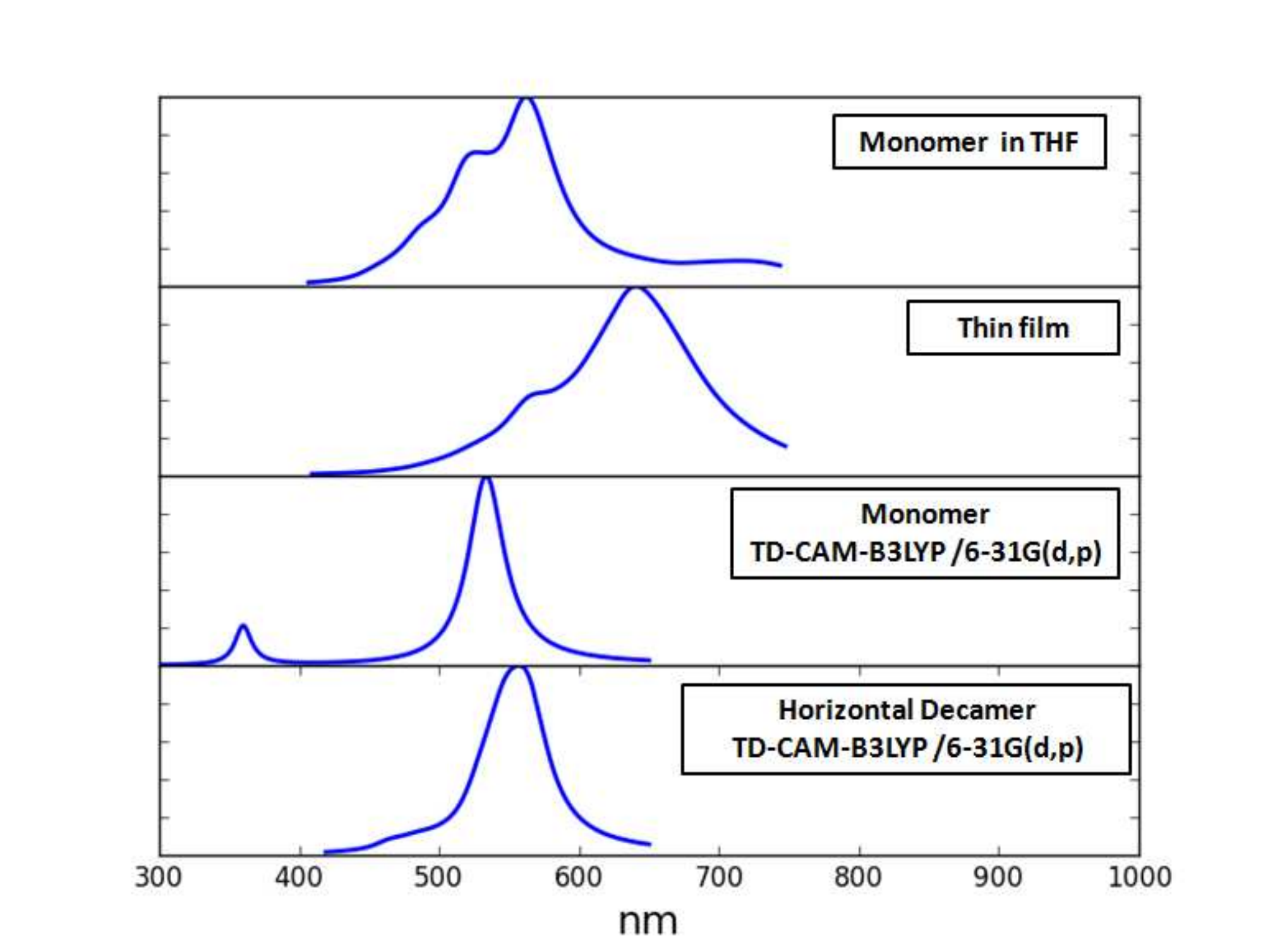} \\
%    & \includegraphics[width=0.43\textwidth]{./graphics/HERRINGBONE/HERRINGBONE_lc_DFTB.eps}
   & \includegraphics[width=0.43\textwidth]{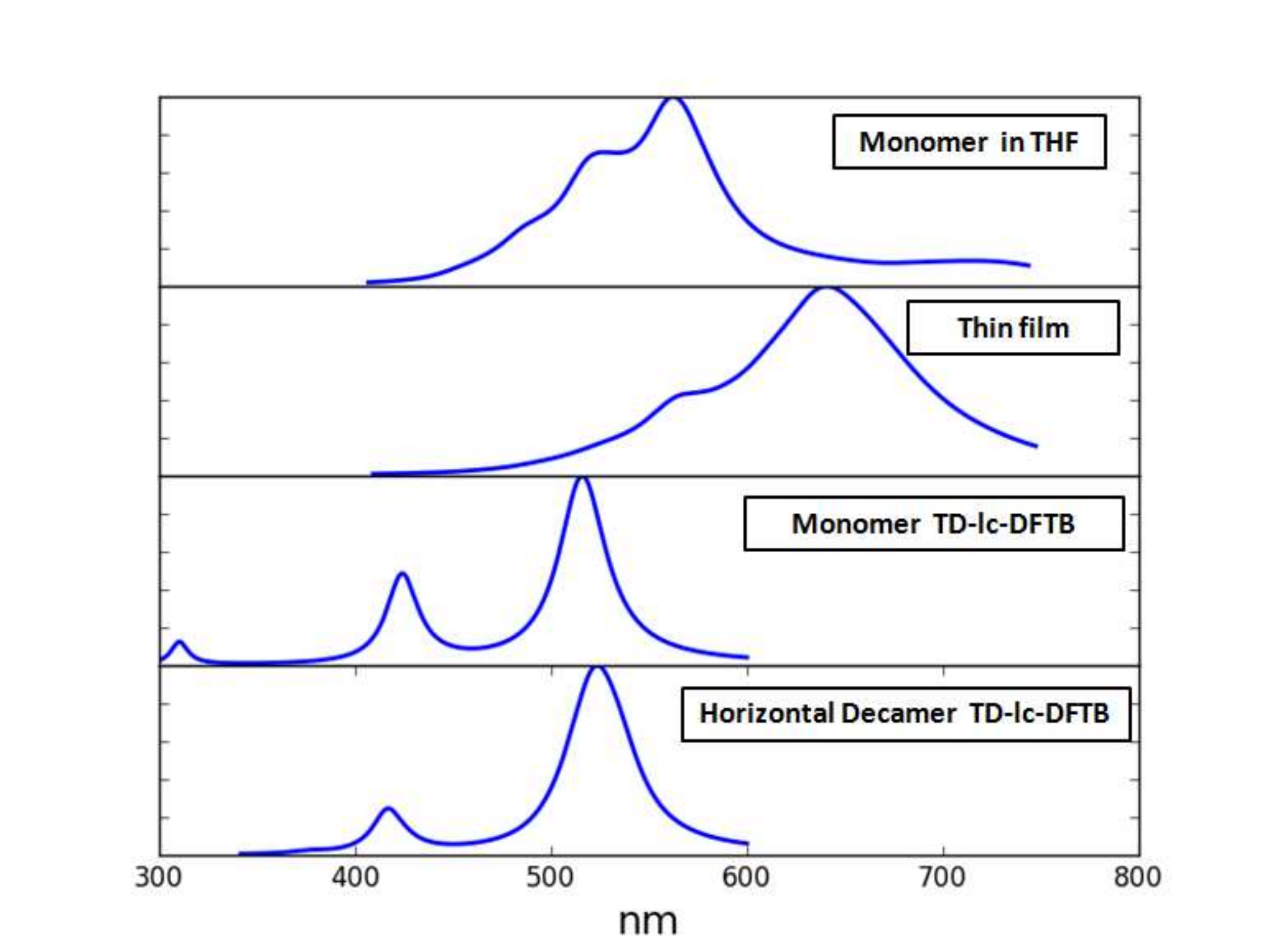}
\end{tabular}
\caption{
\label{fig:excitonshifts}
Comparison of theoretical and experimental exciton shifts: (a) TD-B3LYP/6-31G(d,p), (b) TD-DFTB,
(c) TD-CAM-B3LYP/6-31G(d,p), and (d) TD-lc-DFTB.
}
\end{figure}
% -------------------------------------------------------
The level of agreement with experiment is best judged by Fig.~\ref{fig:excitonshifts}.  Here we see that the TD-B3LYP/6-31G(d,p) and TD-DFTB results are in
resonable qualitative agreement with experiment.  However the TD-CAM-B3LYP/6-31G(d,p) and TD-lc-DFTB results, while in good agreement with each other, do not
at all provide a good description of exciton effects.  We assume that this is
because of the importance of CT which may be over corrected at the 
TD-CAM-B3LYP/6-31G(d,p) and TD-lc-DFTB levels compared with the 
TD-B3LYP/6-31G(d,p) and TD-DFTB levels.

% ========================================================
\subsection{Re-examination of Kasha's Model}
% ========================================================

There is some hope in the literature that you only need to apply Kasha's
original model to a single crystal plane to 
calculate the Davydov splitting of the crystal \cite{P73,YNH+11,MBK+16}.  In their
recent work \cite{MBK+16}, Meyenburg {\em et al.} give the formula [their Eq.~(5) rewritten
in atomic units],
\begin{equation}
  \Delta E = \frac{2 \vert {\vec \mu}^{I0} \vert^2}{\epsilon r^3} \left[ \cos(\phi)
  - 3 \cos(\alpha_1) \cos(\alpha_2) \right] \, ,
  \label{eq:results.1}
\end{equation}
where the angles are defined in Fig.~3 of their paper.  
[Equation~(\ref{eq:results.1}) is a generalization of a formula given
in the paper of Kasha, Rawls, and El-Bayoumi \cite{KRE65} (resulting in Fig.~4 of their
article).]
We translate this the following
relationship between the DS of the herringbone model DS$_{\text{HB}}$ 
and of the parallel stack model DS$_{\text{PS}}$:
\begin{eqnarray}
    \mbox{ DS$_{\text{HB}}$} & = & \mbox{DS$_{\text{PS}}$}
    \left( \frac{\epsilon_{\text{PS}}}{\epsilon_{\text{HB}}}  \right)
    \left( \frac{r_{\text{PS}}}{r_{\text{HB}}} \right)^3 \nonumber \\
    & \times & \left[ \cos(\phi) - 3 \cos(\alpha_1) \cos(\alpha_2) \right] \, .
    \label{eq:results.2}
\end{eqnarray}
The values $\alpha_1 = 98.19^\circ$, $\alpha_2 = 22.98^\circ$, 
$\phi=58.83^\circ$, and $r=4.835$ {\AA} were taken from the experimental crystal structure.
The ratio $\epsilon_{\text{PS}}/\epsilon_{\text{HB}}$ is not something that we can really
determine {\em a priori} and so we just use it as a fitting factor (equal to 10.64 in our
calculations).  We can use this, together with DS$_2$  (from {\em our} model)
to predict the DS$_{\text{HB}}$ that 
we would expect from various models.  These results are shown by arrows in 
Fig.~\ref{fig:Ph_10}. % \ref{fig:Kasha_reality_check}.  
Something very remarkable has happened: The locations of the TD-lc-DFTB, TD-HF, and 
TD-CAM-B3LYP main peaks are well produced, as expected, by the ET peak.  In contrast,
the location of the TD-DFTB, TD-B3LYP, and TD-LDA main peaks has been generated unexpectedly, 
from the CT peak, which was expected to be relatively dark compared to the ET peak.
This illustrates how easy it is to get the ``right answer for the wrong reason.''  In particular,
applying Kasha's original exciton model to results from our exciton model gives results
which seem reasonable (albeit only with a physically-unreasonable value of the dielectric
constant).

% ------------------------------------------------------
\begin{figure}
\begin{tabular}{cc}
a) & \\
% b) & \includegraphics[width=0.25\textwidth]{./graphics/KREgeom2.eps} \\
b) & \includegraphics[width=0.25\textwidth]{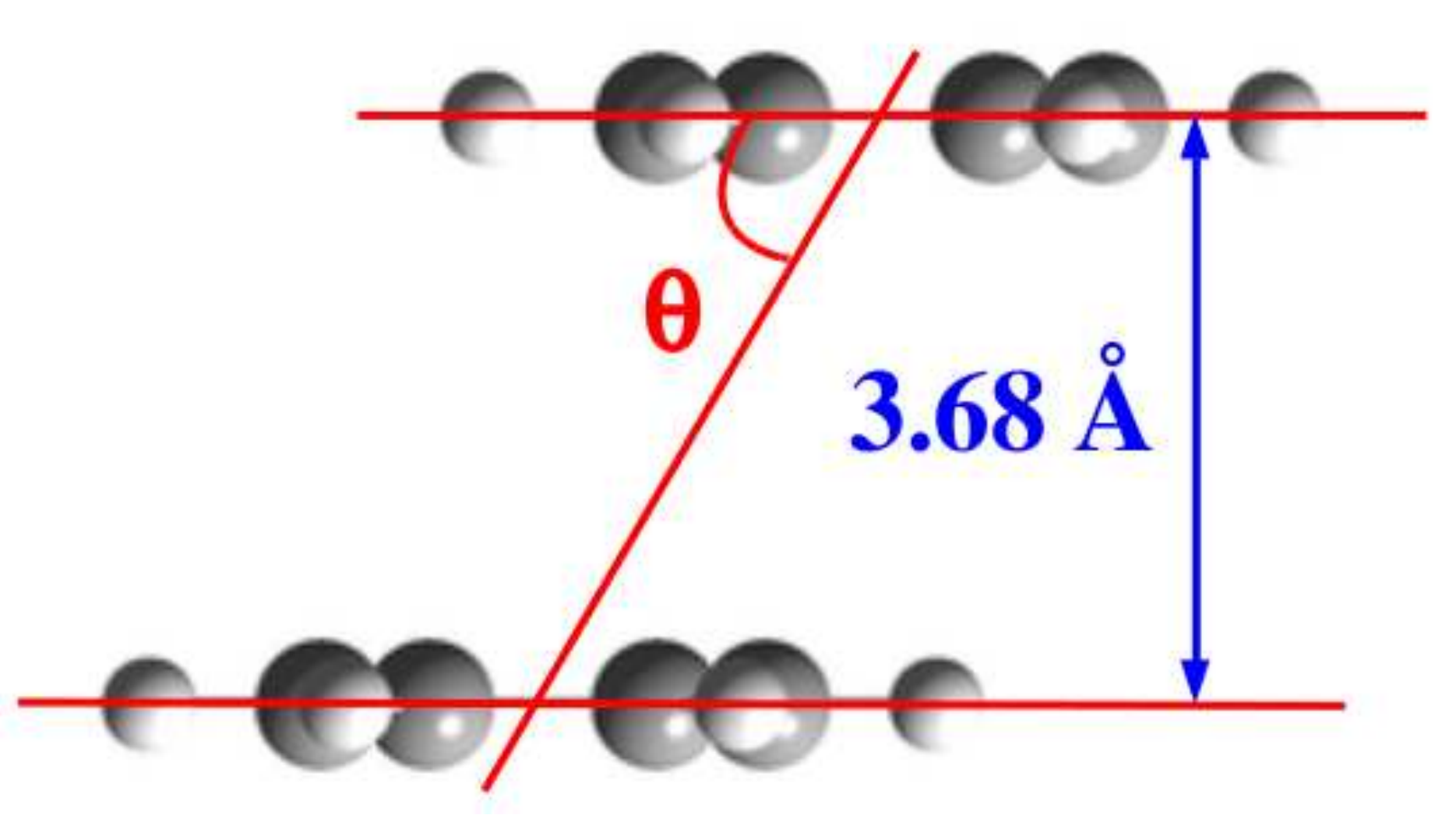} \\
% c) & \includegraphics[width=0.45\textwidth]{./graphics/KREFig4energy.eps} \\
c) & \includegraphics[width=0.45\textwidth]{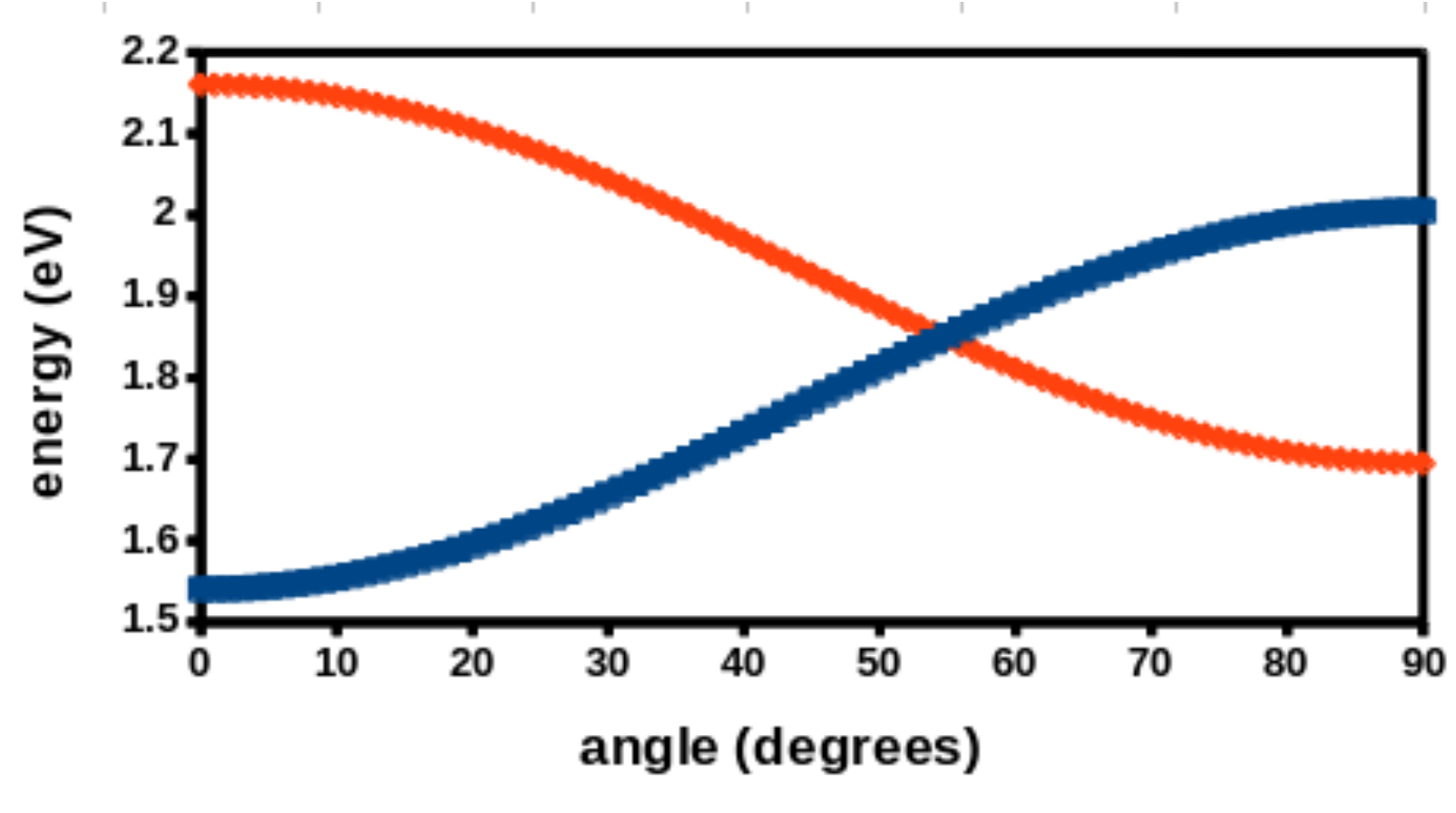} \\
%   & \includegraphics[width=0.35\textwidth]{./graphics/KREFours3Db.eps} \\
   & \includegraphics[width=0.35\textwidth]{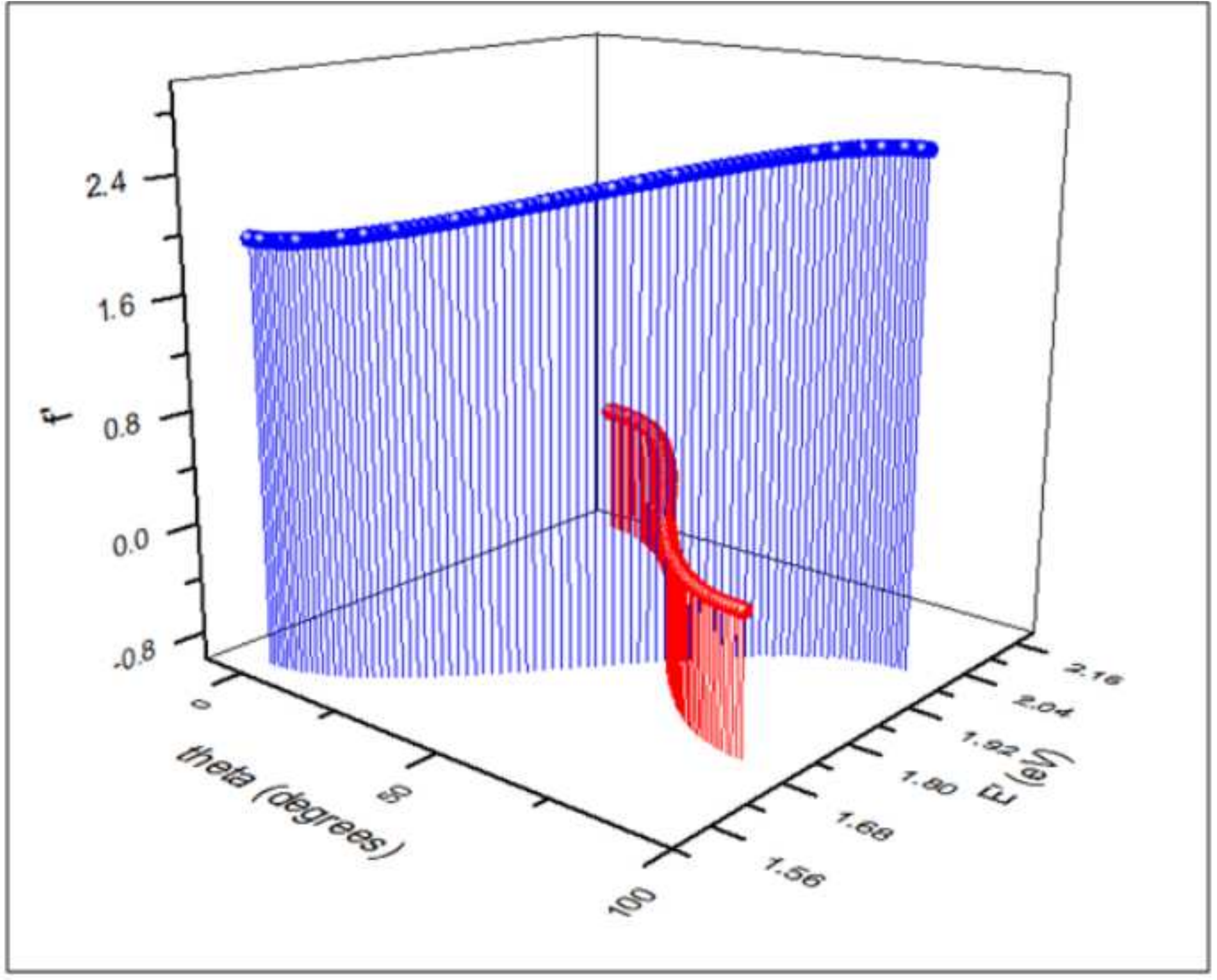} \\
\end{tabular}
\caption{
\label{fig:KREFig4ideal}
Ideal Kasha figures for co-planar inclined transition dipole moments: 
(a) laterally shifted parallel stacked pentacene dimer,
(b) Kasha plot of excitation energies as a function of the angle $\theta$,
(c) 3D plot of Kasha's model for oscillator strengths and excitation energies.
% (c) excitation energy plot for our TD-B3LYP/6-31G(d,p)//B3LYP+D3/6-31G(d,p)
% calculations. 
}
\end{figure}
% -------------------------------------------------------
This leads us to take a closer look at how Kasha's exciton model compares to 
our (TD-)DFT and (TD-)DFTB calculations.  
We have 
done this by looking at two parallel-stacked pentacenes.  This corresponds to a 
simplification of Eq.~(\ref{eq:results.2}), namely
\begin{equation}
  \Delta E = \frac{2 \vert {\vec \mu^{I0}} \vert^2}{r^3} \left(1-3\cos^2 \theta\right)
  \, ,
  \label{eq:conclude.1}
\end{equation}
with $\epsilon=1$.
Figure~\ref{fig:KREFig4ideal} summarizes what we expect to see on the basis of
the original exciton model for excitation energies and oscillator strengths.
This is a simple theory which gives simple curves.
Above $\theta=55^\circ$, the main oscillator strength is in the higher energy (lower
wavelength) ET state; below $\theta=55^\circ$,  the main oscillator strength is in the
lower energy (higher wavelength) ET state which has crossed the CT state.
However the original theory did not anticipate avoided crossings.  
For the herringbone structure, $\theta \approx 60^\circ$ and the prediction is that
the brighter state should be the higher wavelength state.  This is indeed what is
seen in all the calculations in Fig.~\ref{fig:Ph_10}, but the assignment of the
longer wavelength peak depends upon the method.

% ------------------------------------------------------
\begin{figure}
\begin{tabular}{cc}
a) & \\
% b) & \includegraphics[width=0.35\textwidth]{./graphics/Spectra/KREFours2Da.eps} \\
b) & \includegraphics[width=0.35\textwidth]{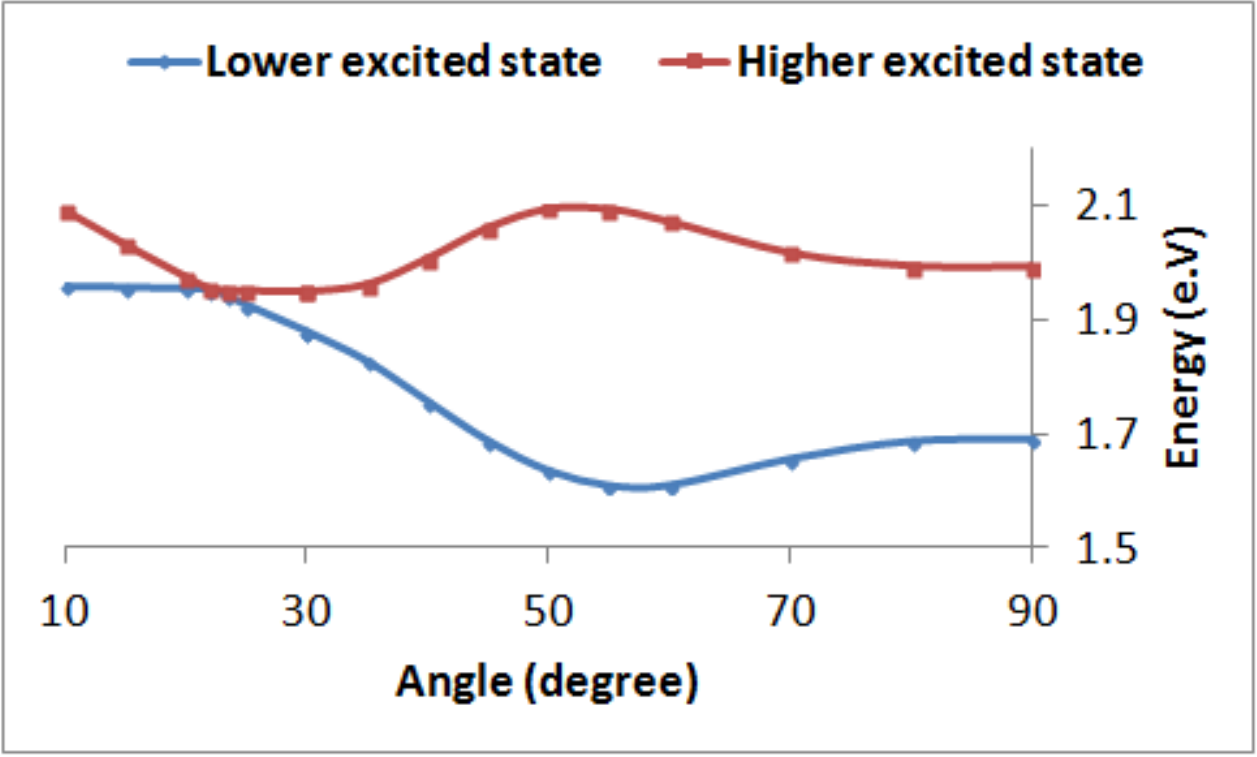} \\
% c) & \includegraphics[width=0.35\textwidth]{./graphics/Spectra/KREFours2Db.eps} \\
c) & \includegraphics[width=0.35\textwidth]{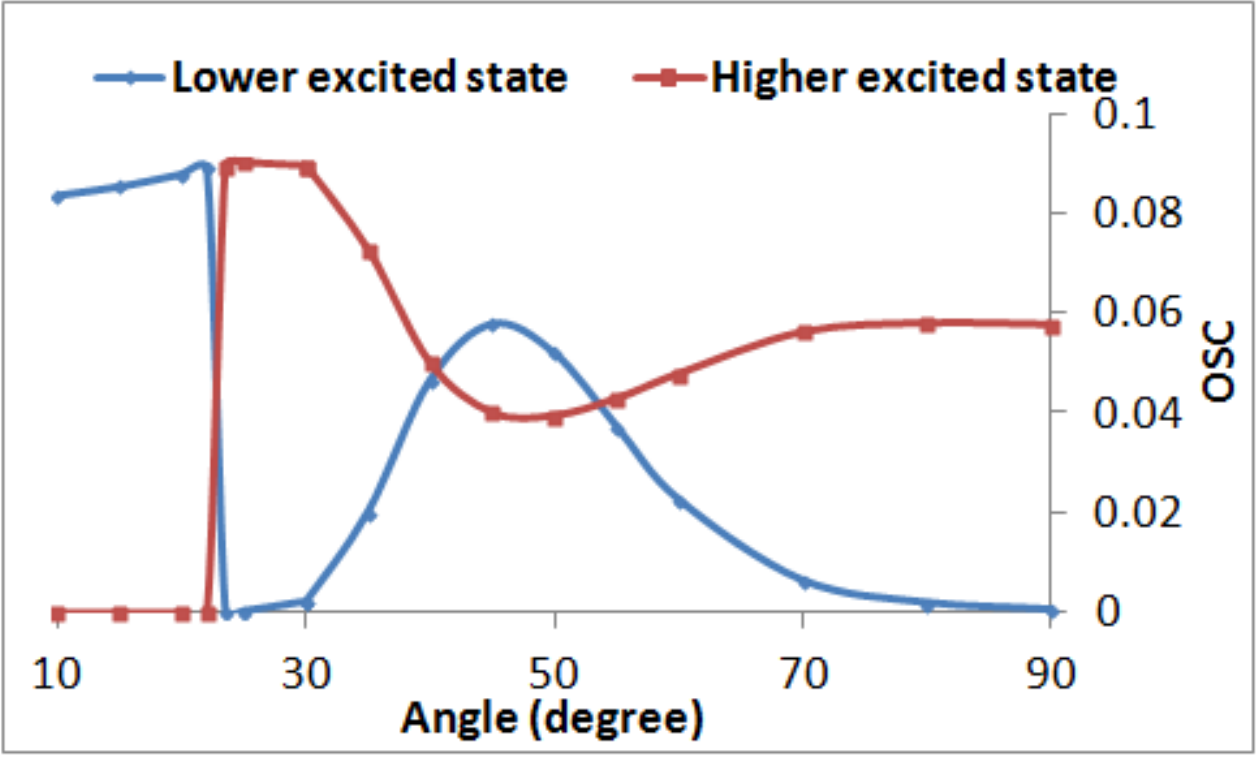} \\
%   & \includegraphics[width=0.35\textwidth]{./graphics/KREFours3Da.eps} \\
   & \includegraphics[width=0.35\textwidth]{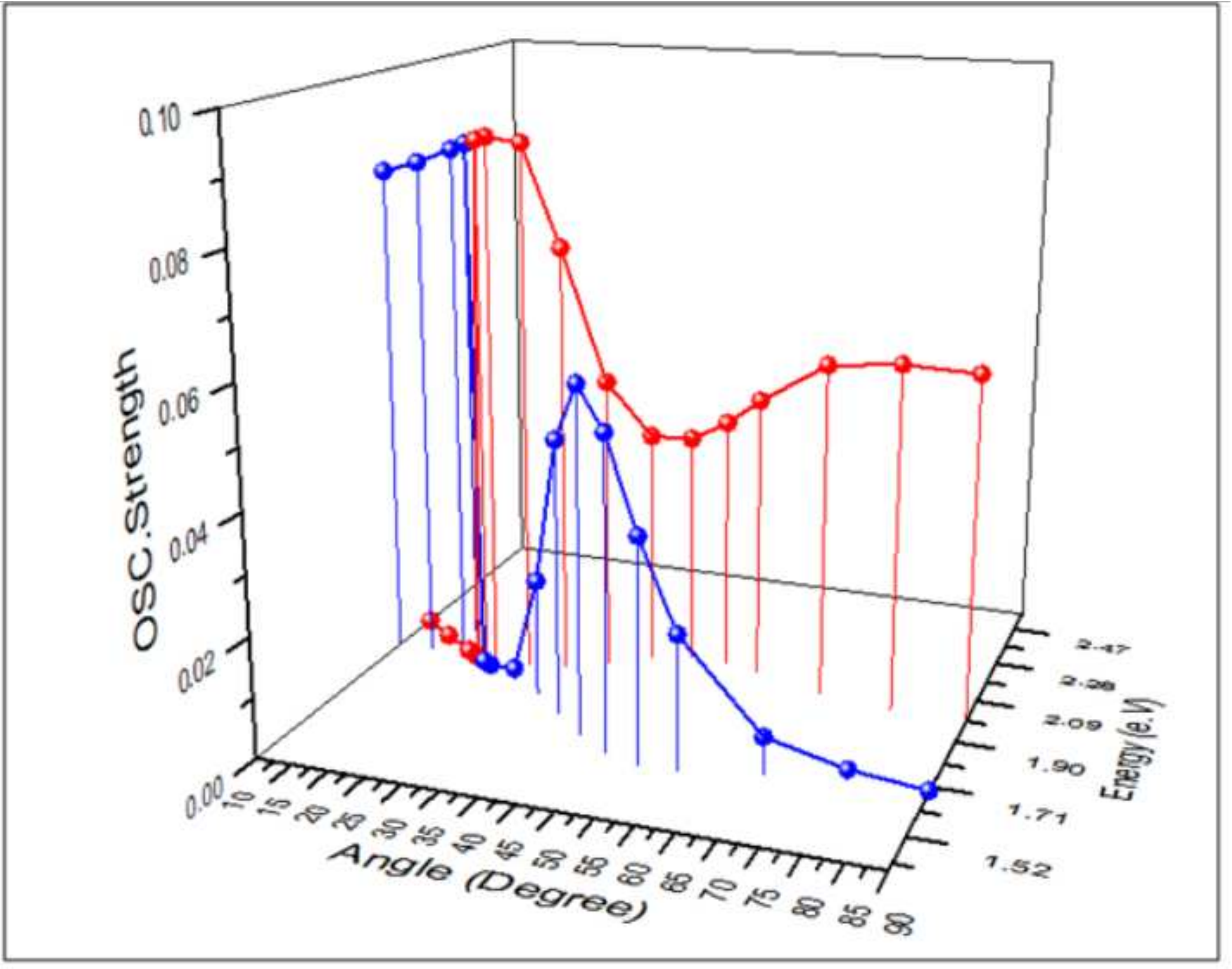} \\
\end{tabular}
\caption{
\label{fig:KREFig4B3LYP}
TD-B3LYP/6-31G(d,p)//B3LYP+D3/6-31G(d,p) Kasha figures for co-planar inclined transition dipole moments: 
(a) excitation energies as a function of the angle $\theta$,
(b) 2D plot of oscillator strengths,
(c) 3D plot of  oscillator strengths and excitation energies,
}
\end{figure}
% -------------------------------------------------------
Let us turn now to Fig.~\ref{fig:KREFig4B3LYP} which shows results from our 
TD-B3LYP/6-31G(d,p)//B3LYP+D3/6-31G(d,p) calculations.  This more realistic calculation
gives more complicated results.  Rather than seeing a simple crossing of energy levels
at about $\theta = 55^\circ$ as in Fig.~\ref{fig:KREFig4ideal}(b), we see evidence of configuration
mixing as the oscillator strength is transfered from one state to antother
with maximum transfer near $\theta = 55^\circ$ (i.e., at $\theta = 45^\circ$) as
the two energy levels mix and move apart at $\theta = 55^\circ$.  This seems even
more clear in the 3D plot [part (c) of Fig.~\ref{fig:KREFig4B3LYP}] where the two curves
behave in a way very consistent with an avoided crossing of two diabatic states 
(one with high and one with low oscillator strength) around $50^\circ$.  Later, at a much
lower value of $\theta=20^\circ$, there appears to be a real crossing, but this is of
little importance for understanding the herringbone results.
Comparing with the TD-DFTB, TD-B3LYP, and TD-LDA herringbone spectra in Fig.~\ref{fig:Ph_10},
we realize that the assignment of the longer wavelength peak as CT is misleading as most
likely there is a great deal of mixing between the ET and CT states.

% ------------------------------------------------------
\begin{figure}
\begin{tabular}{cc}
a) & \\
% b) & \includegraphics[width=0.35\textwidth]{./graphics/Spectra/KREFours2DCAMa.eps} \\
b) & \includegraphics[width=0.35\textwidth]{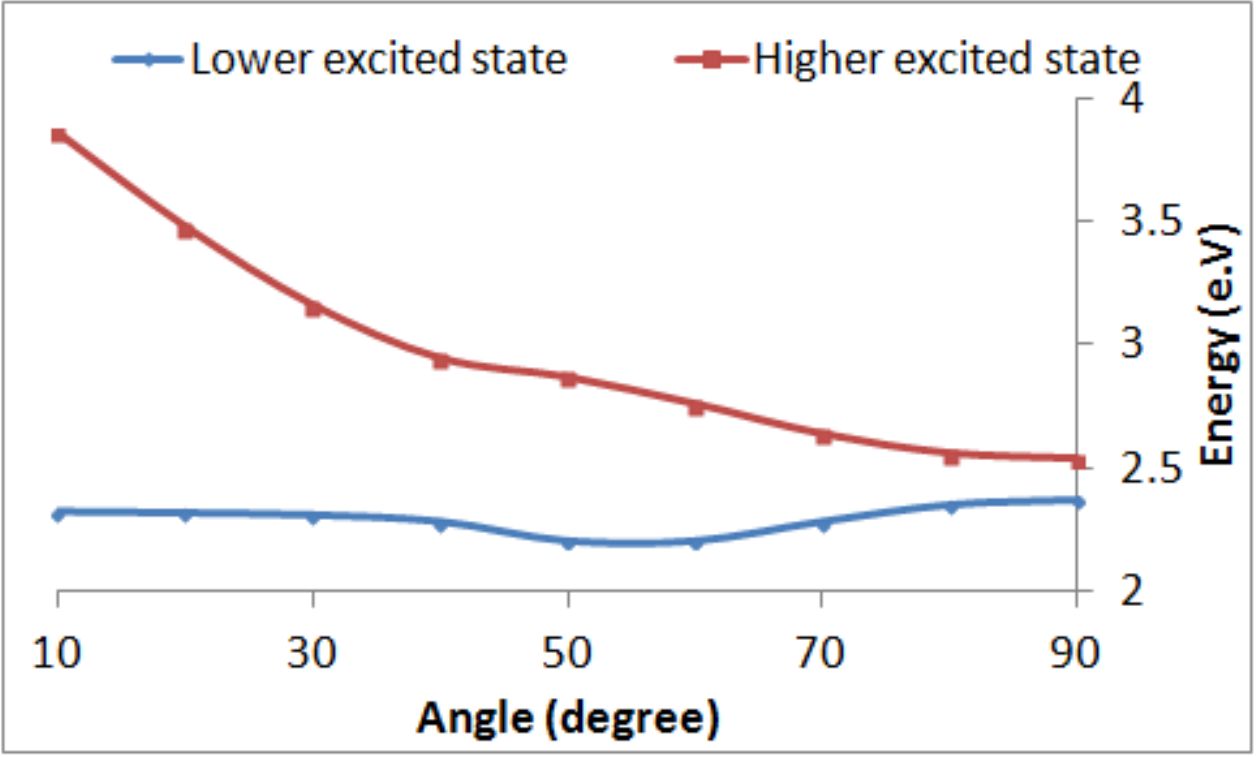} \\
% c) & \includegraphics[width=0.35\textwidth]{./graphics/Spectra/KREFours2DCAMb.eps} \\
c) & \includegraphics[width=0.35\textwidth]{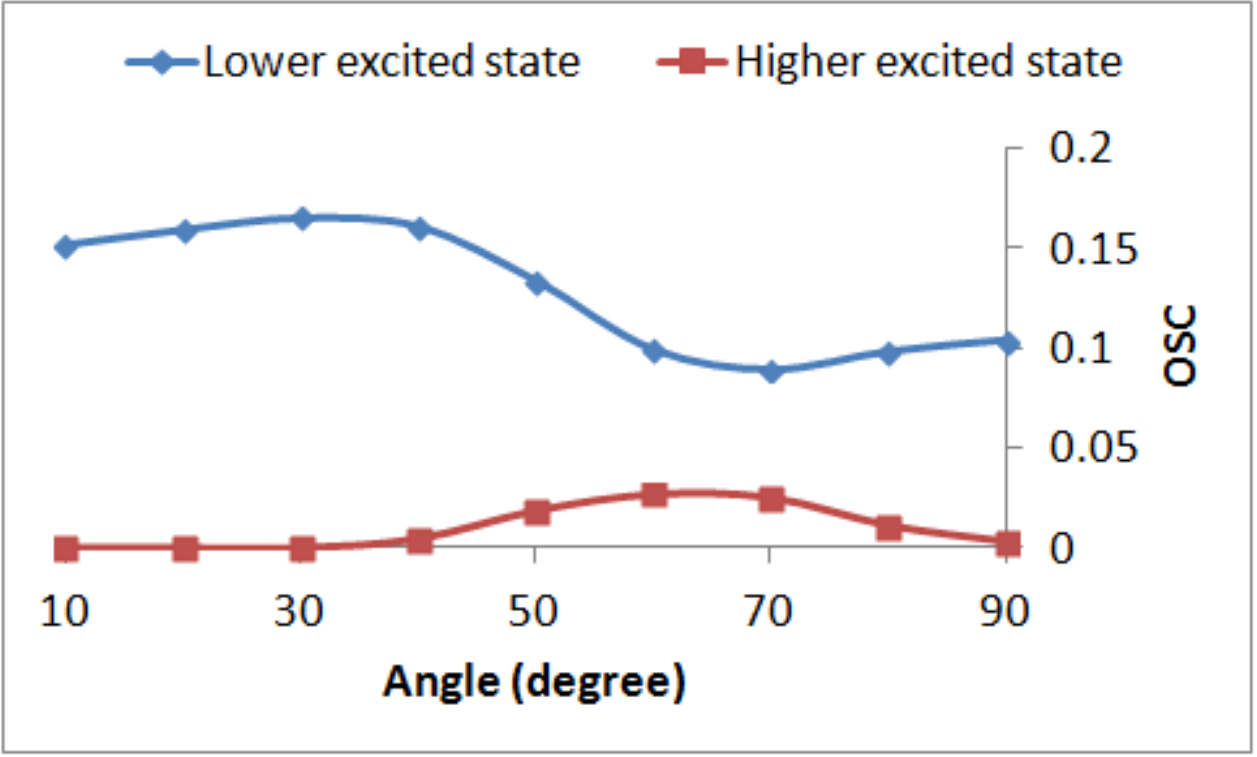} \\
%   & \includegraphics[width=0.35\textwidth]{./graphics/KREFours3DCAMa.eps} \\
   & \includegraphics[width=0.35\textwidth]{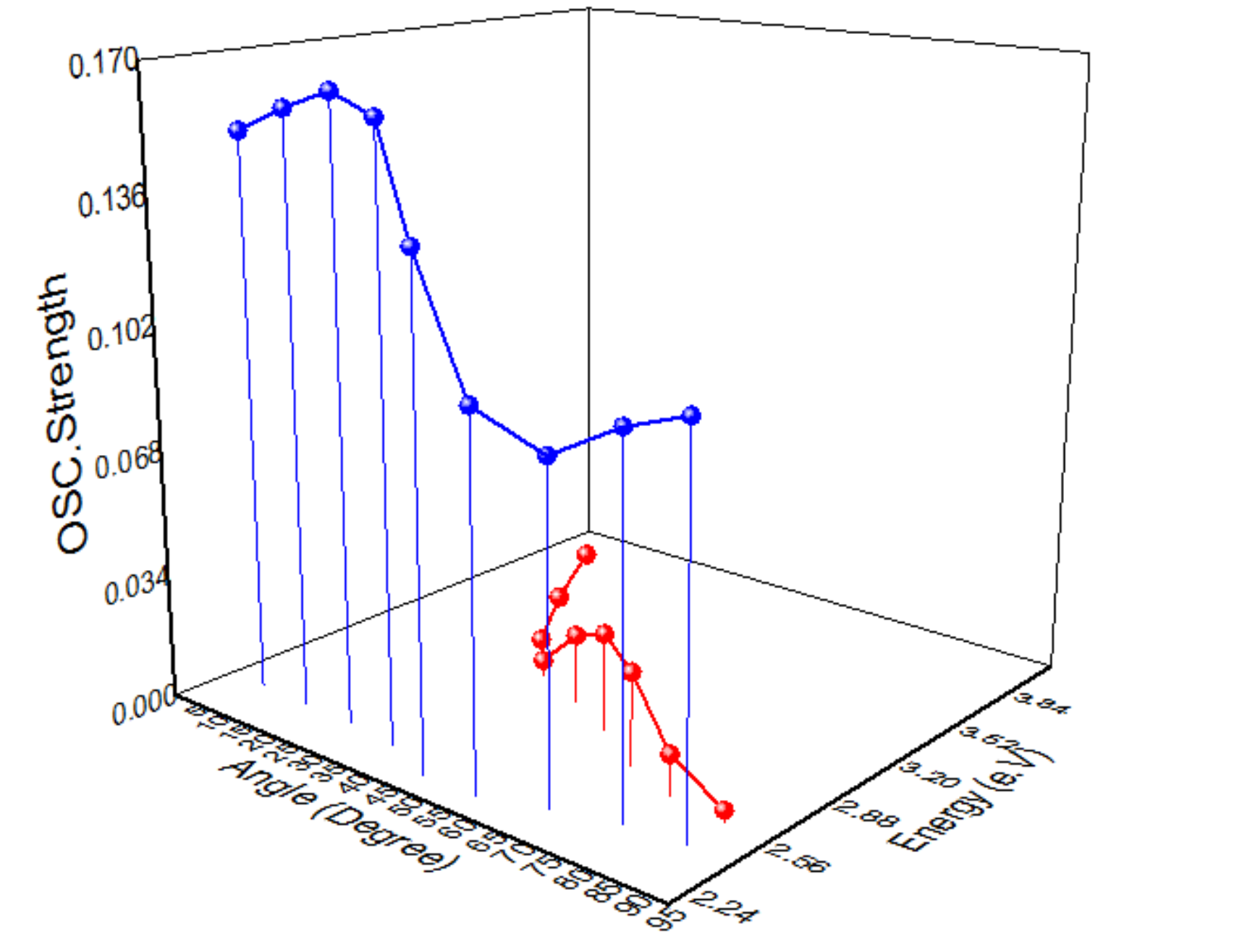} \\
\end{tabular}
\caption{
\label{fig:KREFig4CAMB3LYP}
TD-CAM-B3LYP/6-31G(d,p)//CAM-B3LYP+D3/6-31G(d,p) Kasha figures for co-planar inclined transition dipole moments: 
(a) excitation energies as a function of the angle $\theta$,
(b) 2D plot of oscillator strengths,
(c) 3D plot of  oscillator strengths and excitation energies,
}
\end{figure}
% -------------------------------------------------------
Figure~\ref{fig:KREFig4CAMB3LYP} shows what happens for a RSH.
In particular, TD-CAM-B3LYP/6-31G(d,p)//CAM-B3LYP+D3/6-31G(d,p) results are shown.
There is also some evidence of an avoided crossing here, but the lower energy
(longer wavelength) state keeps most of the oscillator strength.  At $\theta=90^\circ$,
it is of ET type, but shows some mixing between CT and ET type at around
$\theta=65^\circ$. 
Comparing with the TD-lc-DFTB, TD-HF, and TD-CAM-B3LYP herringbone spectra in 
Fig.~\ref{fig:Ph_10}, we realize that the assignment of the longer wavelength peak as 
ET is reasonable in this case.

Despite these criticisms that Kasha's original exciton model is missing many of the subtleties
of our more elaborate calculations, we must conclude that the exciton model allows
us to extrapolate remarkably well from intramolecular interactions between pairs of molecules
to larger aggregates, but only when a RSH is used.  

%%%%%%
% EOF
%%%%%%
% -------------------------------------------------
\section{Conclusion}
\label{sec:conclude}
% \input{conclude.tex}
% ==============================
% File: conclude.tex
% Last update: 4 February 2018
% ==============================

The aggregation of dye molecules leads to additional spectral features which
have long been explained by Kasha's exciton model \cite{KRE65}.  This model
provides a simple explanation of H-bands such as those found in our calculations
of parallel stacked pentacene molecules and of J-bands such as those associated
with the experimentally-known herringbone structure of crystalline pentacene.
We have re-examined exciton effects on spectra in terms of state-of-the-art TD-DFT 
and TD-DFTB calculations.  Our calculations include Grimme's D3 dispersion 
correction which we confirm is absolutely necessary in order to have van der Waals 
binding beyond the LDA level (which, however, ``accidently'' binds when an 
extensive-enough basis set is used).  We have also included some HF exchange
either in the form of the global B3LYP hybrid functional or in the form of
the range-separated CAM-B3LYP hybrid functional.  Corresponding TD-DFTB 
calculations have also been performed with {\sc DFTbaby} which we find does
very well at mimicking the two functionals (depending upon whether or not the
long-range correction is used).

Perhaps not surprisingly we find that the hybrid functionals give better
spectra than the TD-LDA.  We also find that, while the TD-DFTB spectra
have the structural simplicity of STO-3G minimum basis set TD-DFT spectra,
the TD-DFTB spectral peak energies are placed more like the TD-DFT calculations
with larger basis used in the semi-empirical parameterization.  This is reassuring.
However care must be taken when doing TD-DFTB calculations on aggregates to do them
size consistently or results can be misleading --- that is, the size of the active
space must increase in proportion to the number of molecules in the aggregate.

On the other hand, Fig.~\ref{fig:excitonshifts} is definitely telling us that
the Davydov splitting for the herringbone model is reasonable and close to 
the experimental value at the TD-B3LYP and TD-DFTB levels, even if the spectral 
peaks are shifted, but that the Davydov splitting is practically nonexistant
and therefore wrong at the CAM-TD-B3LYP and TD-lc-DFTB levels.  For this reason,
we cannot recommend using either of these latter two approaches for calculating
excitonic effects on the spectra of polyoligocenes and recommend instead the
use of TD-B3LYP  or TD-DFTB carefully carried out on aggregates.

Also, in light of arguments regarding the extent and nature of exciton delocalization
(e.g., Refs.~\onlinecite{ZBCH09,MPD17}), it is interesting that our analysis shows that 
a more sophisticated version of Kasha's exciton model, based only on nearest neighbor
interactions but including both ET and CT as well as avoided crossings, works very well.
In contrast, Kasha's original model without further modification 
fails because of its absence of CT excitations. 
% for displaced parallel pentacene dimers unless 
% RSHs are used because of failure to take proper account of avoided crossings.

It is perhaps not misplaced to say a brief word about some of the many challenges
of describing ET and CT excitons that are important for organic electronics but
which have not been addressed in this paper.  One is the need to take into account
the more extended environment.  This was only included approximately here in our
``Kasha model calculation'' by invoking an environmental dielectric constant.  It
really should be done more correctly and in such a way as to recognize that the effect 
will be different depending upon the ET and CT nature of the exciton.  Furthermore,
as we have seen with the RSH functionals, excitons may be of a strongly mixed ET/CT 
nature whose precise mixture may depend upon environmental effects.  Finally, in many
organic electronic problems there is an interface with local states which must couple
with bulk states.  Finding the correct coupling between these two types of states is
nontrivial, especially if the goal is to describe both states at the same level of theory.
Nevertheless, with all of these caveats, we believe that the present article represents
a helpful step towards the practical investigation of realistic models for organic
electronics.

%%%%%
% EOF
%%%%%
% -----------------------------------------------
\begin{acknowledgments}
% \input{thanks.tex} 
% \begin{verbatim}
% ================================
% File: thanks.tex
% Last update: 25 January 2018
% ================================
% \end{verbatim}

This work is supported, in part, by the ORGAVOLT (ORGAnic solar cell VOLTage by
numerical computation) Grant ANR-12-MONU-0014-02 of the French {\em Agence Nationale de 
la Recherche} ({\em ANR}) 2012 {\em Programme Mod\`eles Num\'eriques}. 
AAMHMD would like to thank the French and Iraq governments
for support through a Campus France scholarship. 
GCC was an American iREU (International Research Experience for
Undergraduates) scholar who acknowledges support from Chemistry (CHE) Grant Number 
1263336 of the American National Science Foundation (NSF) and the 
Labex Arcane ANR-11-LABX-0003-01 French Grant.
We would like to thank Sergei Tretiak and Thomas Niehaus for useful comments on 
our manuscript.
MEC and AAMHMD would like to acknowledge 
helpful discussions with Mathias Rapacioli. 
MEC and AAMHMD would also like to acknowledge 
a useful trip to W\"urzburg funded by the German GRK 2112 Project ``Biradicals.''
MEC thanks Lucia Reining for many profitable discussions about what excitons mean
in different sub-fields of chemical and solid-state physics.

% ====
% EOF
% ====
\end{acknowledgments}

\appendix
% \input{append.tex}
% --------------------------------
\section{DFT}
\label{sec:DFT}
% --------------------------------

Hohenberg-Kohn-Sham DFT \cite{HK64b,KS65} is now so well-known that little needs to be said
about the basics. For those seeking a deeper understanding of the foundations of DFT,
we can recommend Refs.~\onlinecite{PY89,DG90,KH00}.  Our intent here is two fold.
Partly we wish to go beyond what is found in those references and partly we wish to
review that part of the basics which is needed to understand DFTB. 

The fundamental idea of DFT is to replace the real system of $N$ interacting electrons in an
external potential $v_{\text{ext}}$ with a fictitious system of $N$ noninteracting electrons in an
effective potential $v_s$ ($s$ for single electron).  We will designate the orbitals of the
noninteracting system as $\psi_i$ and their associated occupation numbers as $n_i$.
These orbitals are orthonormal, $\langle \psi_i \vert \psi_j \rangle = \delta_{i,j}$.
Normally $n_i$ is zero or one but fractional occupation is also allowed.  
The density matrix for the noninteracting system is then, 
$\gamma(1,2) = \sum_i \psi_i(1) n_i \psi_i^*(2)$,
where the numeral $i=1,2,\cdots$ stands for the space and spin coordinates of the $i$th electron.
The density, $\rho(1) = \gamma(1,1)$, is the diagonal element of the density matrix.  
The electronic energy $E$ is the same for the
real and for the interacting systems.  It may be written as the sum of three terms,
\begin{equation}
  E = E_c + E_H + E_{xc} \, ,
  \label{eq:theory2.4}
\end{equation}
where the core energy $E_c$ is the expectation value of the core hamiltonian, 
$\hat{h}_c = -\frac{1}{2} \nabla^2 + v_{\text{ext}}$,
the Hartree energy (also called the classical coulomb repulsion energy) $E_H$ denotes
the classical repulsion between $\rho$ and itself,
and the remaining terms needed to make the electronic energy {\em exact} are included
in the exchange-correlation (xc) energy, $E_{xc} = E - E_c - E_H$.
This term is approximated in practical calculations.  We will consider these
approximations very soon.  For now, let us note that minimizing $E$ subject to the
orbital orthonormality condition leads to an orbital equation,
\begin{equation}
   \hat{f} \psi_i = \epsilon_i \psi_i \, ,
   \label{eq:theory2.9}
\end{equation}
where $\hat{f} = \hat{h}_c + v_H + v_{xc}$.  Here 
$v_H(1) = \int  \rho(2)/r_{12} \, d2$ is the Hartree potential and 
$v_{xc}(1) = \delta E_{xc}/\delta \rho(1)$ is the xc potential.  Then it is easy to 
see that the effective potential of the noninteracting system is,
$v_s = v_{\text{ext}} + v_H + v_{xc}$.
The orbital equation~(\ref{eq:theory2.9}) must be solved self-consistently
because $\hat{f}$ is orbital dependent.  Once self-consistency has been reached,
then the energy may be calculated either using Eq.~(\ref{eq:theory2.4}) or by
using the equation
\begin{equation}
  E = \sum_i n_i \epsilon_i - \frac{1}{2} E_H + E_{xc} - \int v_{xc}(1) \rho(1) \, d1
  \label{eq:theory2.14}
\end{equation}
This latter form is used as the basis of DFTB.

Thus far the equations are exact, but useless unless approximations are made.
We give a brief review of the approximate functionals used in this paper beginning
from the most well-known and ending with the range-separated hybrids which play a
central role in this paper.

The most common approximation is the local (spin) density approximation (LDA)
which assumes that the xc energy density in a nonhomogeneous system such as an atom,
molecule, or solid, is the same as the xc energy density in the homogeneous
electron gas (HEG),
$ E_{xc}^{\text{LDA}} = \int \epsilon_{xc}^{\text{HEG}}(\rho(1)) \rho(1) \, d1$.
The exchange part of $\epsilon_{xc}^{\text{HEG}}$ has a simple analytic form \cite{D30}.
We will use the Vosko-Wilk-Nusair parameterization of the correlation part of 
$\epsilon_{xc}^{\text{HEG}}$ \cite{VWN80} in the present work.  Strictly speaking
all of our calculations have an explicit dependence on the spin polarization of the
local density.  That is, we are using spin DFT rather than the original DFT which
depended only on the spinless charge density.

The LDA often gives reasonable molecular geometries but is known to overbind.
For this reason, it has been useful to include inhomogeneities
in the density via generalized gradient approximations (GGAs) of the form,
$E_{xc}^{\text{GGA}} = \int \epsilon_{xc}^{\text{HEG}}(\rho(1)) F_{xc}^{\text{GGA}}(\rho(1),x(1)) \rho(1) \, d1$,
where the enhancement factor $F_{xc}$ depends both on the local density and upon the 
local reduced gradient, $x = \vert \vec{\nabla} \rho \vert/\rho^{4/3}$.
Relevant GGAs used in our calculations are Becke's 1988 (B88) exchange GGA \cite{B88,FZ91},
Perdew's 1991 correlation GGA \cite{PCV+92,PCV+93,PBW96,PBW98}, and the Lee, 
Yang, and Parr (LYP) correlation GGA \cite{LYP88}.

Up to this point, the discussion has been limited to pure density functionals --- that is, those
that depend only upon the charge density.  In 1993, Becke introduced some ``exact exchange''
into the xc functional \cite{B93} (exact exchange is HF exchange evaluated with DFT orbitals.)  
He did this based upon an adiabatic connection formalism
and the improvement in computational results was very striking in that time as they 
suggested that such (global) hybrid functionals could provide thermochemical accuracy.  Of course,
this also leaves the framework of formal Hohenberg-Kohn-Sham theory \cite{G01} and we must 
now speak of generalized Kohn-Sham theory.  In particular, the xc energy is now a functional
of the density matrix $E[\gamma]$ rather than just the density ($E[\rho]$) and the xc potential
becomes an xc operator defined by ${\hat v}_{xc} \psi(1) = \int (\delta E_{xc}/\delta \gamma(2,1))
\gamma(1,2) \psi_i(2) \, d2$.
Note that the HF exchange-only operator may be regarded as an extreme case of a
global hybrid: $\delta E_{x}^{\text{HF}}[\gamma]/\delta \gamma(2,1) = 
-\frac{\gamma(1,2)}{r_{1,2}}$.
Equation~(\ref{eq:theory2.14}) still holds but with a nonlocal $\hat{v}_{xc}$.
We use the B3LYP functional in the present work \cite{B3LYP94}.  This is the same
as the B3P91 functional originally introduced by Becke \cite{B93} but with Perdew's 1991
correlation GGA replaced by the LYP correlation GGA.  Specifically,
$E_{xc}  =  E_x^{\text{LDA}} + a_0 \left( E_x^{\text{HF}} - E_x^{\text{LDA}} \right)
+  a_x E_x^{\text{B88}}  +  E_c^{\text{LDA}} + a_c \left( E_c^{\text{LYP}} - E_c^{\text{LDA}} 
\right)$, where the various functionals have been defined above, $a_0=0.20$, $a_x=0.72$, and 
$a_c=0.81$.

% \input{./tables/fun.tex}
% =======================================
% File: fun.tex
% Last modified: 10 February 2018
% =======================================

\begin{table}
\caption{Summary of different functionals used in this work.
See Eq.~(\ref{eq:theory2.24}).
\label{tab:fun}}
\begin{tabular}{ccccc}
\hline \hline
Functional & $\mu$ & $a_0$ & $a_x$ & $a_c$ \\
\hline
HF\footnotemark[1] & $+\infty$ & 1 & 0 & 0 \\
LDA                & 0         & 0 & 0 & 1 \\
B3LYP              & 0         & 0.20 & 0.72 & 0.81 \\
CAM-B3LYP          & 0.33      & 0.19 & 0.46 & 0.81 \\
LRC-LDA            & 0.4       & 0 & 1 & 0 \\
\hline \hline
\end{tabular}
\footnotetext[1]{In the HF case, we also need to drop the $E_c^{\text{VWN80}}$ term.}
\end{table}

We have emphasized that, in TD-DFT, charge-transfer excitations require the use of
RSHs \cite{K17}.  These functionals involve the splitting of the 
electron repulsion into a short-range (sr) and a long-range (lr) part.
% \begin{equation}
%     \frac{1}{r_{12}} = \left( \frac{1}{r_{12}} \right)^{(\text{sr})}
%     + \left( \frac{1}{r_{12}} \right)^{(\text{lr})} \, .
%    \label{eq:theory2.21}
% \end{equation}
For convenience in a Gaussian orbital-based program, the separation is made using the error
function, $\text{erf}(\mu r_{12}) = (2/\sqrt{\pi}) \int_0^{\mu r_{12}} e^{-t^2} \, dt$.
We will be using the Coulomb attenuated model (CAM) B3LYP functional \cite{YTH04} where
$(1/r_{12})^{(\text{sr})}  =  \left\{ 1-[a_0 + a_x \text{erf}(\mu r_{12})]\right\}/r_{12}$ and 
$(1/r_{12})^{(\text{lr})} =  \left[ a_0+a_x \text{erf}(\mu r_{12})\right]/r_{12}$,
The sr part of the functional is treated by DFT while the lr part is treated by HF.
The parameter $\mu$ acts as a range-separation parameter.  The specific form of the CAM-B3LYP 
functional is,
\begin{eqnarray}
  E_{xc} & = & E_x^{\text{sr-LDA}} + a_0 [E_x^{\text{lr-HF}}-E_x^{\text{sr-LDA}}]  \nonumber \\
 & + & a_x E_x^{\text{sr-B88}}  \nonumber \\
 & + & E_c^{\text{VWN80}} + a_c [E_c^{\text{LYP88}}-E_c^{\text{VWN80}}]
  \, ,
  \label{eq:theory2.24}
\end{eqnarray}
The specific parameters of the CAM-B3LYP functional are given in Table~\ref{tab:fun}.  Also
shown in the table are the values of the parameters which give some of the other functionals
used in this paper.
The LRC-LDA functional is the LDA form of the long-range corrected (LRC) functional of
Iikura, Tsuneda, Yanai, and Harao \cite{ITYH01}.  This is given because the RSH DFTB 
used in this paper ({\em vide infra}) is based upon the LRC family of RSH functionals,
rather than upon the CAM-B3LYP form. The often recommended optimally-tuned RSHs 
\cite{SKB09,BLS10,KKK13} have not been used here because of the difficulties encountered
in making a corresponding RSH DFTB and also because of problems encountered when using
optimally-tuned RSHs when calculating potential energy surfaces \cite{KKK13} (needed for
a different part of our on-going projects).

The DFT presented thus far still has one very large failure, namely the lack of van der Waals
(vdW)
forces.  This is particularly important in organic electronics because the organic molecules
in the condensed phase are primarily held together precisely by these forces.  However, to
include vdW forces in {\em ab initio} theory, it is necessary to go beyond HF to at
least second order in many-body perturbation theory (MBPT).  Designing a density functional that
can handle vdW forces has been studied and suggestions usually involve some aspect
of MBPT.  As TD-DFT resembles a MBPT method, it is perhaps not so remarkable that 
vdW coefficients for long-range induced-dipole/induced-dipole vdW forces can be calculated 
reasonably accurately via TD-DFT.  The difficulty is then how to use TD-DFT (or some other
MBPT approach) and make a computationally efficient method.  The present method of choice,
and the one used here, is actually a compromise.  This is Grimme's semi-empirical D3 correction \cite{GAEK10}.  
% It has the semi-emperical form,
% \begin{eqnarray}
%   & & E_{\text{vdW}}  =  \sum_{A,B} \sum_{n=6,8,10,\cdots} s_n 
%      \frac{C_n^{A,B}}
%      {
%        R^n_{A,B} 
%         \left[ 
%               1 + 6 \left( 
%                     \frac{R_{A,B}}
%                     {
%                       s_{r,n} R_0^{A,B}}
%      \right)^{-\alpha_n} \right] }  \nonumber \\
%     & - & \sum_{A,B,C} 
%     \frac{
%          \sqrt{C_6^{A,B} C_6^{A,C} C_6^{B,C}} 
%          \left(
%                  3 \cos \theta_a \cos \theta_b \cos \theta_c + 1 
%          \right)
%          }
%          {
%          \left( R_{A,B} R_{B,C} R_{C,A} \right)^3 
%          \left[ 
%           1 + 6 \left( \frac{\bar{R}_{A,B,C}}
%                         {s_{r,3} R_0^{A,B}}
%                  \right)^{-\alpha_3} 
%          \right] 
%          } \, , \nonumber \\
%   \label{eq:theory2.25}
% \end{eqnarray}
% where the $C_6$ and other vdW coefficients 
whose parameters
are obtained from TD-DFT [see Ref.~\onlinecite{GAEK10}
for a more detailed description of the D3 correction].
Thus this correction may be seen as an interpolation scheme
between DFT and TD-DFT.  However $E_{\text{vdW}}$ enters as a correction which does not
enter into the self-consistent cycle of orbital optimization but instead is added on, after
the fact, as a first-order correction to the self-consistent field energy.
Derivatives of $E_{\text{vdW}}$ are included in force calculations and hence in 
geometry optimizations.

% --------------------------------
\section{TD-DFT}
\label{sec:TD-DFT}
% --------------------------------

TD-DFT is the younger sibling of DFT:  The founding papers of DFT \cite{HK64b,KS65}
were written half a century ago; that of TD-DFT \cite{RG84} a mere 30 years or so ago.
The interested reader can find more information on TD-DFT in the proceedings of two
summer schools on the topic \cite{MUN+06,MNN+11} as well as in two textbooks \cite{MS90,U12} and 
in several recent review articles \cite{BWG05,C09,JWPA09,CH12,AJ13,LJ13,M16}.

The most common application of TD-DFT is to the calculation of electronic absorption spectra
via response theory.  There are several ways to do this, including real-time TD-DFT, but 
the classic approach is to use ``Casida's equation'' 
(see, e.g., pp.~145-153 of the recent textbook, Ref.~\onlinecite{U12}, or the original reference \cite{C95}.)
This method is about 20 years old.  It consists of solving the pseudo-eigenvalue problem,
\begin{equation}
  \left[ \begin{array}{cc} {\bf A} & {\bf B} \\
                           {\bf A}^* & {\bf B}^* \end{array} \right]
  \left( \begin{array}{c} \vec{X} \\ \vec{Y} \end{array} \right) = \omega
  \left[ \begin{array}{cc} {\bf 1} & {\bf 0} \\
                           {\bf 0} & -{\bf 1} \end{array} \right]
  \left( \begin{array}{c} \vec{X} \\ \vec{Y} \end{array} \right) \, ,
  \label{eq:theory2.26}
\end{equation}
where $A_{ia,jb}  =  \delta_{i,j} \delta_{a,b} \left( \epsilon_a - \epsilon_i \right)
                  + K_{ia,jb}$ and $B_{ia,bj}  =  K_{ia,bj}$
and the indices refer to excitations from occupied ($i$ and $j$) to unoccupied ($a$ and $b$)
orbitals.  The coupling matrix is usually evaluated in the adiabatic approximation, 
which leads to $ K_{pq,rs}  =  \int \int \int \int \psi_p^*(1) \psi_q(2) 
            \left( f_H(1,2;3,4) + f_{xc}(1,2;3,4) \right) 
             \times  \psi_r(3) \psi_s^*(4) \, d1 d2 d3 d4$,
where $f_H(1,2;3,4)  =  \delta(1-2) \frac{1}{r_{1,3}} \delta(3-4)$ 
and $f_{xc}(1,2;3,4)  =  \frac{\delta E_{xc}}{\delta \gamma(2,1) \delta \gamma(4,3)}$.

The connection with the formal analysis of the previous section is most easily accomplished within
the Tamm-Dancoff approximation (TDA), ${\bf A} \vec{X} = \omega \vec{X}$.
The TDA makes TD-DFT look like configuration interaction with single excitations (CIS) which
is exactly what it is when the TDA is applied to TD-HF. Many of the strengths and shortcomings
of TD-DFT can be understood when the TDA is applied to the TOTEM to get the singlet excitation
energy, $\omega_S$.  In the case of TD-HF,
\begin{equation}
  \omega_S = \epsilon_L^{\text{HF}} - \epsilon_H^{\text{HF}} + 2 (HL \vert f_H \vert LH) - (HH \vert f_H \vert LL) \, , 
  \label{eq:theory2.30}
\end{equation}
where $(pq \vert f_H \vert rs) = \int \int \psi_p^*(1) \psi_q(1) (1/r_{1,2}) 
\psi_r^*(2) \psi_s(2) \, d1 d2$.
As Koopmans' theorem tells us that the HF orbital energies are better suited for describing 
ionization and electron attachment than for describing excitations, this is not such a good
expression for local excitations.  However it is well suited for describing CT excitations
between two neutral molecules separated by a distance $R$ as Eq.~(\ref{eq:theory2.30}) becomes 
roughly the expected formula, $\omega_S = I-A-(1/R)$,
as the ionization potential, $I \approx -\epsilon_H^{\text{HF}}$,
and the electron affinity, $A \approx -\epsilon_L^{\text{HF}}$,
while $\lim_{R \rightarrow \infty} (HL \vert f_H \vert LH)  =  0$ and 
$\lim_{R \rightarrow \mbox{ large }} (HH \vert f_H \vert LL)  \approx  1/R$.
In contrast, for pure density functionals (LDA and GGAs),
\begin{equation}
  \omega_S = \epsilon_L^{\text{DFT}} - \epsilon_H^{\text{DFT}} + 2 (HL \vert f_H \vert LH) 
  - (HL \vert f_{xc}^{\uparrow \uparrow} - f_{xc}^{\uparrow \downarrow} \vert LH) \, , 
  \label{eq:theory2.36}
\end{equation}
where $(pq \vert f_{xc} \vert rs) = \int \int \psi_p^*(1) \psi_q(1) \left(
\delta^2 E_{xc}/\delta \rho(1) \delta \rho(2) \right)$ 
$\psi_r^*(2) \psi_s(2) \, d1 d2$.
Since pure DFT orbitals see the same potential, and hence the same number of electrons, for
both occupied and unoccupied orbitals, then the orbital energy difference is not a bad first
approximation for local excitation energies.  This provides an intuitive explanation of why
TD-DFT often does better than TD-HF in this case.  But, when we consider charge transfer between two widely
separated molecules, Eq.~(\ref{eq:theory2.36}) becomes
$ \lim_{R \rightarrow \mbox{ large }} \omega_S \approx 
\epsilon_L^{\text{DFT}} - \epsilon_H^{\text{DFT}}$,
which not only has the wrong $R$ dependence but often grossly underestimates the difference between
the ionization potential of one molecule and the electron affinity of the other.

Interestingly, exact TD-DFT with pure density functionals circumvents the CT problem 
in at least two different ways.  The first is by introducing
a complicated frequency dependence into the xc-kernel $f_{xc}(\omega)$ which imitates spatial 
nonlocality at particular values of $\omega$.  In particular, the time-dependent exchange-only
optimized effective potential method gives \cite{GS99,HIG09,CH15}, $f_{x}(\epsilon_a - \epsilon_i) 
= (  ai \vert f_H \vert i a)$.  The second is by introducing compensating peaks in the 
xc-potential in low-density regions between widely-separated donor and acceptor molecules 
\cite{HG12,M17}.

While these formal results are highly interesting and at least the first aspect 
can be implemented in an optimized effective potential formalism, the result is too
involved to be of much use in practical applications.
This suggests that it is better to introduce some exact
exchange into our functionals, either through a global hybrid such as B3LYP or via a RSH such
as CAM-B3LYP.  The astute reader will note that the CAM-B3LYP functional has the wrong asymptotic
behavior for CT excitations.  Nevertheless, it remains a popular compromise for calculating 
excitation energies (e.g., see the DFT popularity poll \cite{DFT2016}).
However the LRC-type of RSH functional does have the correct asymptotic behavior and is
the basis of the lc-DFTB functional described below.

% --------------------------------
\section{DFTB}
\label{sec:DFTB}
% --------------------------------

DFT scales formally as ${\cal O}(N^4)$ with the number of atoms $N$ in the system.
Depending upon the functional, its formal scaling may be reduced to ${\cal O}(N^3)$,
but that still limits the usefulness of DFT.  To go to still larger systems or 
resource intensive dynamics calculations, semi-empirical methods which use only
${\cal O}(N^2)$ integrals are useful.  

Semi-empirical methods have a long history in quantum chemistry.  Originally restricted
to the $\pi$ electrons of conjugated systems, semi-empirical methods had been 
extended to all the valence electrons
of a molecule by 1970 (the date of the classic text of Pople and Beveridge on this 
subject~\cite{PB70}).  Ref.~\onlinecite{BJ05} provides a more up-to-date review of 
semi-empirical methods in quantum chemistry.  A constant question with semi-empirical
methods has been the physical meaning of the parameters used and how to assign them 
values.
% -------------
DFTB is specifically designed to approximate
DFT with no more than two-centers integrals and no more than valence orbitals.  
In so doing, ``the use of [DFT] removed at a stroke much of the problem of fitting
parameters''~\cite{H12}.
% -------------
The result resembles a less accurate (because less rigorous) form of DFT.  Nevertheless
the efficiency of DFTB makes it a highly desirable feature and most quantum chemistry
packages include some form of DFTB.  No attempt is made here to give a thorough
review of all the different flavors of DFTB, but we will review the main points
and refer the reader to the literature for additional information~\cite{KM09,OSHD09,ES14}.  
In particular,
a very nice explanation of DFTB is given in Ref.~\onlinecite{KM09}. 

The original form of DFTB \cite{PFK+95}, was a noniterative one-shot calculation
resembling Hoffmann's extended H\"uckel method \cite{H63a,H64b,H64c,H64d}.  The basis
consists of the valence orbitals of isolated atoms calculated in a confining potential
to ensure that those atomic orbitals remained local.  It is important to keep track of
the atom $I$ on which resides the atomic orbital (AO) $\mu$.  We will denote this basis
function as $\chi_{\mu I} = \chi_{\mu \in I}$,
where the left-hand side is a shorter form of the right-hand side.  
The density is the superposition of atomic densities,
$\rho_0 = \sum_I \rho_I^0$.
The nuclear attraction term is separable,$v_{\text{ext}}[\rho_0] = 
\sum_I v_n[\rho_I^0]$,
and the Hartree potential is also separable, $v_H[\rho_0] = \sum_I 
v_H[\rho_I^0]$,
while the xc potential is assumed separable, $v_{xc}[\rho_0] = 
\sum_I v_{xc}[\rho_I^0]$,
which is a reasonable approximation for pure density functionals (LDA and GGAs).  
The matrix elements of the orbital hamiltonian ($\hat{f}$) are calculated as,
\begin{equation}
  f_{\mu I, \nu J} = \left\{ \begin{array}{ccc} \langle \chi_{\mu I} \vert \hat{t} + 
 v_I^{nH\text{xc}}  \vert \chi_{\nu I} \rangle = \epsilon_I^0 & ; & I= J \\
  \langle \chi_{\mu I} \vert \hat{t} +  v_I^{nH\text{xc}} + v_J^{nH\text{xc}} 
  \vert \chi_{\nu J} \rangle  & ; & I \neq J
  \end{array} \right. \, ,
  \label{eq:theory2.45}
\end{equation}
where $\hat{t} = -(1/2) \nabla^2$ 
is the kinetic energy operator and 
$v_I^{nH\text{xc}} = v_n[\rho_I^0] + v_H[\rho_I^0] + v_{\text{xc}}[\rho_I^0]$.
This is known as the ``potential superposition approximation.''  
A popular alternative is the ``density superposition approximation'' where 
$v_{\text{xc}}[\rho_I^0] + v_{\text{xc}}[\rho_J^0]$
is replaced with $v^{\text{xc}}[\rho_I^0+\rho_J^0]$,
hence reducing reliance on the assumption of a separable xc potential.
Note that Eq.~(\ref{eq:theory2.45}) involves only two-center integrals and
$f_{\mu I, \nu I} = \delta_{\mu,\nu} \epsilon_{\mu I}^0$,
where $\epsilon_{\mu I}$ is the AO energy for the isolated atom.   This suffices for setting
up the matrix form of the orbital equation,
${\bf f} \vec{c}_i = \epsilon_i {\bf s} \vec{c}_i$,
and hence to calculate the band structure (BS) part of the total energy,
$E_{\text{BS}} = \sum_i n_i \epsilon_i$.

  Of course, this neglects important terms on the right-hand side of Eq.~(\ref{eq:theory2.14}),
known in DFTB as the repulsion energy,
$E_{\text{rep}} = -\frac{1}{2} E_H + E_{\text{xc}} 
- \int v_{xc}(1) \rho(1) \, d1$.
It is a fundamental tenant of DFTB that this energy can be expanded as a set of pairwise 
potentials between different atom types,
$E_{\text{rep}} = \sum_{I<J} V_{I,J}(R_{I,J})$.
Finding and tabulating good transferable pair repulsion potentials $V_{I,J}(R_{I,J})$ is 
a major part of DFTB.

An important extension of DFTB is the addition of a self-consistent charge (SCC) term,
$E_{\text{coul}}$, accounting for charge density corrections $\delta \rho$ beyond
the original superposition of atomic densities approximation $\rho_0$.  This correction may
be through second- \cite{EPJ+98} or third-order \cite{GCE11}.  For simplicity, we describe only
the second-order correction here.  We seek a semi-empirical  approximation to
$E_{\text{coul}} = \frac{1}{2} \int \int \delta \rho(1) 
\left( f_H(1,2) + f_{xc}(1,2) \right) \delta \rho(2) \, d1 d2$.
This is accomplished by extensive use of two approximations.  The first is Mulliken's
approximation for use in approximating electron repulsion integrals (ERIs)~\cite{M49},
$\chi_{\mu I}^*(1) \chi_{\nu J}(1) \approx \frac{s_{\mu I,\nu J}}{2} 
  \left( \chi_{\mu I}^*(1) \chi_{\mu I}(1) + \chi_{\nu J}^*(1) \chi_{\nu J}(1) \right)$,
which leads to 
$\psi_r^*(1) \psi_s(1) \approx \sum_{\mu I} q_{\mu I}^{r,s} \chi_{\mu I}^*(1) \chi_{\mu I}(1)$,
where
$q_{\mu I}^{r,s} = (1/2) \sum_{\nu J} \left( c_{\mu I,r}^* s_{\mu I,\nu J} c_{\nu J,s}
  + c_{\mu I,s} s_{\mu I,\nu J} c_{\nu J,r}^* \right)
$
is a Mulliken transition charge.  The second approximation is the gamma 
approximation, 
$\chi_{\mu I}^*(1) \chi_{\mu I}(1) \approx g_I(1)$,
where $g_I$ is an $s$-type function centered on atom $I$.  The restriction to 
$s$-type functions is needed to solve a classic rotational invariance problem 
in semi-empirical theories (pp.~60-63 of Ref.~\cite{PB70}).  (In more recent 
programs, $g_I$ may be replaced
with $s$-type functions $g_{I l}$, allowing $g_{I l}$ to be different for different values
of the angular momentum quantum number of $\chi_{\mu I}$ \cite{FSE+02}.)  Together
Mulliken's ERI approximation and the gamma approximation lead to the auxiliary-function
expansion
$\psi_r^*(1) \psi_s(1) \approx \sum_I q_I^{r,s} g_I(1)$,
where $q_I^{r,s} = \sum_{\mu \in I} q_{\mu I}^{r,s}$.
The name ``gamma approximation'' comes from the integral,
\begin{equation}
  \gamma_{I,J} = \int \int g_I(1) \left( f_H(1,2) + f_{xc}(1,2) \right) g_I(2) d1 d2 \, ,
  \label{eq:theory2.53h}
\end{equation}
where $f_H(1,2)  =  (1/r_{1,2})$ and 
$f_{xc}(1,2)  =  \delta^2 E_{xc}/\delta \rho(1) \delta \rho(2)$,
which means that the density is
$\rho(1)  =  \sum_i n_i \psi_i^*(1) \psi_i(1) \approx \sum_I q_I g_I(1)$,
where the Mulliken charge on atom $I$ is
$q_I = \sum_{\mu \in I} q_{\mu I}^{i,i} n_i$.
The second-order SCC becomes
$E_{\text{coul}}({\bf R}) =  \frac{1}{2} \sum_{I,J} \gamma_{I,J}(R_{I,J}) 
\Delta q_I \Delta q_J$.
where $\Delta q_I$ is the Mulliken charge fluxtuation on atom $I$ because
$\delta \rho(1) \approx \sum_I \Delta q_I g_I(1)$.

Sometimes the diagonal
% --------------------------------
elements $\gamma_{I,I}$ are % may be 
calculated using some variation on Pariser's 
observation \cite{P53} that it should be equal to the difference between the ionization potential 
and electron affinity of atom $I$.  This suggests that DFTB may work particularly well for 
calculating ionization potentials and electron affinities and this does seem to be the case
\cite{DCT+15}, although the cited reference points out that DFTB was not parameterized to fit a single
specific property, but rather to behave like DFT across a broad range of properties.  

The SCC correction to the orbital hamilonian matrix,
$f_{\mu I, \nu J} = \frac{s_{\mu I,\nu J}}{2} \sum_K 
\left( \gamma_{J,K} + \gamma_{K,I} \right) \Delta q_K$,
is obtained in the usual way by variational minimization of the energy,
$E = E_{\text{BS}} + E_{\text{rep}} + E_{\text{coul}} + E_{\text{vdW}}$.
Note that we have added Grimme's D3 vdW correction to the DFTB energy 
formula as this is an important addition used in the present paper.  
Adding in this vdW term requires no essential changes in the DFTB formalism.

The construction of global hybrid and RSH versions of DFTB is now straightforward with 
appropriate modifications of the gamma integral.  For example, to make the HF form of
DFTB, both the xc parts of $E_{\text{BS}}[\rho_0]$ and of $E_{\text{rep}}[\delta \rho]$
must be replaced by the semi-empirical forms:
\begin{eqnarray}
  \gamma_{I,J} & = & \int \int g_I(1) f_H(1,2) g_J(2) \, d1 d2 \, , \nonumber \\
  J & = & \sum_{i,j} n_i (ii \vert f_H \vert jj) n_j = \sum_{I,J} q_I \gamma_{I,J} q_J
  \, , \nonumber \\
  K & = & \sum_{i,j} n_i (ij \vert f_H \vert ji) n_j = \sum_{I,J} \gamma_{I,J}
  \left( \sum_{i,j} n_k q_I^{i,j} q_J^{j,i} n_j \right) \, . \nonumber \\
  \label{eq:theory2.57b}
\end{eqnarray}
The first version of RSH DFTB seems to be that of Niehaus and Della Sella \cite{ND12}.
% --------------------------------------------
This was followed by an implementation by Niehaus \cite{LAN15} (see also Ref.~\onlinecite{VKK+18}) and by Humeniuk and 
Mitri\'{c} \cite{HM15,HM17}.  We will be using the latter form which was parameterized 
to behave like a LRC version of their DFTB method which, itself, was parameterized
to behave like PBE DFT.  We will refer to the resultant method as lc-DFTB
following Humenuik and Mitri\'{c} and we will see that it behaves not unlike 
CAM-B3LYP DFT.  The long-range $\gamma$-matrix is given by,
$\gamma^{\text{lr}}_{I,J}(R_{I,J};\mu) = \text{erf} \left( \mu R_{I,J}
  \right) \gamma_{I,J}(R_{I,J})$,
where $\mu = 1/R_{\text{max}}$ is the usual range-separation parameter.
Humeniuk and Mitri\'{c} also neglect the long-range contribution to
$E_{BS}$ on the grounds that the zero-order system ``already accounts
for all interactions between electrons in the neutral atoms'' \cite{HM17}.

% --------------------------------
\section{TD-DFTB}
\label{sec:TD-DFTB}
% --------------------------------
   
Niehaus {\em et al.} were the first to extend DFTB to TD-DFTB  
\cite{TSD+01,FSE+02,HNWF07,N09,DAF+13}.
This is greatly facilitated by the observation that the TD-DFT coupling matrix is 
already approximated in the $E_{\text{coul}}$ term of TD-DFTB.  This allows the
${\bf A}$ and ${\bf B}$ matrices in Casida's equation to be written out in 
TD-DFTB form.  
The TD-DFTB coupling matrix is given by
$K_{pq,rs} = \sum_{I,J} q_I^{p,q} \gamma_{I,J} q_J^{s,r}$,
with $\gamma_{I,J}$ defined as in Eq.~(\ref{eq:theory2.53h}).  For a detailed 
treatment of spin and separation into singlet and triplet blocks, see e.g., 
Ref.~\cite{NSD+01}.  The dipole matrix elements needed to calculate oscillator
strengths are calculated as
$\langle \psi_p \vert \vec{r} \vert \psi_q \rangle = 
\sum_I \vec{R}_I q_I^{p,q}$.
Other implementations include that of Ref.~\onlinecite{TSZ+11}.  
In this work we used the TD-lc-DFTB described in Refs.~\onlinecite{HM15,HM17}.

%%%%%%
% EOF
%%%%%%

% -----------------------------------------------
% \newpage
% \singlespace
% \bibliographystyle{myaip}
% \bibliography{refs}

% -----------------------------------------------
% \clearpage
% \input{tables.tex}
% -----------------------------------------------
% Please compile a list of all figure captions on a separate page:
% \input{figures.tex}
% ------------------------------------------------
\end{document}